\documentclass[prd,aps,nofootinbib,preprintnumbers,showpacs,superscriptaddress,twocolumn]{revtex4-1}
\usepackage{latexsym}
\usepackage{amssymb}
\usepackage{amsfonts}
\usepackage{amsmath}
\usepackage{bm}
\usepackage{graphicx}
\usepackage{color}
\usepackage{subfigure}
\usepackage{times}
\usepackage{units}
\usepackage{hyperref}

\newcommand{\be}{\begin{equation}}
\newcommand{\ee}{\end{equation}}
\newcommand{\bea}{\begin{eqnarray}}
\newcommand{\eea}{\end{eqnarray}}
\newcommand{\beqn}{\begin{eqnarray*}}
\newcommand{\eeqn}{\end{eqnarray*}}
\newcommand{\ba}{\begin{align}}
\newcommand{\ea}{\end{align}}

\DeclareMathOperator{\tr}{\ensuremath{\mathrm{tr}}}

\newcommand{\ie}{\textit{i.e.,}~}
\newcommand{\eg}{\textit{e.g.,}~}
\newcommand{\ms}{{\rm ms}}

\def\ms{{\rm ms}}

\begin{document}

%%%%%%%%%%%%%%%%%%%%%%%%%%%%%%%%%%%%%%%%%%%%%%%%%%%%%%%%%%%%%%%%%%%%%%%

\title{General-relativistic resistive magnetohydrodynamics in three
  dimensions: Formulation and tests}

%%%%%%%%%%%%%%%%%%%%%%%%%%%%%%%%%%%%%%%%%%%%%%%%%%%%%%%%%%%%%%%%%%%%%%%

\author{Kyriaki \surname{Dionysopoulou}}
\affiliation{Max-Planck-Institut f\"ur Gravitationsphysik, Albert-Einstein-Institut, Potsdam, 14476, Germany}

\author{Daniela \surname{Alic}}
\affiliation{Max-Planck-Institut f\"ur Gravitationsphysik, Albert-Einstein-Institut, Potsdam, 14476, Germany}

\author{Carlos \surname{Palenzuela}}
\affiliation{Canadian Institute for Theoretical Astrophysics, Toronto, Ontario M5S 3H8, Canada}

\author{Luciano \surname{Rezzolla}}
\affiliation{Max-Planck-Institut f\"ur Gravitationsphysik, Albert-Einstein-Institut, Potsdam, 14476, Germany}
\affiliation{Institute f\"ur Theoretische Physik, Frankfurt am Main 60438, Germany}
\affiliation{Department of Physics and Astronomy, Louisiana State University, Baton Rouge, Louisiana 70803, USA}

\author{Bruno \surname{Giacomazzo}}
\affiliation{JILA, University of Colorado and National Institute of Standards and Technology, 440 UCB, Boulder, Colorado 80309, USA}

\begin{abstract}
We present a new numerical implementation of the general-relativistic
resistive magnetohydrodynamics (MHD) equations within the
\texttt{Whisky} code. The numerical method adopted exploits the
properties of implicit-explicit Runge-Kutta numerical schemes to treat
the stiff terms that appear in the equations for large electrical
conductivities. Using tests in one, two, and three dimensions, we show
that our implementation is robust and recovers the ideal-MHD limit in
regimes of very high conductivity. Moreover, the results illustrate
that the code is capable of describing scenarios in a very wide range
of conductivities. In addition to tests in flat spacetime, we report
simulations of magnetized nonrotating relativistic stars, both in the
Cowling approximation and in dynamical spacetimes. Finally, because of
its astrophysical relevance and because it provides a severe testbed
for general-relativistic codes with dynamical electromagnetic fields,
we study the collapse of a nonrotating star to a black hole. We show that also in this case our results on the quasinormal mode frequencies of the excited electromagnetic fields in the Schwarzschild
  background agree with the perturbative studies within $0.7\%$ and
  $5.6\%$ for the real and the imaginary part of the $\ell=1$ mode
  eigenfrequency, respectively. Finally we provide an estimate of the
  electromagnetic efficiency of this process.

\end{abstract}


\maketitle

%%%%%%%%%%%%%%%%%%%%%%%%%%%%%%%%%%%%%%%%%%%%%%%%%%%%%%%%%%%%%%%%%%%%%%%%

\section{Introduction}
\label{sec:intro}

Magnetic fields play an important role in several astrophysical
scenarios, many of which involve also the presence of compact objects
such as neutron stars (NSs) and black holes (BHs), whose accurate
description requires the numerical solution of the equations of
general relativistic magnetohydrodynamics (GRMHD). 

In most of these phenomena, such as for the interior dynamics of
magnetized stars, or for the accretion of matter onto BHs, the
electrical conductivity of the plasma is extremely high and the
ideal-MHD approximation, in which the conductivity is actually assumed
to be infinite, represents a very good approximation. In this case,
the magnetic flux is conserved and the magnetic field is frozen in
the fluid, being simply advected with it. Following this
approximation, several numerical codes solving the equations of
general-relativistic ideal-MHD have been developed over the years
\cite{Komissarov1999, Koide98, DelZanna2003, Gammie03, Anninos05c,
  Duez05MHD0, Shibata05b, Anton06, Neilsen2005, DelZanna2007,
  Giacomazzo:2007ti, Farris08, Zink2011}. By construction, therefore,
the ideal-MHD equations neglect any effect of resistivity on the
dynamics. In practice, however, even in the scenarios mentioned above,
there will be spatial regions with very hot plasma where the
electrical conductivity is finite and the resistive effects, most
notably, magnetic reconnection, will occur in reality. Such effects
are expected to take place, for example, during the merger of two
magnetized NSs or of binary system composed by a NS and a BH, or near
the accretion disks of active galactic nuclei (AGN), and could provide an important
contribution to the energy losses from the system.

The importance of resistivity effects can be easily deduced from the
evolution of a current sheet in high but finite conductivity. Under these
conditions, several instabilities can take place in the plasma and
release substantial amounts of energy via magnetic
reconnection~\cite{Biskamp1986}, as frequently observed, for example, in
solar flares~\cite{Parker1987}. The study of reconnection in relativistic
phenomena is instead important to try to explain the origin of flares in
relativistic sources, such as blazars~\cite{Giannios2009} and
magnetars~\cite{Lyutikov2006}. It is not surprising then that several
groups have developed in the recent years numerical codes to solve the
equations of special and general relativistic resistive
MHD~\cite{Komissarov2007, Palenzuela:2008sf, Dumbser2009, Zenitani2010,
  Takamoto2011b, Zanotti2011b, Bucciantini2012}.

There are several processes involving compact objects, such as NSs and
BHs, where resistive effects could play an important role. These include
the interaction of the magnetospheres of two NSs in a binary before the
merger, the stability of the magnetosphere that may be produced around
the hypermassive neutron star (HMNS), or the stability of the magnetic
field within the torus that will be produced once the HMNS collapses to a
black hole. In all of these scenarios, the ideal-MHD limit may not be
sufficient to study those physical processes which involve reconnection
or the presence of anisotropic resistivities. So far, the problem of
dealing with regions which are magnetically dominated (i.e. with small
ratio of fluid pressure over magnetic pressure) has been avoided by
burying the magnetic fields inside the stars, where the ideal-MHD limit
is a very good approximation~\cite{Anderson2008, Etienne08,
  Giacomazzo:2009mp, Chawla:2010sw, Giacomazzo:2010, Rezzolla:2011,
  Etienne2012} and therefore neglecting any effect that could come from
the magnetic field evolution in the NS magnetosphere. To improve on the
ideal-MHD description, it is possible to employ the equations of general
relativistic resistive MHD. These equations provide a complete MHD
description and a mathematical framework that can be used to study both
the regions with a high conductivity, such as the NS interior, the
magnetosphere (if present), accretion disks, etc, and regions with small
conductivity, such as the magnetosphere exterior in
electrovacuum. Moreover, when the conductivity is set to zero, Maxwell
equations in vacuum are recovered~\cite{Palenzuela:2008sf}, thus allowing
for the study of the magnetic field evolution also well outside the NS
magnetosphere. This is particularly important, since several recent works
have claimed that the interaction of magnetic fields surrounding BNS and
NS-BH systems may lead to strong electromagnetic
emissions~\cite{McWilliams2011}, and even affect the dynamics of these
systems (see~\cite{Piro2012}, but also~\cite{Lai2012} for a different
conclusion). In order to verify such predictions, it is therefore
important to be able to accurately follow the dynamics of the magnetic
fields in the region surrounding these compact binaries and this cannot
be done in the limit of ideal-MHD. Last but not least, binary mergers are
also thought to be behind the central engine of short gamma-ray bursts
(GRBs)~\cite{Paczynski86,Eichler89,Narayan92,Rezzolla:2011} and the
accurate study of the magnetic field both before and after merger could
provide insights on current observations.

We present the first fully general-relativistic resistive MHD code in
a 3+1 decomposition of spacetime. We extended the ideal GRMHD
\texttt{Whisky} code to include the general relativistic version of
the resistive MHD formalism presented in
Ref.~\cite{Palenzuela:2008sf}. This new version of the
~\texttt{Whisky} code can handle different values of the conductivity
going from the ideal MHD limit (for very high conductivities) to
resistive and electrovacuum regimes (obtained respectively with low
and zero conductivity)\footnote{Hereafter, when referring to
  conductivity we will effectively refer to \emph{isotropic}
  conductivity.}. The code implements state-of-the-art numerical
techniques and has been tested in both fixed and dynamical
spacetimes. In particular we show the first fully general relativistic
simulation of a magnetized NS collapse to BH using resistive MHD to
accurately follow the dynamics of magnetic fields both inside and
outside the NS.

The paper is organized as follows. In Sec.~\ref{math} we describe
the general relativistic resistive MHD equations, in
Sec.~\ref{numerical} the main numerical methods used to solve them,
and in Sec.~\ref{tests} our numerical tests. In Sec.~\ref{conclusions}
we summarize and conclude.

Throughout this paper we use a spacelike signature of $(-,+,+,+)$ and
a system of units in which $c=G=M_\odot =1$. We also note that a factor $2.03\times 10^{5}\,{\rm s}^{-1}$ is needed to convert the conductivity $\sigma$ to cgs units, while a factor $1.456\times 10^{3}\,{\rm cm}^{5}\,{\rm g}^{-1}\,{\rm s}^{-2}$ is needed to convert the polytropic $K$ to cgs units. Greek indices are taken
to run from 0 to 3, Latin indices from 1 to 3 and we adopt the
standard convention for the summation over repeated indices.

%%%%%%%%%%%%%%%%%%%%%%%%%%%%%%%%%%%%%%%%%%%%%%%%%%%%%%%%%%%%%%%%%%%%%%%%

\section{Mathematical Setup}
\label{math}

We next describe our extension of the special-relativistic resistive MHD
formalism presented in Ref.~\cite{Palenzuela:2008sf} to a general
relativistic MHD framework. A similar (but independent) extension has
been presented recently in~\cite{Bucciantini2012}, which describes the
first 3+1 general-relativistic resistive MHD implementation in fixed
spacetimes.

%%%%%%%%%%%%%%%%%%%%%%%%%%%%%%%%%%%%%%%%%%%%%%%%%%%%%%%%%%%%%%%%%%%%%%%%

\subsection{The magnetohydrodynamic equations}

The complete set of relativistic MHD equations result from the
combination of the conservation of rest mass
\begin{eqnarray}\label{baryon_conservation}
  \nabla_{\mu} ( \rho u^{\mu} ) = 0,
\end{eqnarray}
and the conservation of energy and momentum 
\begin{eqnarray}
\nabla_{\nu} T^{\mu\nu} = 0. \label{Tconservation}
\end{eqnarray}
The stress-energy tensor for a magnetized perfect fluid is given by
\begin{eqnarray}
\label{stress-energy-perfectfluid}
   T_{\mu \nu} &\equiv& \left[ \rho (1 + \epsilon) + p \right] u_{\mu} u_{\nu} 
+ p g_{\mu \nu} + {F_{\mu}}^{\lambda} F_{\nu \lambda}
   \nonumber\\ 
& & 
     - \frac{1}{4} g_{\mu \nu} ~ F^{\lambda \alpha} F_{\lambda \alpha},
\end{eqnarray}
where the rest mass density $\rho$, the specific internal energy
$\epsilon$, the pressure $p$ and the velocity $u^{\mu}$ describe the
state of the fluid, and are usually referred to as the ``primitive''
variables. We write the pressure $p$ as a function $p=p(\rho,\epsilon)$ and it is a property of the type
of fluid considered.

The velocity of the fluid can be decomposed as
\begin{eqnarray}\label{velocity_decomposition}
   u^{\mu} = W (n^{\mu} + v^{\mu}),
\end{eqnarray}
where $v^{\mu}$ corresponds to the three-dimensional velocity measured by
Eulerian observers moving along a four-vector $n_{\mu}$ normal to the
spacelike hypersurface in a 3+1 decomposition of spacetime (i.e.,
$v^{\mu} n_{\mu} = 0$). Notice that the time component of the
four-velocity is not independent due to the normalization relation
$u^{\mu} u_{\mu} = -1$, so that
\begin{eqnarray}\label{velocity_decomposition2}
   W & \equiv & - n_{\mu} u^{\mu} = (1 - v_i v^i)^{-1/2}, \nonumber\\
  u^{i} &=& W \left(v^{i} - \frac{\beta^i}{\alpha}\right),
\end{eqnarray}
where $W$ is the Lorentz factor.

The 3+1 decomposition of the conservation laws \eqref{Tconservation},
\eqref{stress-energy-perfectfluid} provides the evolution equations
for the fluid variables $D, U, S_{i}$, which come from the following
projections of the stress-energy tensor
\begin{eqnarray}
\label{Tmunu_decomposition2}
   D &\equiv& \rho W ,~~~ \\
   U &\equiv& h W^2 - p + \frac{1}{2} (E^2 + B^2) ,~~~ \\
   S_{i} &\equiv& h W^2 v_{i} + \epsilon_{ijk} E^j B^k ,~~~ \\
   S_{ij} &\equiv& h W^2 v_{i} v_{j} + \gamma_{ij} p 
 -E_i E_j - B_i B_j + \nonumber \\ 
& & \frac{1}{2} \gamma_{ij} (E^2 + B^2)~~,
\end{eqnarray}
where $\gamma_{ij}$ is the usual spatial part of the metric and where
we have introduced the specific enthalpy $h=\rho (1+\epsilon) +
p$. The conserved rest-mass density $D$, the energy density $U$ and
the momentum $S_{i}$ are usually referred to as the ``conserved''
quantities since they can be shown to satisfy conservation laws in
flat spacetimes~\cite{Banyuls97}. In general, it is more convenient to
describe the energy conservation in terms of the quantity $\tau = U -
D$, which allows to recover the Newtonian limit of the energy
density. 

%%%%%%%%%%%%%%%%%%%%%%%%%%%%%%%%%%%%%%%%%%%%%%%%%%%%%%%%%%%%%%%%%%%%%%%%

\subsection{The Maxwell equations}

Given a four-metric tensor $g_{\mu\nu}$, the dynamics of the
electromagnetic fields is described by the extended Maxwell
equations~\cite{Dedner:2002,Palenzuela:2008sf}
\begin{eqnarray}
  \nabla_{\nu} (F^{\mu \nu} + g^{\mu \nu} \psi) &=& I^{\mu} - \kappa
  n^{\mu} \psi ,
\label{Maxwell1} \\
 \nabla_{\nu} (^*F^{\mu \nu} + g^{\mu \nu} \phi) &=& - \kappa n^{\mu}
 \phi ,
\label{Maxwell2} 
\end{eqnarray}
where $F^{\mu \nu}$ is the Maxwell tensor, $^{*\!}F^{\mu \nu}$ is the
Faraday tensor, $I^{\mu}$ is the electric current and $\phi,\,\psi$ are
two auxiliary scalar variables added to the Maxwell equations to
  control the constraints for the magnetic and electric part,
  respectively. We note that the use of Eq.~\eqref{Maxwell2} allows us not to use other constraint-preserving approaches such as the constrained-transport schemes described in~\cite{Toth2000}. In vacuum or
highly magnetized plasmas, where the electric and magnetic
susceptibilities of the medium vanish, the Faraday tensor can be written
as the dual of the Maxwell tensor,
\begin{eqnarray}
^{*\!}F^{\mu \nu} &=& \frac{1}{2}
\epsilon^{\mu\nu\alpha\beta} F_{\alpha\beta},
\end{eqnarray}
with $\epsilon^{\mu\nu\alpha\beta} \equiv \eta^{\mu\nu\alpha\beta} /
{\sqrt{-g}}$, with $\eta^{\mu\nu\alpha\beta}$ being the Levi-Civita
symbol, and $g$ the determinant of the four-metric. These tensors can be
decomposed in terms of the electric and magnetic fields measured by an
observer moving along a normal direction $n^{\nu}$ as
\begin{eqnarray}
F^{\mu\nu} &=& n^{\mu} E^{\nu} - n^{\nu} E^{\mu} +
\epsilon^{\mu\nu\alpha\beta} B_{\alpha} n_{\beta}, \\
^{*\!}F^{\mu\nu} &=& n^{\mu} B^{\nu} - n^{\nu} B^{\mu} -
\epsilon^{\mu\nu\alpha\beta} E_{\alpha} n_{\beta}.
\end{eqnarray}

Following the same decomposition, the electric current $I^{\mu}$ can
be written as
\begin{eqnarray}\label{current1}
I^{\mu} = n^{\mu} q + J^{\mu},
\end{eqnarray}
where $q$ and $J^{\mu}$ are the charge density and the current for an
observer moving along $n^{\mu}$, respectively. Using these definitions
and performing a 3+1 decomposition of the Eqs.~\eqref{Maxwell1},
\eqref{Maxwell2}, \eqref{current1} with respect to the normal vector
$n^{\nu}$, we arrive to the following evolution equations:
\begin{eqnarray}
  (\partial_t - {\cal L}_{\beta}) E^{i} &-& \epsilon^{ijk} \nabla_j (\alpha B_k) 
   + \alpha \gamma^{ij} \nabla_j \psi =  \nonumber\\ 
& & \alpha \tr K E^i - \alpha J^i, 
\label{maxwellext_3+1_eq1a} \\
  (\partial_t - {\cal L}_{\beta}) \psi &+& \alpha \nabla_i E^i = 
    \alpha q -\alpha \kappa \psi,
\label{maxwellext_3+1_eq1b} \\
  (\partial_t - {\cal L}_{\beta}) B^{i} &+& \epsilon^{ijk} \nabla_j (\alpha E_k) 
   + \alpha \gamma^{ij} \nabla_j \phi = \nonumber\\ 
& & \alpha \tr K B^i, 
\label{maxwellext_3+1_eq1c} \\
  (\partial_t - {\cal L}_{\beta}) \phi &+& \alpha \nabla_i B^i = 
   -\alpha \kappa \phi,
\label{maxwellext_3+1_eq1d}
\end{eqnarray} 
where the scalar fields $\phi, \psi$ measure the deviation from the
constrained solution. More specifically, the scalar $\phi$ drives the
solution of Eq.~\eqref{maxwellext_3+1_eq1d} towards the zero-divergence
condition $\nabla_i B^i=0$, while the scalar $\psi$ drives the solution
of Eq.~\eqref{maxwellext_3+1_eq1b} towards the condition $\nabla_i
E^i=q$. This driving is exponentially fast and over a timescale
$1/\kappa$. This approach, named hyperbolic divergence cleaning in the
context of ideal MHD, was proposed in Ref.~\cite{Dedner:2002} as a simple
way of solving the Maxwell equations and enforcing the conservation of
the divergence-free condition for the magnetic field and has been
extended to the resistive relativistic case in
Ref.~\cite{Palenzuela:2008sf}. More on the notation of
Eqs.~\eqref{maxwellext_3+1_eq1a}--\eqref{maxwellext_3+1_eq1d}:
$\mathcal{L}$ is the Lie derivative along the shift vector $\beta^i$,
while $\alpha$ is the lapse function in a standard $3+1$ decomposition of
spacetime~\cite{Alcubierre:2008}.

A consequence of the Maxwell equations is the conservation of electric charge
\begin{eqnarray}
\nabla_{\mu} I^{\mu} = 0,
\end{eqnarray}
which provides an evolution equation for the charge density
\begin{eqnarray}\label{charge}
(\partial_t - \mathcal{L}_{\beta}) q + \nabla_i (\alpha J^i) = \alpha\, K q.
\end{eqnarray}
%

%%%%%%%%%%%%%%%%%%%%%%%%%%%%%%%%%%%%%%%%%%%%%%%%%%%%%%%%%%%%%%%%%%%%%%%%

Finally, a relation for the current as a function of the other fields is
needed in order to close the system. Ohm's law provides a prescription
for the spatial conduction current. For simplicity, and because we are
here interested mostly in idealized tests, we will consider here an
isotropic scalar Ohm law
\begin{eqnarray}
\label{Ohm_law}
J^i = q v^i + W \sigma [E^i + \epsilon^{ijk} v_j B_k - (v_k E^k) v^i],
\end{eqnarray}
where the conductivity $\sigma$ is chosen to be either a constant or a
function of the rest-mass density. We note that this prescription is far
from being realistic and normally a more general, tensorial conductivity
prescription $\sigma_{\mu\nu}=\sigma_{\mu\nu}(D,E,B)$ is to be sought,
starting from microphysical considerations (see~\cite{Andersson2012} for
a recent discussion).

%%%%%%%%%%%%%%%%%%%%%%%%%%%%%%%%%%%%%%%%%%%%%%%%%%%%%%%%%%%%%%%%%%%%%%%%

\subsection{The full set of evolution equations}

Combining the MHD and Maxwell equations we obtain the following set of
evolution equations, which we write in flux-conservative form as
\begin{widetext}
\begin{eqnarray}
\partial_t (\sqrt{\gamma} B^i) &+& \partial_k(-\beta^k \sqrt{\gamma} B^i
+ \alpha \epsilon^{ikj} \sqrt{\gamma} E_j) = -\sqrt{\gamma} B^k
(\partial_k \beta^i) - \alpha \sqrt{\gamma} \gamma^{ij} \partial_j
\phi , \label{magnetic_evol}\\
\partial_t (\sqrt{\gamma} E^i) &+& \partial_k(-\beta^k \sqrt{\gamma} E^i
- \alpha \epsilon^{ikj} \sqrt{\gamma} B_j) = -\sqrt{\gamma} E^k
(\partial_k \beta^i) - \alpha \sqrt{\gamma} \gamma^{ij} \partial_j \psi
- \alpha \sqrt{\gamma} J^i,\label{electric_evol} \\
\partial_t \phi &+& \partial_k (-\beta^k \phi + \alpha B^k) = -\phi
(\partial_k \beta^k) + B^k (\partial_k \alpha) - \frac{\alpha}{2}
(\gamma^{lm} \partial_k \gamma_{lm}) B^k - \alpha \kappa \phi,\label{psib_evol} \\
\partial_t \psi &+& \partial_k (-\beta^k \psi + \alpha E^k) = -\psi
(\partial_k \beta^k) + E^k (\partial_k \alpha) - \frac{\alpha}{2}
(\gamma^{lm} \partial_k \gamma_{lm}) E^k + \alpha q - \alpha \kappa
\psi,  \\
\partial_t (\sqrt{\gamma} q) &+& \partial_k [\sqrt{\gamma} (-\beta^k q
+ \alpha J^k)] = 0, \label{charge_evol}\\
\partial_t (\sqrt{\gamma} D) &+& \partial_k [\sqrt{\gamma} (-\beta^k D
+ \alpha v^k D)] = 0, \label{rest_mass_evol}\\
\partial_t (\sqrt{\gamma} \tau) &+& \partial_k \{\sqrt{\gamma}
[-\beta^k \tau + \alpha ( S^k - v^k D)]\} = \sqrt{\gamma} (\alpha
S^{lm} K_{lm} - S^k \partial_k \alpha), \label{energy_density_evol}\\
\partial_t (\sqrt{\gamma} S_i) &+& \partial_k [\sqrt{\gamma}
(-\beta^k S_i + \alpha  S^k{}_i)] = \sqrt{\gamma} \left[\frac{\alpha}{2}
S^{lm} \partial_i \gamma_{lm} + S_k \partial_i \beta^k - (\tau +
D) \partial_i \alpha \right]. \label{momentum_evol}
\end{eqnarray}
\end{widetext}
%

%%%%%%%%%%%%%%%%%%%%%%%%%%%%%%%%%%%%%%%%%%%%%%%%%%%%%%%%%%%%%%%%%%%%%%%%

\section{Numerical Setup}
\label{numerical}

This new version of the \texttt{Whisky} code implements several
numerical methods that have been successfully used in its ideal-MHD
version~\cite{Giacomazzo:2007ti,Giacomazzo:2010}, but it also
implements new numerical algorithms which are instead needed in order
to handle the evolution in time of the resistive MHD equations. Here
we briefly summarize the numerical methods that are in common with the
ideal-MHD version of \texttt{Whisky}~\cite{Pollney:2007ss,
  Giacomazzo:2007ti,Giacomazzo:2009mp, Giacomazzo:2010}, while in the
following section we provide a more detailed description of the new
algorithms that have been implemented.

The evolution of the spacetime is obtained using the \texttt{CCATIE}
code, a three-dimensional finite-differencing code providing the
solution of a conformal traceless formulation of the Einstein
equations~\cite{Pollney:2007ss}. The general-relativistic RMHD
Eqs.~\eqref{magnetic_evol}-\eqref{psib_evol} and~\eqref{rest_mass_evol}-\eqref{momentum_evol} are solved instead using high-resolution shock-capturing
schemes (HRSC)~\cite{Toro99}. As its ideal-MHD counterpart, also the
\texttt{WhiskyRMHD} code implements several reconstruction methods,
such as total-variation-diminishing (TVD) methods,
essentially-non-oscillatory (ENO) methods~\cite{Harten87} and the
piecewise parabolic method (PPM)~\cite{Colella84}. The Harten--Lax--van
Leer--Einfeldt (HLLE) approximate Riemann solver~\cite{Harten83} has
been used to compute the fluxes in all the results presented
here. Since the code is based on the \texttt{Cactus}~\cite{cactus_url}
computational framework, it can also use adaptive mesh refinement
(AMR) via the \texttt{Carpet} driver~\cite{Schnetter-etal-03b}.

%%%%%%%%%%%%%%%%%%%%%%%%%%%%%%%%%%%%%%%%%%%%%%%%%%%%%%%%%%%%%%%%%%%%%%%%

\subsection{IMEX Runge-Kutta methods}

The general-relativistic RMHD equations in high-conductivity media
contain stiff terms which make the time evolution with an explicit
time integrator very inefficient, if not impossible. The prototype of
the stiff system of partial differential equations can be written as
\begin{eqnarray}
\label{stiff_equation}
    \partial_t {\bf U} = F({\bf U}) + \frac{1}{\varepsilon} R({\bf U}),
\end{eqnarray}
where {\bf U} is the state vector composed of all the evolution variables in Eqs.~\eqref{magnetic_evol}--\eqref{momentum_evol}, and $\varepsilon \equiv 1/\sigma >0$ is the relaxation time. In the
limit of $\varepsilon \rightarrow \infty$, the second term on the
right-hand side of Eq.~\eqref{stiff_equation} becomes negligible and the system
is then hyperbolic with a spectral radius $c_h$ (i.e., with $c_h$
being the absolute value of the maximum eigenvalue). In the opposite
limit of $\varepsilon \rightarrow 0$ the first term on the right-hand
side of Eq.~\eqref{stiff_equation} vanishes and the system is clearly
stiff, since the timescale $\varepsilon$ of the relaxation (or stiff
term) $R({\bf U})$ is very different from the speeds $c_h$ of the
hyperbolic (or non-stiff) part $F({\bf U})$.

Stiff systems of this type can be solved efficiently by a combination
of implicit and explicit time integrators. In particular, the IMEX
Runge-Kutta scheme consists in applying an implicit discretization to
the stiff terms and an explicit one to the non-stiff terms. When
applied to the system (\ref{stiff_equation}) it takes the
form~\cite{pareschi_2005_ier}
\begin{eqnarray}\label{IMEX}
   {\bf U}^{(i)} = {\bf U}^n &+& \Delta t \sum_{j=1}^{i-1} {\tilde{a}}_{ij} F({\bf U}^{(j)}),
\nonumber \\
     &+& \Delta t  \sum_{j=1}^{N} a_{ij} \frac{1}{\varepsilon} R({\bf U}^{(j)}) \\
  {\bf U}^{n+1} = {\bf U}^n &+& \Delta t \sum_{i=1}^{N} {\tilde{\omega}}_{i} F({\bf U}^{(i)})
     + \Delta t  \sum_{i=1}^{N} \omega_{i} \frac{1}{\varepsilon} R({\bf U}^{(i)}),
\nonumber
\end{eqnarray}
where ${\bf U}^{(i)}$ are the auxiliary intermediate values of the
Runge-Kutta time integrator. The matrices $\tilde{A}= (\tilde{a}_{ij})$,
$\tilde{a}_{ij} = 0$ for $j \geq i$ and $A= (a_{ij})$, are $N \times N$
matrices such that the resulting scheme is explicit in $F$ and implicit
in $R$. An IMEX Runge-Kutta scheme is characterized by these two matrices
and the coefficient vectors $\tilde{\omega}_i$ and $\omega_i$, \eg,
$\tilde{\omega}_3=(0,1,0,0)$ and $\omega_3=(0,1-a,a,0)$. Since the
simplicity and efficiency of solving the implicit part at each step is of
great importance, it is natural to consider diagonally-implicit
Runge-Kutta schemes for the stiff terms, i.e., ($a_{ij}=0$ for $j
> i$). The matrices of coefficients are reported in Table \ref{SSP3-433}.

\begin{table}[h]
\caption{Tableau for the ``explicit'' matrix $\tilde A$ (left) and for
  the ``implicit'' matrix $A$ (right) in a IMEX-SSP3(4,3,3) L-stable
  scheme.}
\begin{minipage}{1.4in}
\begin{tabular} {c c c c c c}
 0   & \vline & 0  &  0  &  0  & 0 \\
 0   & \vline & 0  &  0  &  0  & 0 \\
 1   & \vline & 0  &  1  &  0  & 0 \\
 1/2 & \vline & 0  & 1/4 & 1/4 & 0 \\
\hline 
   & \vline &  0 & 1/6 & 1/6 & 2/3 \\
\end{tabular}
\end{minipage}
\begin{minipage}{1.8in}
\begin{tabular} {c c c c c c}
  $a$   & \vline & $a$  &  0  &  0   & 0 \\
 0   & \vline & $-a$  &  $a$  &  0   & 0 \\
 1   & \vline & 0  &  $1-a$  &  $a$  & 0 \\
 1/2 & \vline & $b$ & $c$ & $1/2-b-c-a$ & $a$ \\
\hline 
   & \vline &  0 & 1/6 & 1/6 & 2/3 \\
\end{tabular}
\end{minipage}
\begin{eqnarray}
 a &=& 0.24169426078821~,~b = 0.06042356519705~,~ \nonumber \\
 c &=& 0.12915286960590 \nonumber
\end{eqnarray}
\label{SSP3-433}
\end{table}

Our approach to the solution of the potentially stiff set of
general-relativistic RMHD equation consists therefore in the use of the
IMEX RK method introduced above, with a third-order RK integrator. For
the particular set of Eqs.  \eqref{magnetic_evol}--\eqref{momentum_evol},
the evolved fields can be split into stiff terms ${\bf V}= \{ ~ \sqrt{\gamma}~E^i \}$ and
into nonstiff terms ${\bf W} = \{ ~ \sqrt{\gamma}~B^i, \psi, \phi, ~ \sqrt{\gamma}~q, ~ \sqrt{\gamma}~\tau, ~ \sqrt{\gamma}~S_i, ~ \sqrt{\gamma}~D \}$.

The evolution of the electric field \eqref{electric_evol} can become
stiff depending on the value of the conductivity $\sigma =
1/\varepsilon$ in the Ohm law \eqref{Ohm_law}. Its right-hand side can
be split into potentially stiff terms and regular ones,
\begin{eqnarray}\label{split}
    \partial_t (\sqrt{\gamma} E^i) &=& F^i_E + R^i_E ,
\end{eqnarray}
where we have introduced the factor $1/\varepsilon$ on the definition of
$R^i_E$ and
\begin{eqnarray}\label{stiff_part}
F^i_E &=& - \partial_k [-\beta^k \sqrt{\gamma} E^i - 
\alpha \epsilon^{ikj} \sqrt{\gamma} B_j]
   -\sqrt{\gamma} E^k (\partial_k \beta^i) - \nonumber\\
& & \alpha \sqrt{\gamma} \gamma^{ij} \partial_j \psi - 
\alpha \sqrt{\gamma} q v^i, \\ 
R^i_E &=& -\alpha \sqrt{\gamma} W \sigma 
         \left[ E^i + \epsilon^{i j k} v_{j} B_{k}
                         - (v_{k} E^{k}) v^{i} \right].
\end{eqnarray}  

In order to evolve this system numerically, the fluxes $\{ F_{\tau},
F_{S^i}, F_D \}$ have to be computed at each substep. This implies
that the primitive quantities $\{ \rho, ~ p, ~ v^i, ~ E^i , ~B^i\}$
have to be recovered from the conserved fields $\{ ~ \sqrt{\gamma}~D, ~ \sqrt{\gamma}~\tau , ~ \sqrt{\gamma}~S_i,
~ \sqrt{\gamma}~E^i, ~ \sqrt{\gamma}~B^i \}$. With the exception of
very simple EOSs, this recovery cannot be done analytically and it is
instead necessary to solve a set of algebraic equations via some
root-finding iterative procedure, which we will describe below.

Before that, we note that the solution of the conserved quantities $\{
~ \sqrt{\gamma}~D, ~ \sqrt{\gamma}~\tau , ~ \sqrt{\gamma}~S_i, ~\sqrt{\gamma}~B^i\}$ at time $t=(n+1)\Delta t$ is
obtained by simply evolving the Eqs. \eqref{rest_mass_evol}-\eqref{momentum_evol}, \eqref{magnetic_evol}. However, the same
procedure for the electric field leads only to an approximate solution
$\{ \tilde{E}^i \}$ containing only the explicit terms. The full
solution, involving also the potentially stiff terms, requires
therefore the inversion the implicit equation \eqref{electric_evol},
which depends on the velocity $v^i$ and the fields $\{ B^i,
\tilde{E}^i \}$. In the case of the scalar Ohm law~\eqref{Ohm_law},
the stiff part is linear in $E^i$, so a simple analytic inversion can
be performed 
\begin{eqnarray}\label{invert_matrix_E2}
   E^i = {\bf M}^{-1}(v^j) ~
   [ \tilde{E}^i + \bar{\sigma}~S_E(v^j,B^j) ],
\end{eqnarray}
where $\bar{\sigma} \equiv a_{ii} ~ \Delta t ~ \alpha ~ W ~ \sigma$ and
the inversion matrix is given by
\be
{\bf M}\!\! =\!\! \left[\begin{array}{ccc}
1 + \bar{\sigma} (1 - v_x v^x)  \!&\! - \bar{\sigma} (v_y v^x) \!&\!
- \bar{\sigma} (v_z v^x) \\ \\ - \bar{\sigma} (v_x v^y)  \!&\! 1 + \bar{\sigma} (1 - v_y
v^y) \!&\! - \bar{\sigma} (v_z v^y) \\ \\ - \bar{\sigma} (v_x v^z)
\!&\! - \bar{\sigma} (v_y v^z)  \!&\! 1 + \bar{\sigma} (1 - v_z v^z)  
\end{array}\right].
\label{eq:inversion}
\ee
The recovery procedure is similar to the one presented in
Ref.~\cite{Palenzuela:2008sf} and can be summarized in the following
steps:
\begin{enumerate}

\item Consider an initial guess for the electric field. Some possible
  options are its value in the previous time step, its approximate
  value in the current time step $\tilde{E}^i$, or the ideal MHD value
  $E^i = - \epsilon^{ijk} v_j B_k$, where $v_j$ is the velocity in the
  previous time level.

\item Subtract the electromagnetic field contributions from the
  conserved fields, namely, compute
\begin{eqnarray}
\tilde{\tau} &=& \tau - \frac{1}{2}(E^2 + B^2), \\ \tilde{S_i} &=& S_i
- \epsilon_{ijk} E^j B^k.
\end{eqnarray}

\item Perform the recovery as in the nonmagnetized case: The EOS can
  be used to write the pressure as a function of the conserved
  quantities and the unknown $x = h W^2$, so that the definition of
  $\tau$ can be written as
\begin{eqnarray}
f(x) &=& \left(1 - \frac{\Gamma - 1}{W^2 \Gamma} \right) x +
\left( \frac{\Gamma - 1}{\Gamma W} - 1 \right) D \nonumber\\
& & + \frac{\Gamma - \Gamma_p}{\Gamma (\Gamma_p -1)} K
\left( \frac{D}{W} \right)^{\Gamma_p} - \tilde{\tau},
\end{eqnarray}
which must vanish for the physical solutions. Here $\Gamma_p$ and
$\Gamma$ are the adiabatic indices corresponding to an ideal gas and a
polytropic EOS, respectively, while $K$ is the polytropic constant. By
setting $\Gamma = 1$ we recover the simple polytropic EOS, while the
ideal EOS can be recovered by setting $\Gamma_p = \Gamma$.

\item A solution of the function $f(x)=0$ can be found numerically by
  means of an iterative Newton-Raphson solver. The initial guess for
  the unknown $x$ is given by the previous time step.

\item After each step of the Newton-Raphson solver, update the values
  of the fluid primitives
\begin{eqnarray}
v_i &=& \frac{\tilde{S_i}}{x}\,,~~~ 
W^2 = \frac{x^2}{x^2 - \tilde{S}^2}\,,~~~ 
\rho = \frac{D}{W}\,,~~~ \\
p &=& \frac{\Gamma - 1}{\Gamma} \left( \frac{x}{W^2} - \rho \right) 
+ \frac{(\Gamma_p - \Gamma) K \rho^{\Gamma_p}}{\Gamma (\Gamma_p - 1)}\,.
\end{eqnarray}
and then invert the electric field according to
\eqref{invert_matrix_E2}.

\item Iterate the steps 2--5 until the difference between two
  successive values of $x$ and the electric field fall below a given
  threshold, usually of the order of $10^{-10}$.

\end{enumerate}

The electric charge density is a nonstiff evolution variable and can
either be computed using the evolution equation~\eqref{charge_evol} or
using the constraint,
\begin{equation}
\label{eq:charge_divE}
q = \nabla_i E^i\,.
\end{equation} 
Note that this latter approach is considerably simpler and avoids the
complications arising from the large gradients of the currents across,
for instance, a stellar surface; for this reason it is the one we adopt
in our evolutions. In particular, the system of equations we are solving
includes Eqs.~\eqref{magnetic_evol}--\eqref{psib_evol}
and~\eqref{rest_mass_evol}--\eqref{momentum_evol}, and therefore the state
vector ${\bf U}$ is composed of variables with stiff terms ${\bf V}=\{
~ \sqrt{\gamma}~E^i \}$ and nonstiff terms ${\bf W} = \{ ~ \sqrt{\gamma}~B^i, \phi, ~ \sqrt{\gamma}~\tau, ~ \sqrt{\gamma}~S_i, ~ \sqrt{\gamma}~D \}$ in
their evolution equations. If useful, the total electric charge can be
computed through a volume integral of the electric charge density in the
same manner as for the total rest mass.

The previous procedure converges quickly in the high-conductivity regions
if the ideal MHD solution is chosen as an initial guess, and in the
intermediate conductivity regions if the initial guess is given by the
approximate electric field $\tilde{E}^i$. In general, $\lesssim 5$
iterations are sufficient for intermediate conductivities, while
$\lesssim 70$ iterations are usually necessary in the regions with high
conductivity. In those situations when the convergence occasionally
fails, \eg near stellar surface, we treat the corresponding cell as if
belonging to the atmosphere. All in all, the solution of the resistive
MHD equations for a nontrivial fully general-relativistic test such as
an oscillating magnetized star, is about three times more expensive than the
equivalent simulation run within an ideal-MHD approach.

\section{Numerical Tests and Results}
\label{tests}

In this extended Section we report the numerical results obtained in
one-, two- and three-dimensional tests, which confirm that our
implementation is correct and provides the expected results in a large
range of conductivities. More specifically, the one-dimensional tests
involve: (i) a large-amplitude circularly polarized (CP)
Alfv\'en wave to validate that our implementation matches the
ideal-MHD results in the high conductivity regime, (ii) the
evolution of a self-similar current sheet, which tests our
implementation in the intermediate conductivity regime, and (iii)
a collection of shock-tube tests involving a range of uniform and
nonuniform conductivities. In these particular tests we also examine
the zero-conductivity regime, where the electromagnetic fields are
expected to follow the vacuum Maxwell equations and hence behave as
propagating waves.

Following the one-dimensional tests, we then present two and
three-dimensional tests, which include the standard cylindrical and
spherical explosion tests, which we consider in the case of very large
conductivities in order to test the ideal-MHD limit of our
equations. Finally, in addition to the tests above, which are performed
in a flat spacetime, we have performed three different sets of
simulations involving spherical magnetized stars in general
relativity. The first setup consists in a spherical (TOV) star with
prescribed magnetic fields confined initially in the interior of the
star. The second set involves the evolution of a magnetized star with
initial data generated by the \texttt{LORENE} library and having a
dipolar magnetic field that extends also outside the star. As a
conclusive three-dimensional test we consider the gravitational collapse
of a nonrotating star to a black hole, where the initial data is again
generated by the \texttt{LORENE} library~\cite{lorene41}.

\begin{figure}
\centering
\includegraphics[width=0.4\textwidth]{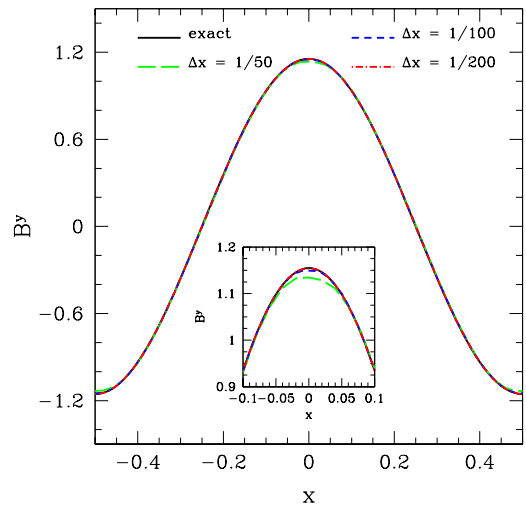}
  \caption{{\em Circularly polarized Alfv\'en wave.} $B^y$ component
    of the magnetic field for three different resolutions $\Delta
    x=\{1/50,1/100,1/200\}$, together with the exact initial solution
    (black solid line). Clearly, the numerical solution provided by
    the resistive MHD implementation and the exact one overlap for a
    uniform conductivity $\sigma=10^6$ and the highest resolution.}
  \label{fig:alfvenwaves}
\end{figure}

With the exception of the collapsing star, where we have used a
polytropic EOS, all simulations reported here have employed an ideal
gas ($\Gamma$-law) EOS 
\begin{equation}
  p = \rho \epsilon (\Gamma-1)\,,
  \label{eq:gamma-law}
\end{equation}
with $\Gamma = 2$ for the one-dimensional tests and $\Gamma=4/3$ for the
two and three-dimensional tests. In addition, for the evolution of the
stable magnetized stars we have adopted a $\Gamma=2$. As mentioned above,
the collapse of the unstable magnetized star has been followed using a
polytropic EOS, $p=K \rho^{\Gamma}$, with $\Gamma=2$. Finally, to ensure
a divergence-free magnetic field with our implementation of the
hyperbolic divergence-cleaning approach, we have set the damping
coefficient $\kappa$ to be one everywhere.

%%%%%%%%%%%%%%%%%%%%%%%%%%%%%%%%%%%%%%%%%%%%%%%%%%%%%%%%%%%%%%%%%%%%%%%%

\subsection{One-dimensional test problems}\label{sec:OneDtests}

%%%%%%%%%%%%%%%%%%%%%%%%%%%%%%%%%%%%%%%%%%%%%%%%%%%%%%%%%%%%%%%%%%%%%%%%

\subsubsection{Circularly polarized Alfv\'en waves}\label{sec:Alfvenwave}

The present test has been discussed in detail in
Ref.~\cite{DelZanna2007} and it computes the propagation of a
large-amplitude circularly polarized Alfv\'en wave through a uniform
background magnetic field $B_0$. For the purpose of this test, we set
a very high conductivity $\sigma=10^6$ in order to recover the
ideal-MHD limit.  Since the propagating wave is expected to be the
advected initial profile, it is convenient to apply periodic boundary
conditions and compare the evolved profile after one full period with
the initial one, in order to check the accuracy of our implementation.

In particular, we consider a CP Alfv\'en wave with a normalized
amplitude $\eta_A$ traveling along positive $x$-axis, in a uniform
background magnetic field $B_0$ with components
\begin{equation}
B^i=\{B_0,\eta_A B_0 \cos [k(x-v_A t)],\eta_A B_0 \sin [k(x-v_A t)]\}.
\label{eq:alfvenwave_initialdata}
\end{equation}
For simplicity, we take $v^x=0$ and write the remaining velocity
components as
\begin{align}
v_y &= -v_A B^y/B_0\,, & v_z &= -v_A B^z/B_0\,,
\end{align}
where
\begin{align}
v_A^2\!=\!\frac{2B_0^2}{\rho h+B_0^2(1+\eta_A^2)}\!\!
 \left[\!1\!+\!\sqrt{1\!-\!\left(\!\frac{2\eta_A B_0^2}
               {\rho h+B_0^2(1+\eta_A^2)}\!\right)^2}\!
\right]^{\!-1}\!\!\!\!\!\!.
\end{align}
By setting $\rho=p=\eta_A=1$ and $B_0=1.1547$, we fix the Alfv\'en
velocity to $v_A=0.5$. Therefore, in a computational domain centered
at $x=0$ with $x\ \in [-0.5,0.5]$, we expect the wave to return to its
initial position after one full period $t=L/v_A=2$. The comparison of
the numerical solution with the initial condition
\eqref{eq:alfvenwave_initialdata} at $t=0$ gives us a measure of the
error.

\begin{figure}
  \centering
  \includegraphics[width=0.4\textwidth]{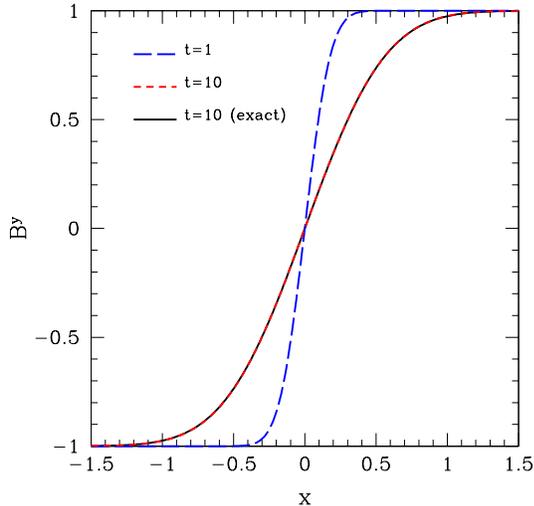}
  \caption{{\em Self-similar current sheet.} $B^y$ component of the
    magnetic field at the initial $t=1$ and final time $t=10$. The
    exact solution at $t=1$ is shown with a dashed blue line. The
    solution given by the analytic expression \eqref{eq:currentsheet}
    at $t=10$ (black solid line) is indistinguishable from the
    numerical solution obtained form the resistive MHD equations (red
    dashed line).}
  \label{fig:currentsheet}
\end{figure}

In principle, the resistive MHD formalism would allow us to recover
the ideal-MHD limit only for an infinite conductivity. In practice,
however, the use of a conductivity as large as $\sigma=10^6$ is
sufficient to obtain a solution that converges to the ideal-MHD one
with increasing resolution. As a result, we have chosen to perform
simulations with a uniform conductivity of $\sigma=10^6$, using the
following resolutions: $\Delta x=\{1/50,1/100,1/200\}$.

In Fig.~\ref{fig:alfvenwaves} we show the component $B^y$ at time $t =
2$, corresponding to one full period. By superimposing the results at $t
= 2$ with the initial data at $t = 0$, it is evident that the numerical
solution of the resistive MHD equations tends to the ideal-MHD exact
solution for a high-enough conductivity and resolution\footnote{We recall
  that here the solution is referred to as exact because it is the
  solution of the exact Riemann problem, although it is still a numerical
  solution with a nonzero error; see discussion
  in~\cite{Giacomazzo:2007ti}.}. We have used both a linear
reconstruction with monotonized-central (MC) slope limiter and the second
order PPM reconstruction. The numerical solution converges to the exact
one at second order when using PPM reconstruction and at second order
with the linear reconstruction, exactly the same convergence rates than
with the original ideal MHD system implemented in \texttt{WhiskyMHD}.

%%%%%%%%%%%%%%%%%%%%%%%%%%%%%%%%%%%%%%%%%%%%%%%%%%%%%%%%%%%%%%%%%%%%%%%%

\subsubsection{Self-similar current sheet}\label{sec:CurrentSheet}

We next consider a test that involves the evolution of a self-similar
current sheet, as proposed in Ref.~\cite{Komissarov2007}. This setup
is useful for testing codes which solve the resistive MHD equations
with a moderate conductivity regime, which we set to be $\sigma=100$.

\begin{figure}
  \centering
  \includegraphics[width=0.4\textwidth]{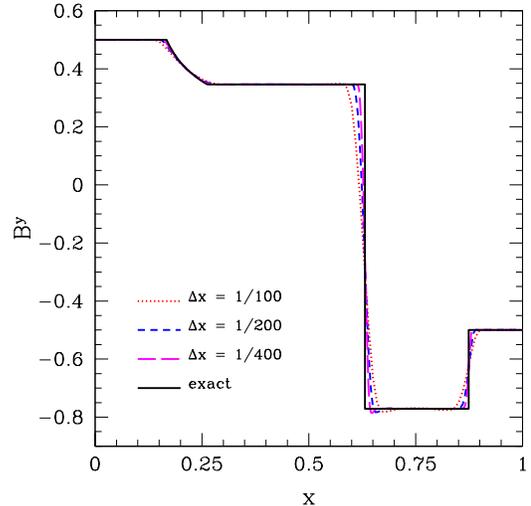}
  \caption{{\em Shock-tube tests.} $B^y$ component of the magnetic field
    at $t=0.4$ for different resolutions $\Delta x =\{1/100, 1/200,
    1/400\}$. The highest resolution $\Delta x = 1/400$ matches the
    exact ideal-MHD solution remarkably well.}
  \label{fig:condconv}
\end{figure}

In practice, the initial data consists in a magnetic field solely in
the $y$-direction which changes sign in a thin current layer. Provided
that the initial solution is in equilibrium (i.e., the pressure and
density are constant, and the velocity is zero) and that the magnetic
pressure is much smaller than the fluid pressure everywhere, then the
evolution of the magnetic field is given by the simple diffusion
equation $\partial_t B^y-(1/\sigma)\ \partial^2_x B^y=0$, which will be
responsible for the diffusive expansion of the current layer in response to the physical
resistivity (we are also assuming that $E^i=0=\partial_t E^i$). Under
these simplified assumptions, the analytical solution of the diffusion
equation is given, for $t>0$, by 
\begin{equation}
B^y(x,t) = B_0 \operatorname{Erf}
\left(\frac{1}{2}\sqrt{\frac{\sigma}{\xi}}\right)\,,
\label{eq:currentsheet}
\end{equation}
where $\xi \equiv t/x^2$ and $\rm{Erf}$ is the error
function. Clearly, as the evolution proceeds, the current layer
expands in a self-similar fashion.

Following~\cite{Palenzuela:2008sf,Komissarov2007}, we use as initial data
the analytic solution \eqref{eq:currentsheet} at $t=1$ and set the
density and pressure to be uniform with $\rho=1$ and $p=50$ respectively,
while keeping the components of the electric field and velocity to zero
initially\footnote{Note that~\eqref{eq:currentsheet} is an exact solution
  only in the limit of infinite pressure~\cite{Bucciantini2012}.}. In our
calculations we have used a computational domain with extents $x=y=z\ \in
[-5,5]$ with a resolution of $\Delta x = 1/200$. Furthermore, a linear
reconstruction method was adopted with the further application of the MC
limiter.

In Fig.~\ref{fig:currentsheet} we present the results we obtained by
solving numerically the resistive MHD equations and the comparison
with the exact solution \eqref{eq:currentsheet} at $t=10$ (black solid
line). Clearly, the numerical solution (red dashed line) is
indistinguishable from the analytic one, thus providing convincing
evidence that the code can accurately describe resistive evolutions
with intermediate values of the conductivity.

%%%%%%%%%%%%%%%%%%%%%%%%%%%%%%%%%%%%%%%%%%%%%%%%%%%%%%%%%%%%%%%%%%%%%%%%

\begin{figure}
  \centering
  \includegraphics[width=0.4\textwidth]{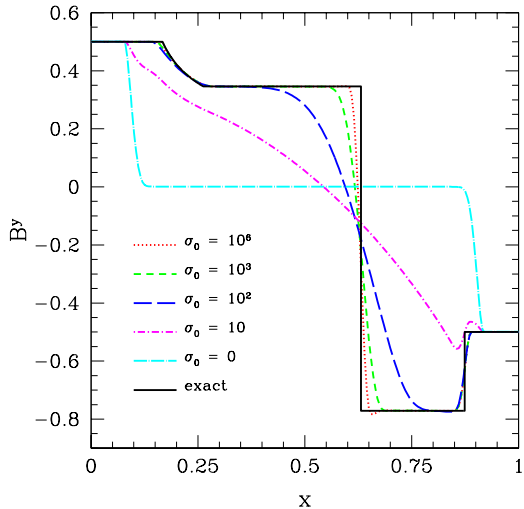}
  \caption{{\em Shock-tube tests.} $B^y$ component of the magnetic
    field for conductivities $\sigma_0=\{0,10,10^2,10^3,10^6\}$ at
    $t=0.4$ and resolution $\Delta x = 1/200$. For $\sigma_0=0$ the
    magnetic field is governed by a wavelike equation, corresponding
    to the solution of the Maxwell equations in vacuum.}
  \label{fig:condBy}
\end{figure}

\begin{figure*}
  \includegraphics[width=0.4\textwidth]{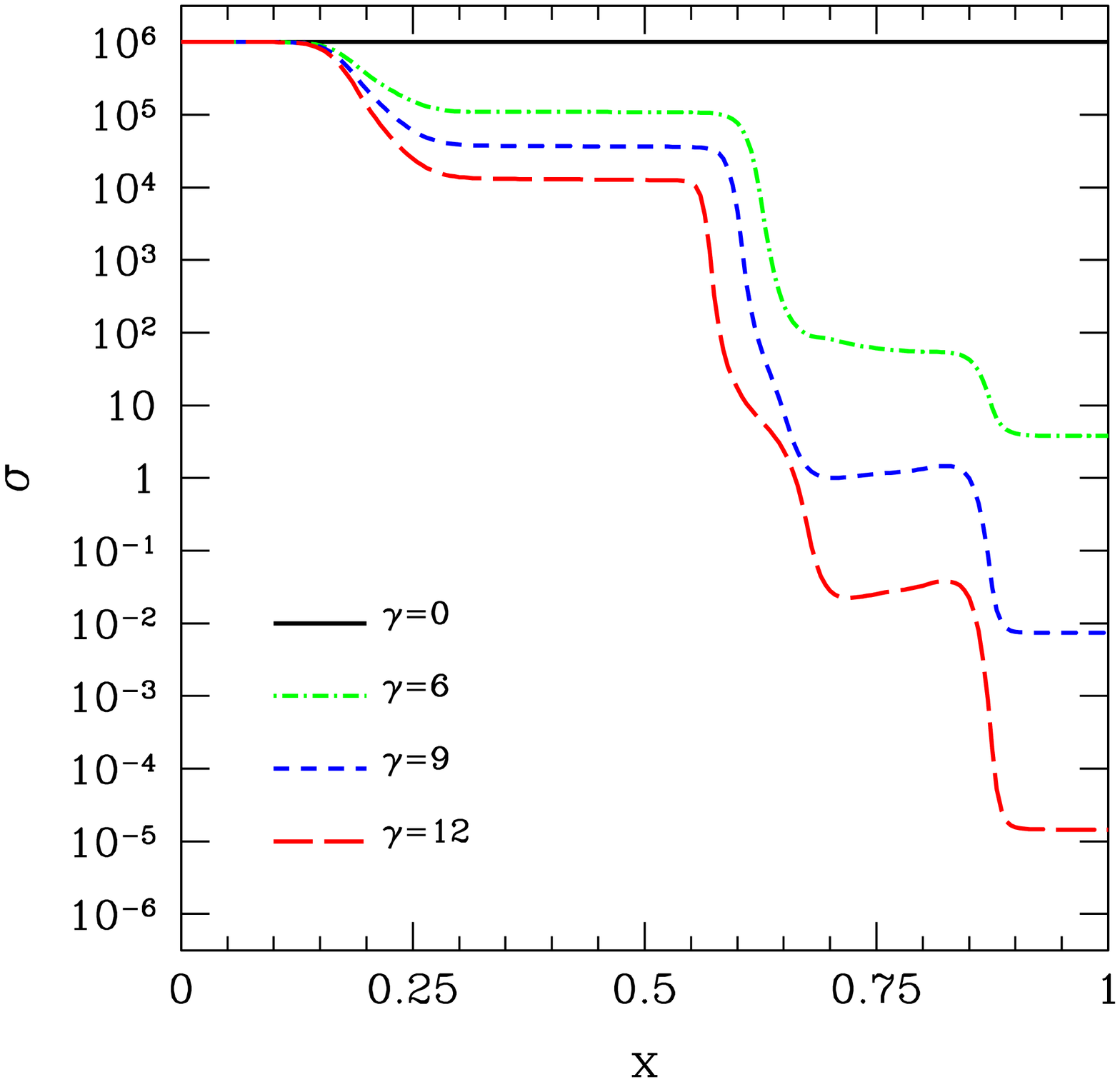}
  \hskip 1.0cm
  \includegraphics[width=0.4\textwidth]{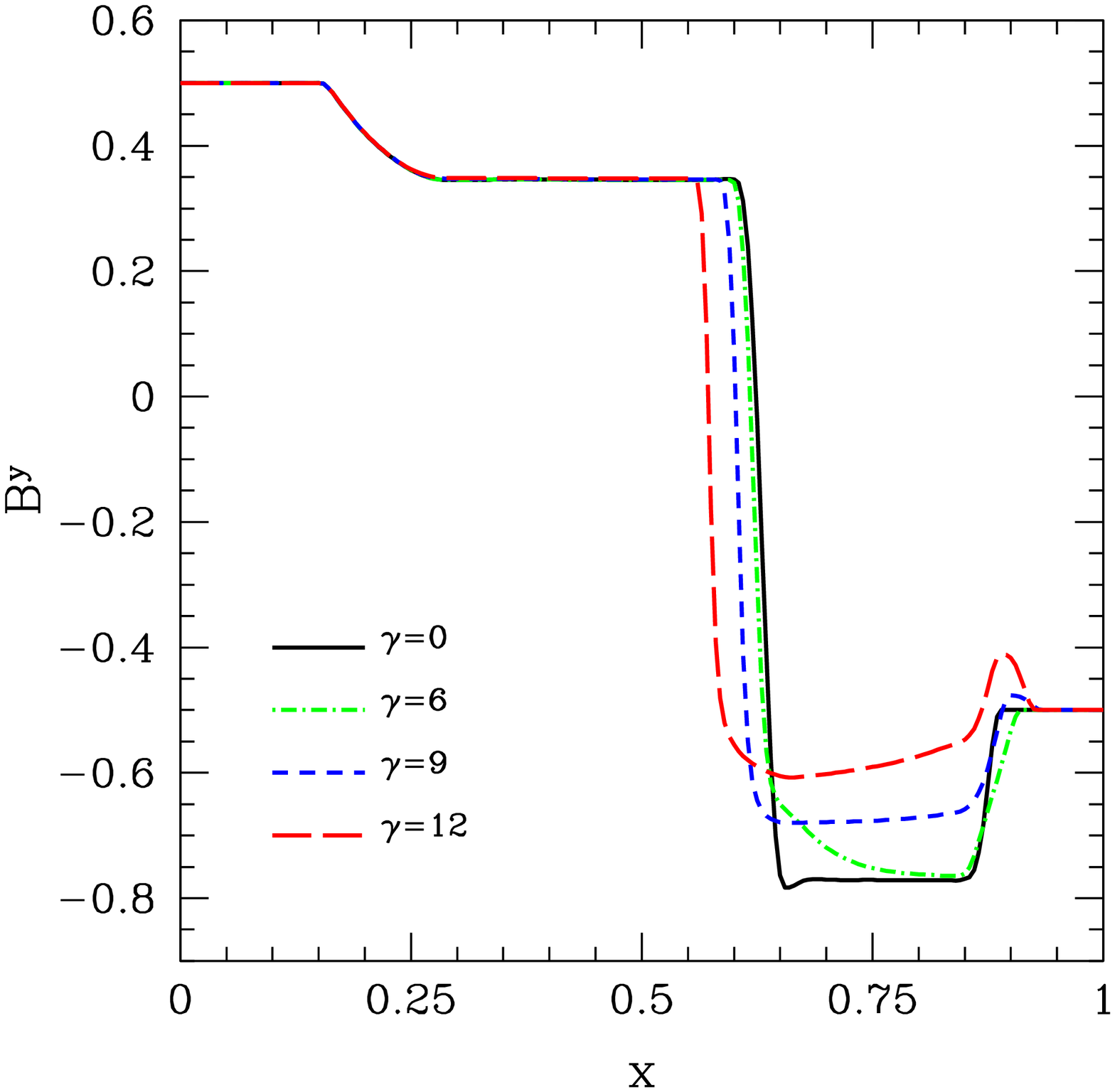}
  \caption{{\em Shock-tube tests.} The left panel shows the
    conductivity profile at $t=0.4$ for nonuniform conductivity with
    different power laws, i.e., $\gamma=\{0,6,9,12\}$. The $\gamma=0$
    case corresponds to the high-conductivity regime of the resistive
    MHD equations. The right panel reports instead the $B^y$ component
    of magnetic field for the same initial conditions as in the left
    one. The leftmost region tends to the ideal MHD solution, while
    the rightmost tends to the vacuum solution for $\gamma=12$.}
  \label{fig:condvarsigma}
\end{figure*}

\subsubsection{Shock-Tube tests}\label{sec:ShocktubeTests}

We next consider the shock-tube test presented in~\cite{Brio1988} and
then modified in~\cite{Giacomazzo:2005jy} to validate our code in the
ideal-MHD limit. The cases under investigation involve the numerical
evolution of discontinuous initial data for a variety of uniform and
density-dependent conductivities parametrized by a reference conductivity
$\sigma_0$. More specifically, the initial data consists of a
discontinuity at $x=0.5$ and left ($L$) and right ($R$) states given by
\begin{eqnarray*}
(\rho_L,\ p_L,\ B^y_L) &=& (1.0,\ 1.0,\ 0.5) \,,\\
(\rho_R,\ p_R,\ B^y_R) &=& (0.125,\ 0.1,\ -0.5)\,,
\end{eqnarray*}
while all other variables are set to zero. The ideal-MHD evolution of
the aforementioned setup with $B^x=0$ leads to two fast waves, one
rarefaction propagating to the left and a shock propagating to the
right of the discontinuity. The solution of this test in the ideal-MHD
limit exists and is found in the exact ideal-MHD Riemann solver
provided by Ref.~\cite{Giacomazzo:2005jy}. For the rest of the
one-dimensional tests, any comparison between the solution of the
resistive MHD equations in the high-conductivity regime and the exact
solution of the ideal-MHD equations is performed with data obtained
from the publicly available code~\cite{Giacomazzo:2005jy}. All tests
have been performed employing a linear reconstruction method with
further application of the MC slope limiter.

As a first setup of our shock-tube tests, we consider the case of a
uniform high conductivity $\sigma=\sigma_0=10^6$ and, in analogy with
the Alfv\'en-wave test in the high-conductivity regime, we verify that
the solution of the coupled Maxwell-hydrodynamics equations tends to
the ideal-MHD exact solution~\cite{Giacomazzo:2005jy} as the
resolution is increased. Figure~\ref{fig:condconv} reports the
magnetic field component $B^y$ at $t=0.4$ for the three resolutions
$\Delta x=\{1/100,1/200,1/400\}$ considered. The high-resolution
result matches the exact ideal-MHD solution so well that is difficult
to distinguish them, thus providing the first evidence that our
implementation is robust also in the presence of discontinuities.

As a second setup of the shock-tube tests, we consider the case in
which the conductivity is still uniform in space, but of different
strength. In particular, we perform the same test for
$\sigma=\{0,10,10^2,10^3,10^6\}$, while keeping the resolution fixed
at $\Delta x = 1/200$. Figure~\ref{fig:condBy} reports different
solutions of the magnetic-field component $B^y$ given by the resistive
MHD equations with different values of $\sigma_0$. It is important to
note here that the solutions change smoothly from the ideal-MHD
solution computed for $\sigma_0=10^6$, to the wavelike solution for
$\sigma_0=0$, which corresponds to the propagation of a discontinuity
at the speed of light, corresponding to a solution of the vacuum
Maxwell equations. The ability of treating the two extreme behaviours
of the Maxwell-MHD equations via a resistive treatment is an
essential feature of our approach and a fundamental one in the
description of the dynamics of magnetized binary neutron stars.

As a final setup our of our suite of shock-tube test, we have
considered the same initial data but now prescribed a nonuniform
conductivity given by the expression
\begin{equation}
\label{eq:varconductivity}
  \sigma = \sigma_0 \left(\frac{D}{D_0}\right)^{\gamma}\,,
\end{equation}
where, however, $D_0 = 1$ (since $v=0$, $\rho=1$) and $\gamma$ is an integer exponent we vary in the range $\gamma \in
[0,12]$. Thes prescription above introduces nonlinearities with
respect to the conserved rest-mass density $D$ and provides an
intuitive way of tracking the dense fluid regions. It leads to low
values of the conductivity in places were the plasma is tenuous and
high values in more dense regions, which will prove very useful later
on when evolving magnetized stars. 

\begin{figure*}
    \includegraphics[width=0.45\textwidth]{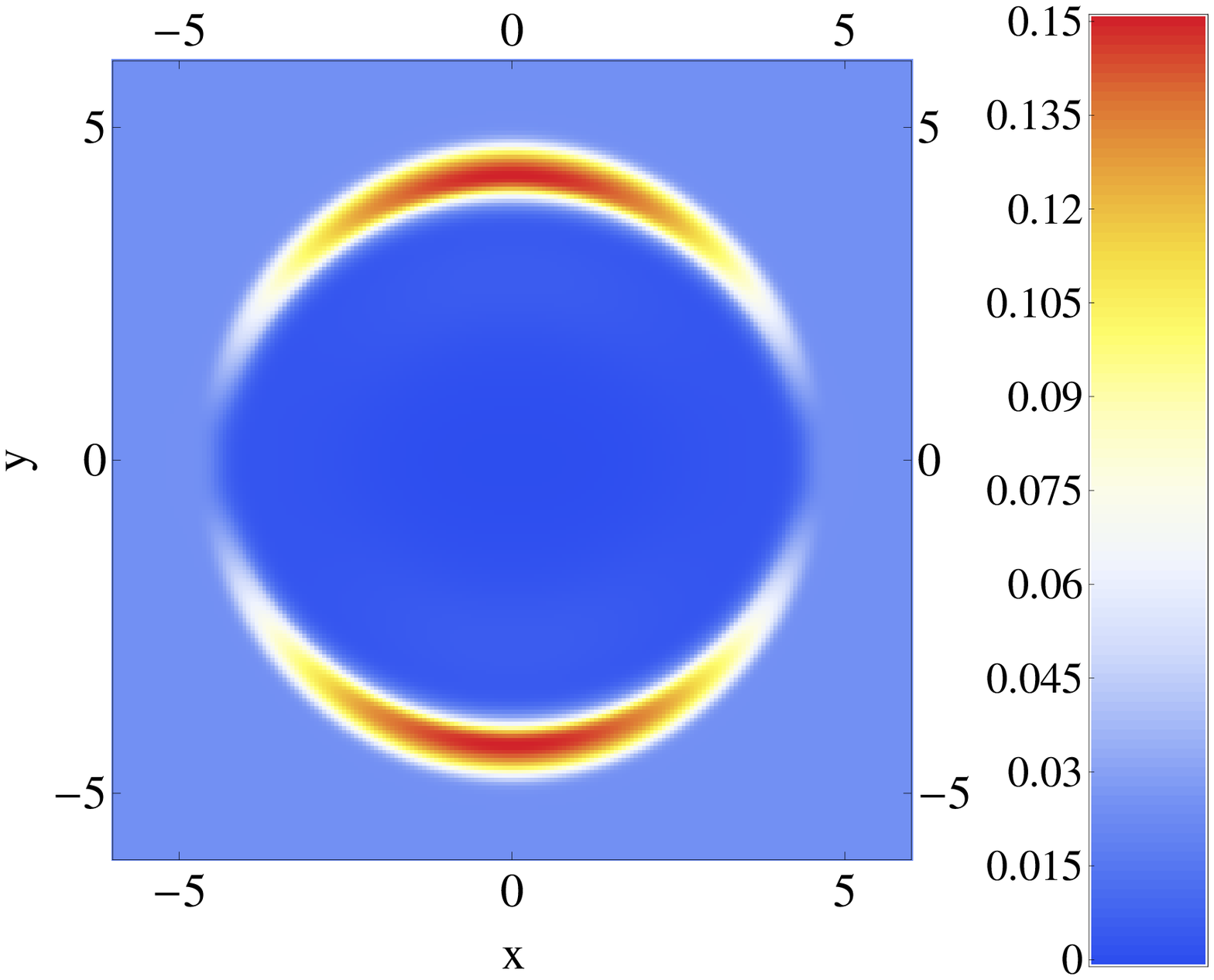}
    \hskip 0.5cm
    \includegraphics[width=0.45\textwidth]{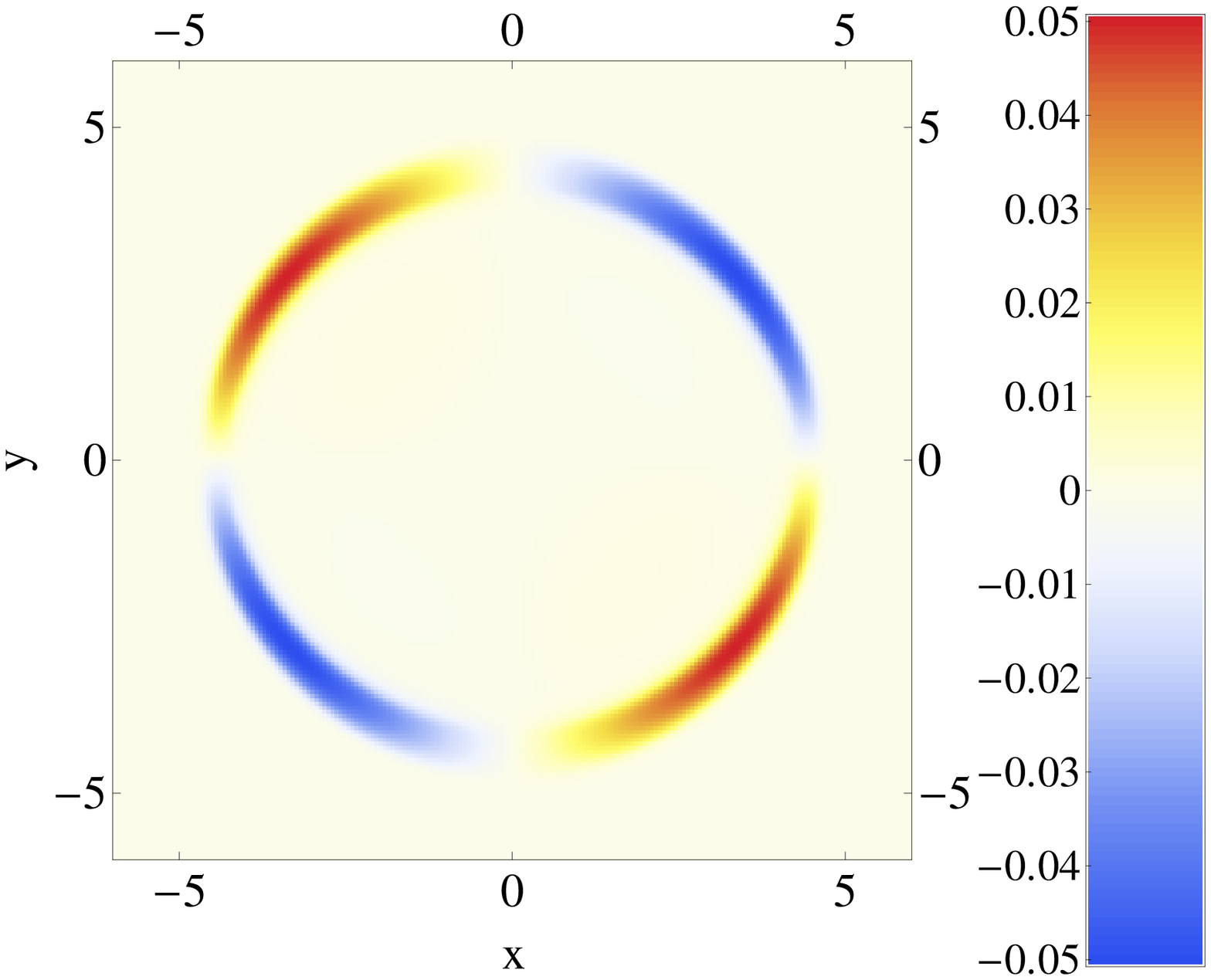}
    \caption{\textit{Left panel:} Snapshot of the magnetic field
      component $B^x$ in the $(x,y)$ plane at $t=4.0$; \textit{Right
        panel:} Snapshot of the magnetic field component $B^y$ in the
      $(x,y)$ plane at $t=4.0$.}
\label{fig:cylexpl}
\end{figure*}

Following~\cite{Palenzuela:2008sf}, we adopt the same initial data as
before, however this time we change the exponent $\gamma$ of
Eq. \eqref{eq:varconductivity} while maintaining the value of
conductivity to $\sigma_0=10^6$.

The results of this last test are reported in the left panel of
Fig.~\ref{fig:condvarsigma}, which shows the profile of the
conductivity at $t=0.4$ for different values of the power-law
exponent, i.e., $\gamma=\{0,6,9,12\}$. Clearly, the conductivity
follows the evolution of the rest-mass density, with a left-going
rarefaction wave and right-going shock. It is interesting to note that
our approach is able to track even very large variations in the
conductivity, with jumps as large as eleven orders of magnitude across
the computational domain. The right panel of
Fig.~\ref{fig:condvarsigma}, on the other hand, reports instead the
magnetic field-component $B^y$ at $t=0.4$ for the same initial
conditions. As imposed by Eq. \eqref{eq:varconductivity}, the
solution in the leftmost part of the computational domain, where the
rest-mass density is very high, is controlled by a very high
conductivity, which tends to $\sigma_0=10^6$. In turn, this implies
that the solution for the magnetic field should approach the ideal-MHD
limit in that region. On the other hand, in the rightmost region,
where the rest-mass density is very low, the conductivity is
correspondigly small and tending to zero for high values of
$\gamma$. In such regions, therefore, the magnetic field is expected
to behave as a wave, thus explaining the appearance of a moving peak
for $\gamma=12$.

Overall, this suite of shock-tube tests, demonstrates that our
numerical implementation is able to treat both uniform and
nonuniform conductivity profiles in one dimensional tests,
independently of the steepness of the profiles and even in the
presence of shocks.

\begin{figure*}
  \includegraphics[width=0.4\textwidth]{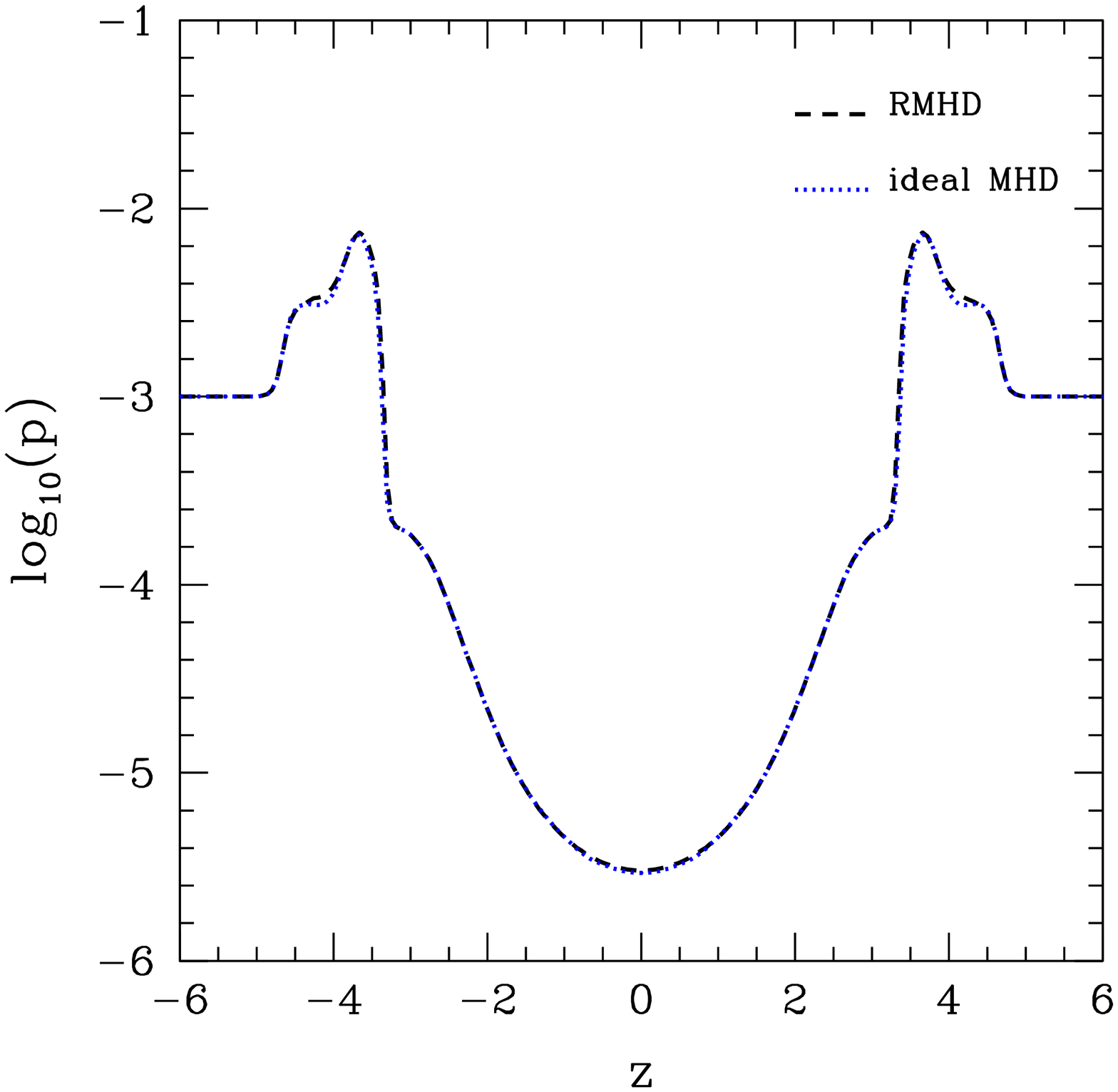}
  \hskip 1.0cm
  \includegraphics[width=0.4\textwidth]{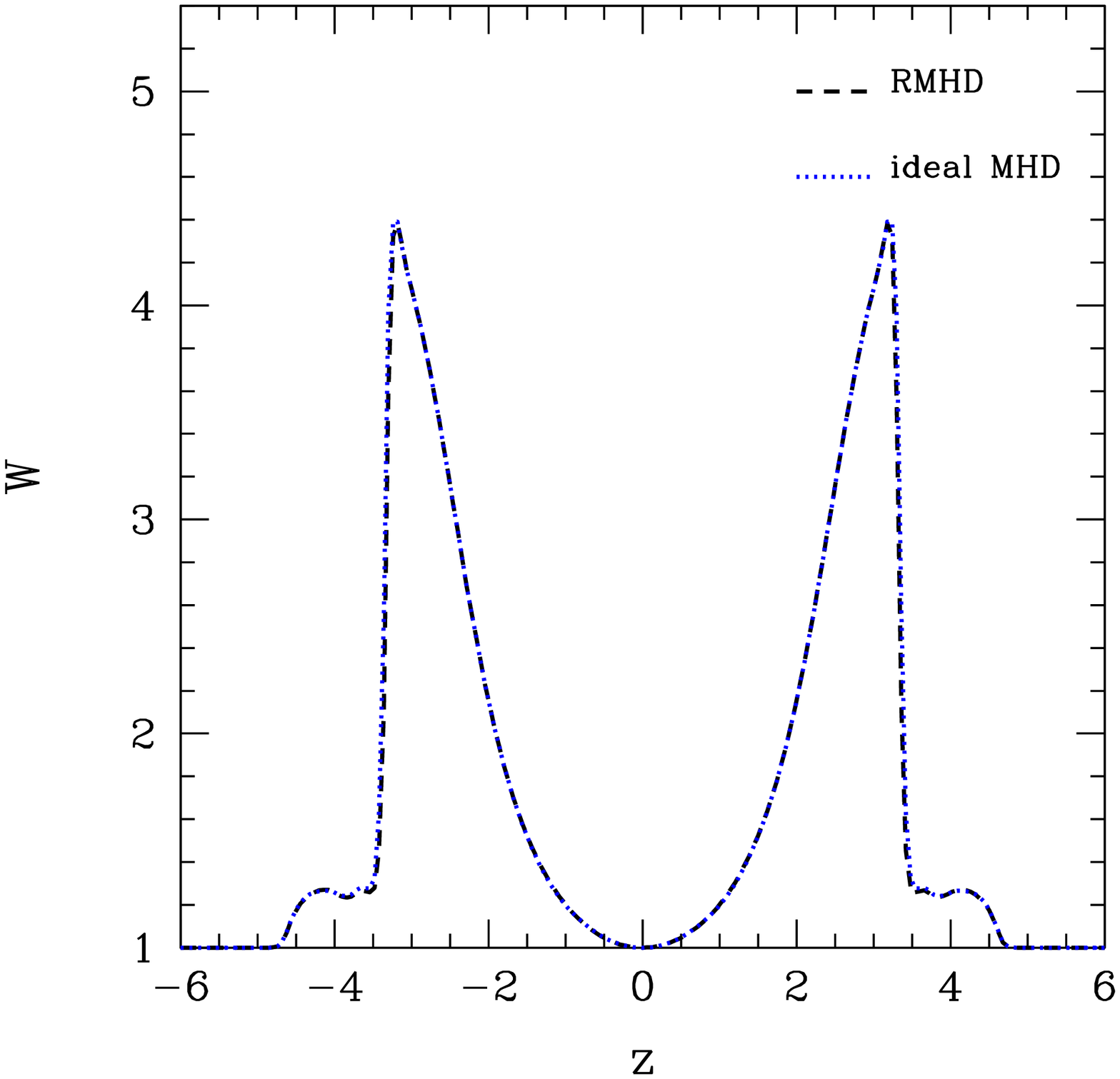}
  \caption{\textit{Left panel:} One-dimensional cuts along the
    $z$-direction and at $t=4.0$ of the pressure. The black dashed
    line corresponds to the resistive code (the \texttt{WhiskyRMHD}
    code), while the blue dotted line corresponds to the ideal-MHD
    code, (the \texttt{WhiskyMHD} code). \textit{Right panel:} The
    same as in the left panel but for the Lorentz factor. }
        \label{fig:sphexpl1d}
\end{figure*}

\subsection{Multidimensional tests}

We now focus on multidimensional tests that involve shocks in several
directions, such as the two-dimensional cylindrical explosion and the
three-dimensional spherical explosion test suggested in
Ref.\cite{Komissarov1999}. Despite the fact that there is no
analytical solution for any of these tests, even in the ideal-MHD
case, the symmetries of the problem can be of great help in verifying
that the numerical implementation is correct and that it preserves the
expected symmetries. Our approach in these tests will be therefore
that of comparing the solution of the same multidimensional test as
obtained with the ideal-MHD code presented in~\cite{Giacomazzo:2007ti}
and our new resistive \texttt{WhiskyRMHD} code in the limit of very
high conductivities. The initial electric field is computed in such a
way that it satisfies the ideal-MHD condition, i.e.,
$E^i=-\epsilon^{ijk}v_j B_k$, and all the tests have been performed
adopting a linear reconstruction method and the minmod slope limiter.

%%%%%%%%%%%%%%%%%%%%%%%%%%%%%%%%%%%%%%%%%%%%%%%%%%%%%%%%%%%%%%%%%%%%%%%%

\subsubsection{Cylindrical blast wave}

In the two-dimensional cylindrical blast-wave problem, we adopt a
square domain with 200 grid cells per direction, in a range of
$(-6.0,6.0) \times (-6.0,6.0)$. The setup of the problem consists of
three regions. The innermost region with $0\leq r\leq0.8$, for which
the pressure and the density are set to $p=1$, $\rho=0.01$,
respectively, the intermediate region which extends from $0.8<r<1.0$
where $r \equiv (x^2+y^2)^{1/2}$ both the pressure and the density
exponentially decrease, and the outermost region which is filled with
an ambient plasma with $p=0.001$, $\rho=0.001$ and occupies the domain
$1.0\leq r \leq 6.0$. The initial magnetic field is along the
$x$-direction with an initial magnetic field strength of $B_0=0.05$.

The numerical results are presented in Fig.~\ref{fig:cylexpl}, where
we show that the magnetic field solution is regular everywhere and
that there are no visible artifacts that could indicate a possible
symmetry error in our implementation. Furthermore, when
one-dimensional cuts of the resistive solution are plotted against the
ideal-MHD solution obtained with the code presented in
\cite{Giacomazzo:2007ti}, the agreement is extremely good (this is not
shown in Fig.~\ref{fig:cylexpl}).

%%%%%%%%%%%%%%%%%%%%%%%%%%%%%%%%%%%%%%%%%%%%%%%%%%%%%%%%%%%%%%%%%%%%%%%%

\subsubsection{Spherical blast wave}

In the three-dimensional spherical blast-wave problem, the grid
structure is similar, but the domain is now within the ranges
$(-6.0,6.0) \times (-6.0,6.0)\times(-6.0,6.0)$. The problem setup
consists of the same three regions as in the cylindrical blast wave
problem, although here the radius $r$ refers to the spherical-polar
radial coordinate, and not to the cylindrical radius, i.e., $r \equiv
(x^2+y^2+z^2)^{1/2}$.

The corresponding solution of the spherical blast-wave problem in the
$(x,y)$ plane is essentially identical to the one already reported in
Fig.~\ref{fig:cylexpl} and for this reason we do not show it here. What
we do show in Fig.~\ref{fig:sphexpl1d}, however, are one-dimensional cuts
along the $z$ direction of the pressure $p$ and of the Lorentz factor $W$
as computed with the ideal-MHD code (blue dotted line) and the resistive
MHD code (black dashed line). This comparison, which is not expected to
be exact given that the resistivity is large but not infinite, provides
convincing evidence of the ability of our implementation to accurately
describe higher-dimensional discontinuous flows in the high-conductivity
regime (the relative difference in the solutions is at most, \ie, at the
shock, of $\sim 7\%$ and of $\sim 0.1\%$ on average).

%%%%%%%%%%%%%%%%%%%%%%%%%%%%%%%%%%%%%%%%%%%%%%%%%%%%%%%%%%%%%%%%%%%%%%%%

\subsection{Nonrotating magnetized stars}

In the following section we present the numerical results obtained from
the evolution of nonrotating spherical stars in the presence of
electromagnetic fields and for a variety of conductivities. Since our
stars are nonrotating there are no charges to support the development of
a magnetosphere. Therefore, describing the exterior as an electrovacuum
is a valid approximation for a magnetized star that has lost its
magnetosphere. In order to model both the interior and the exterior of
the star, we prescribe a spatial dependence of the electrical
conductivity such that the ideal-MHD limit is recovered in the deep
interior of the star (which is expected to be an excellent conductor) and
such that the electrovacuum limit is recovered outside the star, where
the density and the isotropic conductivity is expected to be negligibly
small.

This behaviour can be easily achieved assuming that the conductivity
tracks the (conserved) rest-mass density, thus insuring a smooth
transition between the two regimes. In practice, we have experimented
with functional prescriptions of the type
\begin{equation}
   \sigma = \sigma_0 \max\left[ \left( 1 - D_{\rm atmo}/D \right), 0 
 \right]^2\,,
   \label{eq:sigma}
\end{equation}
where $\sigma \simeq \sigma_0$ is the conductivity in the regions of
large rest-mass density ($\sigma = \sigma_0$ at the stellar center) and
$\sigma=0$ in the atmosphere, where we set the conserved rest-mass
density to its uniform value $D=D_{\rm atmo}$. In our calculations we
normally set $\sigma_0 = 10^6$ in dimensionless code units, or equivalently in cgs units $\sigma_0=2.03\times 10^{11}\,{\rm s}^{-1}$, and
$D_{\rm atmo}$ to be about ten orders of magnitude smaller than the value
of $D$ at the center of the star. Furthermore, in the atmosphere we set
the fluid velocity to zero and since $\sigma=0$ there, the electric and
magnetic fields are evolved via the Maxwell equations with zero currents
(electrovacuum). The charges in the atmosphere are computed again using
Eq.~\eqref{eq:charge_divE} and lead to a net electric flux which is
extremely small when compared to the magnetic flux (see comment on the
right panel of Fig.~\ref{fig:collapse_QNM}).

\begin{table*}
 \begin{tabular}{|l|cccccc|cccc|}
\hline
\hline
Star type & $M_{\rm{ADM}}\ [M_{\odot}]$ & $M_b \ [M_{\odot}]$ & $R_{\rm{eq}}\ [{\rm km}]$ & $K$ & $\Gamma$ & $B_c\ [G]$ & $Number of\ {\rm levels}$ & $N$ & $N_{\rm star}$ &
$R_\mathrm{out}\ [{\rm km}]$ \\
\hline
Confined\ fields & $1.40$  & $1.51$ & $12.00$ & $100.0$ &~~ $2$ ~~& ~~$10^{12}$~~& $4$ &~ $80, 120, 160$ ~&~ $56, 80, 112$ & $141.81$ \\
Extended\ fields & $1.33$  & $1.37$ & $32.56$ & $372.0$ &~~ $2$ ~~& ~~$2.4 \times 10^{14}$ & $4$ &~ $120$ ~&~ $84$~~       & $354.51$ \\
Unstable\ model  & $2.75$  & $2.89$ & $16.30$ & $364.7$ &~~ $2$ ~~& ~~$5 \times
10^{15}$   & $5$ &~ $272$ ~&~ $216$ & $241.07$ \\
\hline
\hline
\end{tabular}
\caption{Properties of the magnetized star models used in the
  simulations. The columns report: the ADM and baryon masses in units of solar
  masses $M_{\rm{ADM}}$ and $M_{\rm b}$ respectively, the circumferential equatorial radius
  of the star in kilometers $R_{\rm{eq}}$,
  the polytropic constant $K$, the polytropic index $\Gamma$, the
  value of the magnetic field in Gauss at the center of the star
  $B_c$, the number of refinement levels, the number of grid points on
  the finest level $N$, the number of grid points across the star
  $N_{\rm star}$ for the different resolutions considered, the
  computational grid outer boundary in kilometers $R_\mathrm{out}$.}
\label{tab:grid}
\end{table*}

\begin{figure*}
  \includegraphics[width=0.33\textwidth]{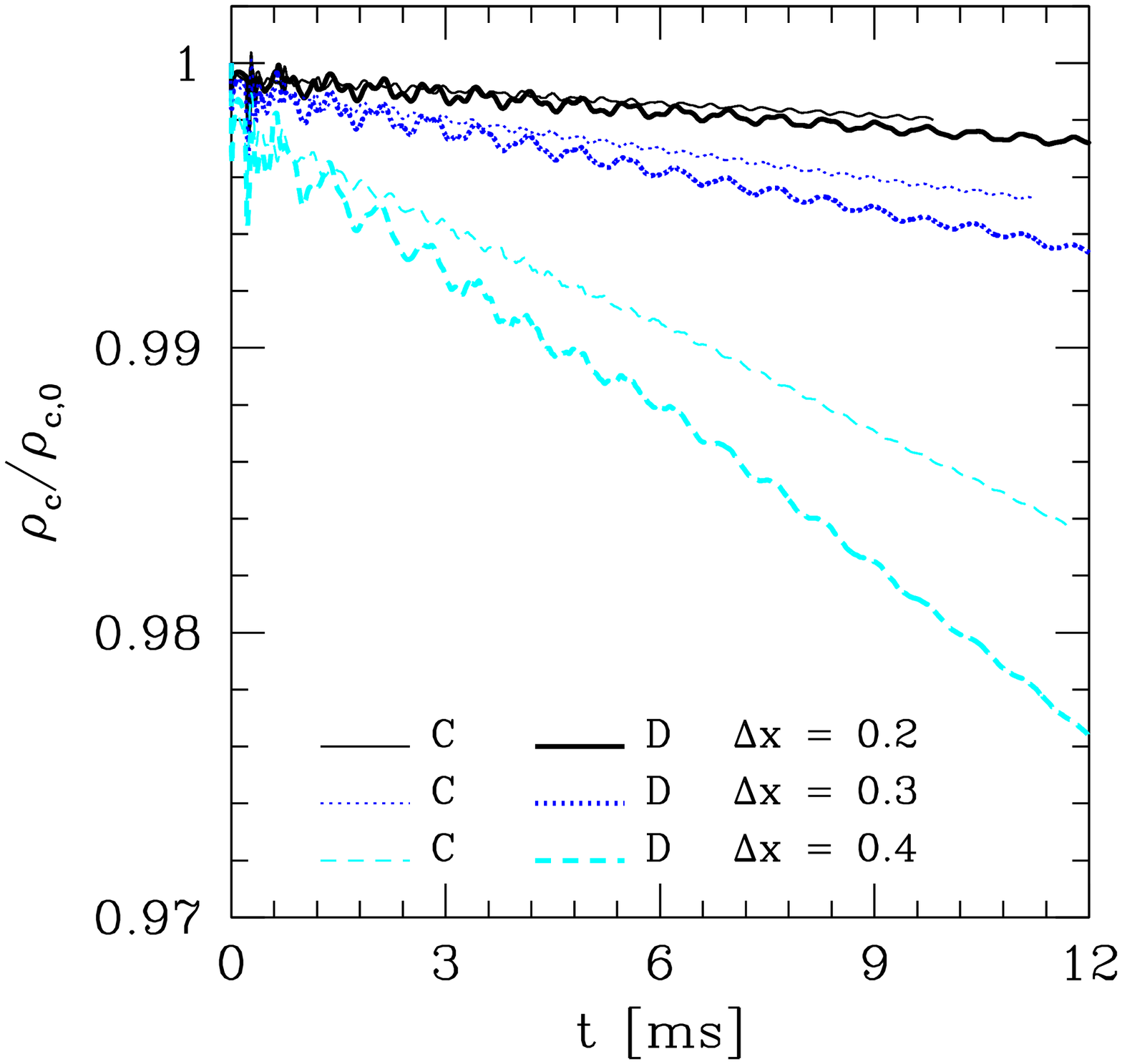}
  \includegraphics[width=0.33\textwidth]{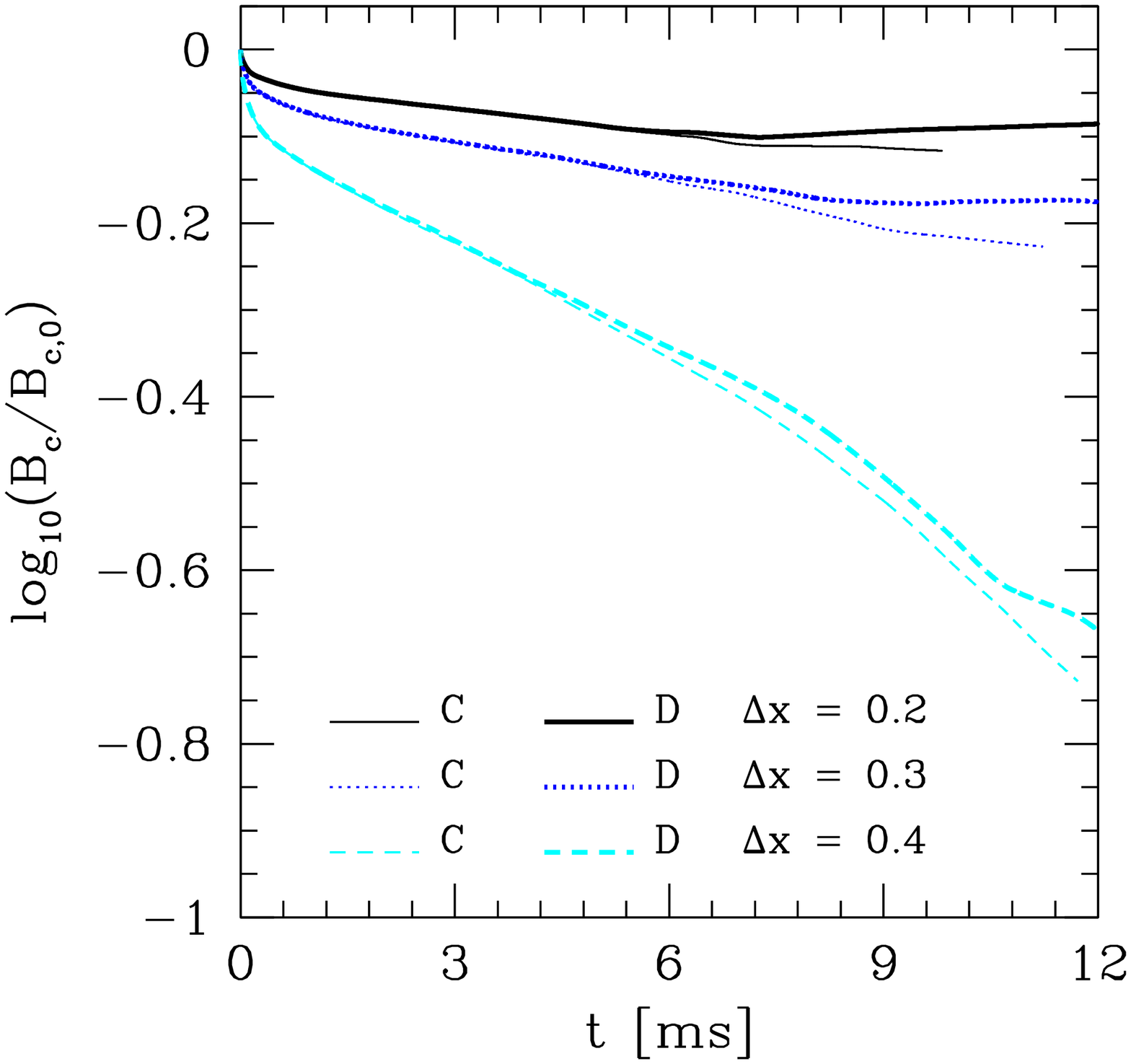}
  \includegraphics[width=0.33\textwidth]{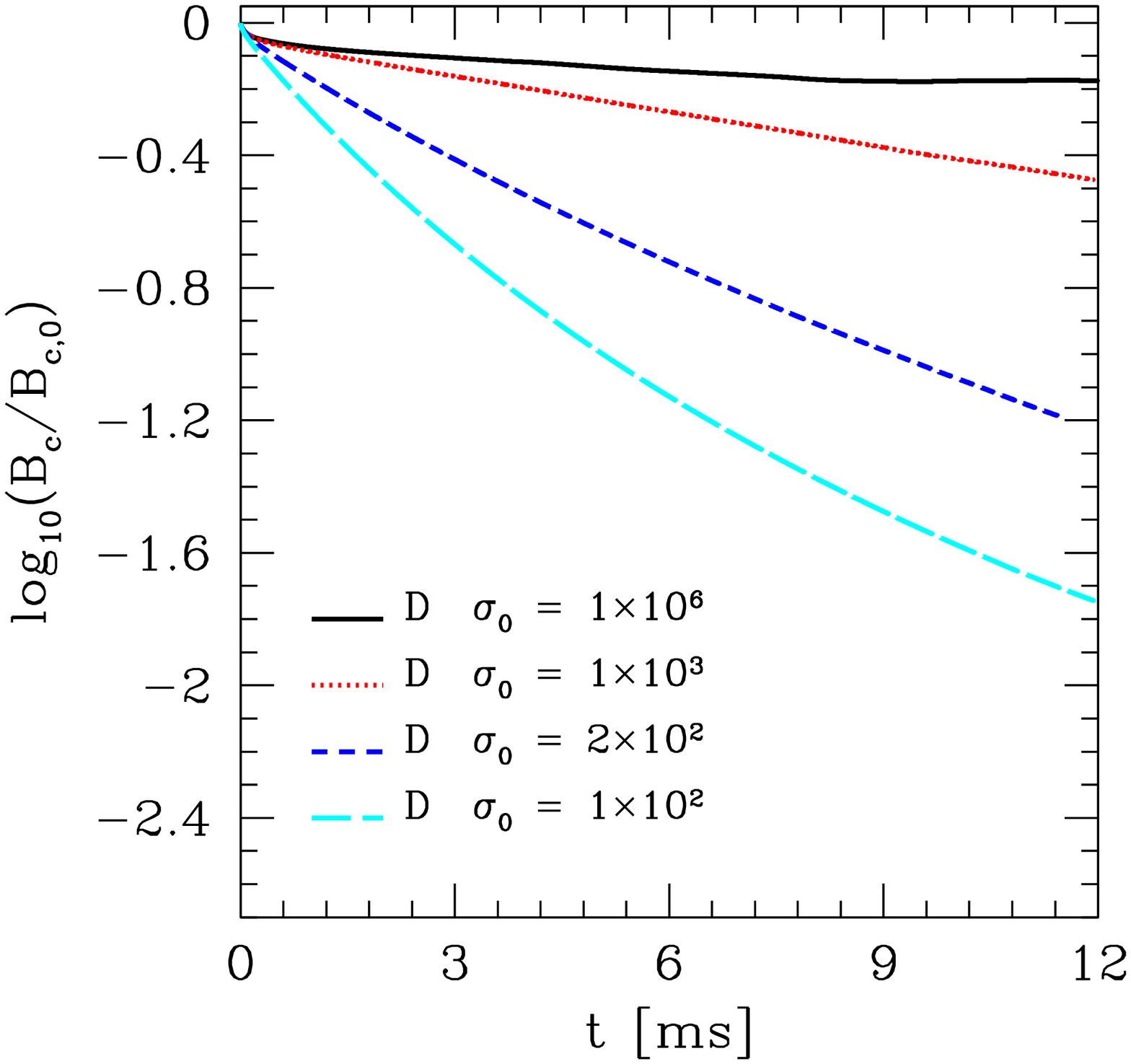}
  \caption{\textit{Left panel:} Evolution of the central rest-mass
    density of a nonrotating magnetized star for both the Cowling
    approximation (C, thin lines) and a dynamical spacetime (D, thick
    lines). Different line types mark different resolutions: dashed
    light blue $\Delta x=0.443$ km, dotted dark blue $\Delta x=0.295$ km,
    continuous black $\Delta x=0.222$ km. \textit{Middle panel:} The
    same as the left one but for the central magnetic
    field. \textit{Right panel:} The same as the middle one but
    different values of the conductivity $\sigma_0$. All lines refer
    to a resolution of $\Delta x=0.222$ km.}
\label{fig:rhoc} 
\end{figure*}

This nonuniform conductivity prescription allows us to provide
effective boundary conditions at the surface of the star for the
exterior electrovacuum solution similar to those in
Refs.~\cite{Baumgarte02b2,Lehner2011}, but without the limitations of
using an analytical solution for the interior of the star or the
further complications of finding a suitable matching between the
electromagnetic fields of the interior ideal-MHD solution and the
exterior one. All the simulations reported hereafter have been
performed adopting the PPM reconstruction scheme, for relativistic
stars whose initial properties are summarized in Table~\ref{tab:grid}.

\subsubsection{Stable star with confined magnetic fields}

For the sake of simplicity, we consider as initial data spherical stars
in equilibrium to which a poloidal magnetic field confined to the stellar
interior is superimposed (see, e.g.,~\cite{Duez:2005cj, Shibata:2005mz,
  Duez:2006qe}). While the hydrodynamical quantities are consistent
solutions of the Einstein equations, the magnetic field is added
\textit{a posteriori}, with a consequent violation of the Einstein constraint
equations at the initial time~\cite{Alcubierre:2008}. In practice,
however, this violation is always very small, even for the largest
fields, and is quickly dominated by the violations introduced by the
standard evolution.

The toroidal vector potential that generates the poloidal interior
magnetic field is expressed as~\cite{Giacomazzo:2007ti}
\begin{equation}
A_{\phi} = r^2  \max\left[A_b(P-P_{\rm cut}), 0\right]^2\,,
\end{equation}
where $P_{\rm{cut}}$ is about $4\%$ the central pressure $P_c$. The star,
initially computed with a polytropic EOS with $\Gamma=2$, $K=100$, has a
gravitational mass $M=1.40 M_{\odot}$ and is endowed with a poloidal
magnetic field of strength $B_c=10^{12}\ G$ at the center of the star and
$\beta \equiv p_{\rm mag}/p = 4.49\times10^{-13}$, with $p_{\rm mag}$ the
magnetic pressure. The magnetic field in the atmosphere is initially
zero. For all of the evolutions presented hereafter we have used an
ideal-fluid EOS with $\Gamma=2$.

We first examine the evolution of the magnetized star in the fixed
spacetime of the initial solution (Cowling-approximation). In the left
panel of Fig.~\ref{fig:rhoc} we show with thin solid, dashed and
dotted lines the evolution of the central rest-mass density normalized
to its initial value $\rho_{c,0}$ in thin colored lines. The tests
were performed using three spatial resolutions of $\Delta
x=\{0.443, 0.295, 0.222\}$ km, corresponding, respectively, to
$N=\{80,120,160\}$ points across the finest AMR grid, which extends up
to $R_{\rm out}= \pm 17.72$ km. As customary in this type of tests, stellar
oscillations are triggered by the truncation error and their amplitude
decreases as the numerical resolution is increased. The importance of
the test rests, therefore, in the calculation of the eigenfrequencies of
the oscillations, which we find to be in very good agreement (within $0.14\%$ precision for the f mode) with
those computed via perturbative analyses (not shown here) and with
other hydrodynamics and ideal-MHD codes~\cite{Baiotti03a,
  Giacomazzo:2007ti}. In addition, a comparison with the ideal-MHD
code \cite{Giacomazzo:2007ti} shows a similar agreement in the
evolution of the rest mass density, indicating that the oscillations
are tracked correctly by our resistive MHD implementation.

\begin{figure}
  \includegraphics[width=0.45\textwidth]{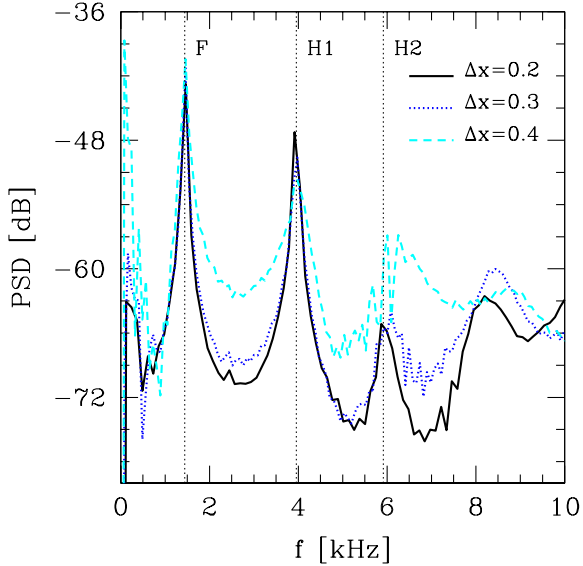}
  \caption{Power spectral density of a full general-relativistic
    evolution of the central rest-mass density for a stable star with
    confined magnetic fields. Different line types refer to different
    resolutions. Shown with dotted vertical lines are the
    eigenfrequencies obtained from linear perturbation theory.}
\label{fig:tovmhdfourierrho}
\end{figure}

We next examine the same scenario, but in a fully dynamical spacetime and
find also in this case a very good agreement with the ideal MHD
solution. Still in the left panel of Fig. \ref{fig:rhoc} we report with
thick solid, dashed and dotted lines the evolution of the central
rest-mass central density in a dynamical spacetime for different
resolutions. As well known from perturbation theory, the eigenfrequencies
of oscillations are in this case lower but what is relevant to note is
that the secular evolution in both the fixed and dynamical spacetimes are
very similar, with variations in the central density that is less than a
couple of percent over tens of dynamical timescales.

\begin{figure*}
 \begin{center}
 \includegraphics[angle=0,width=0.33\textwidth]{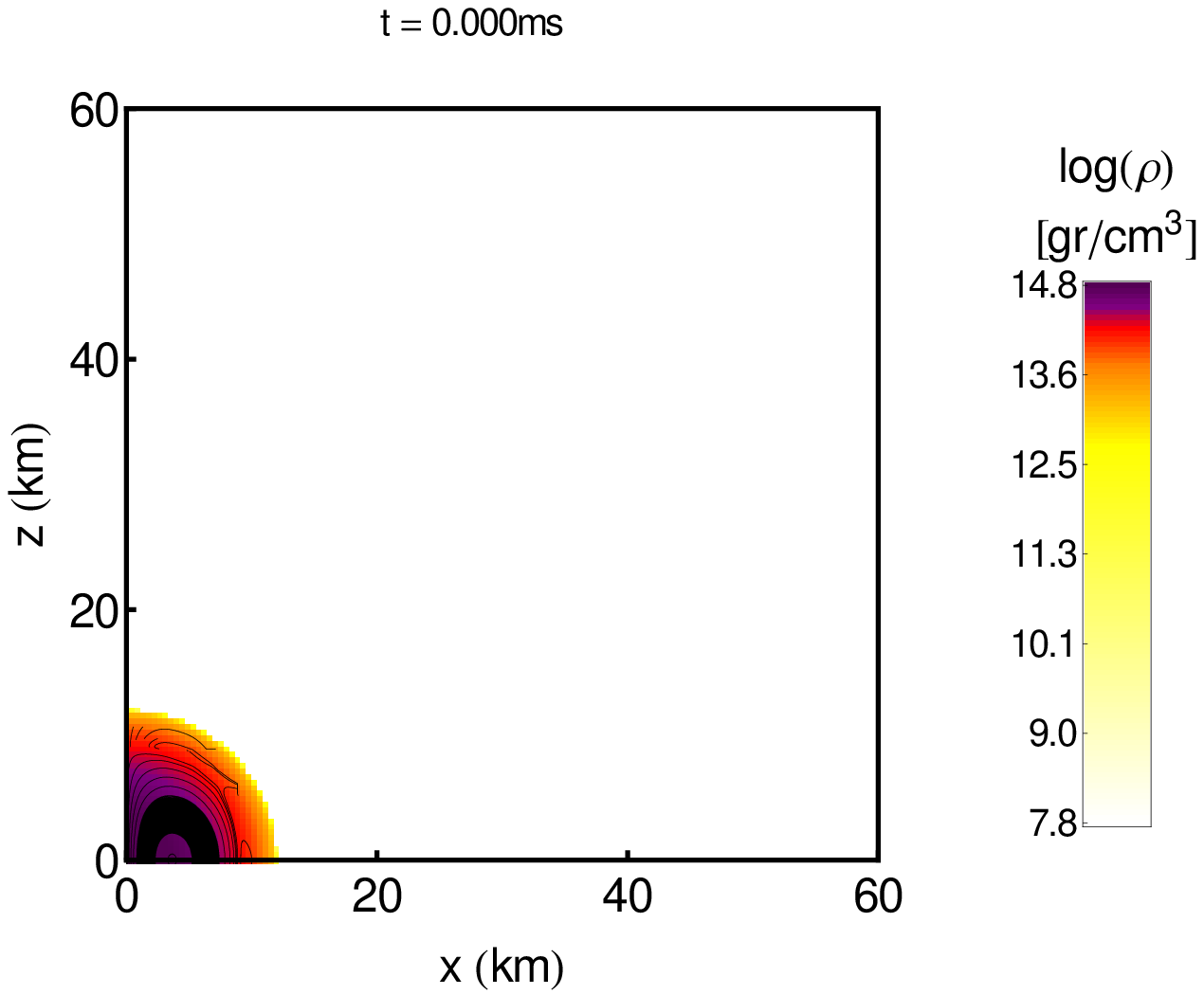}
 \includegraphics[angle=0,width=0.33\textwidth]{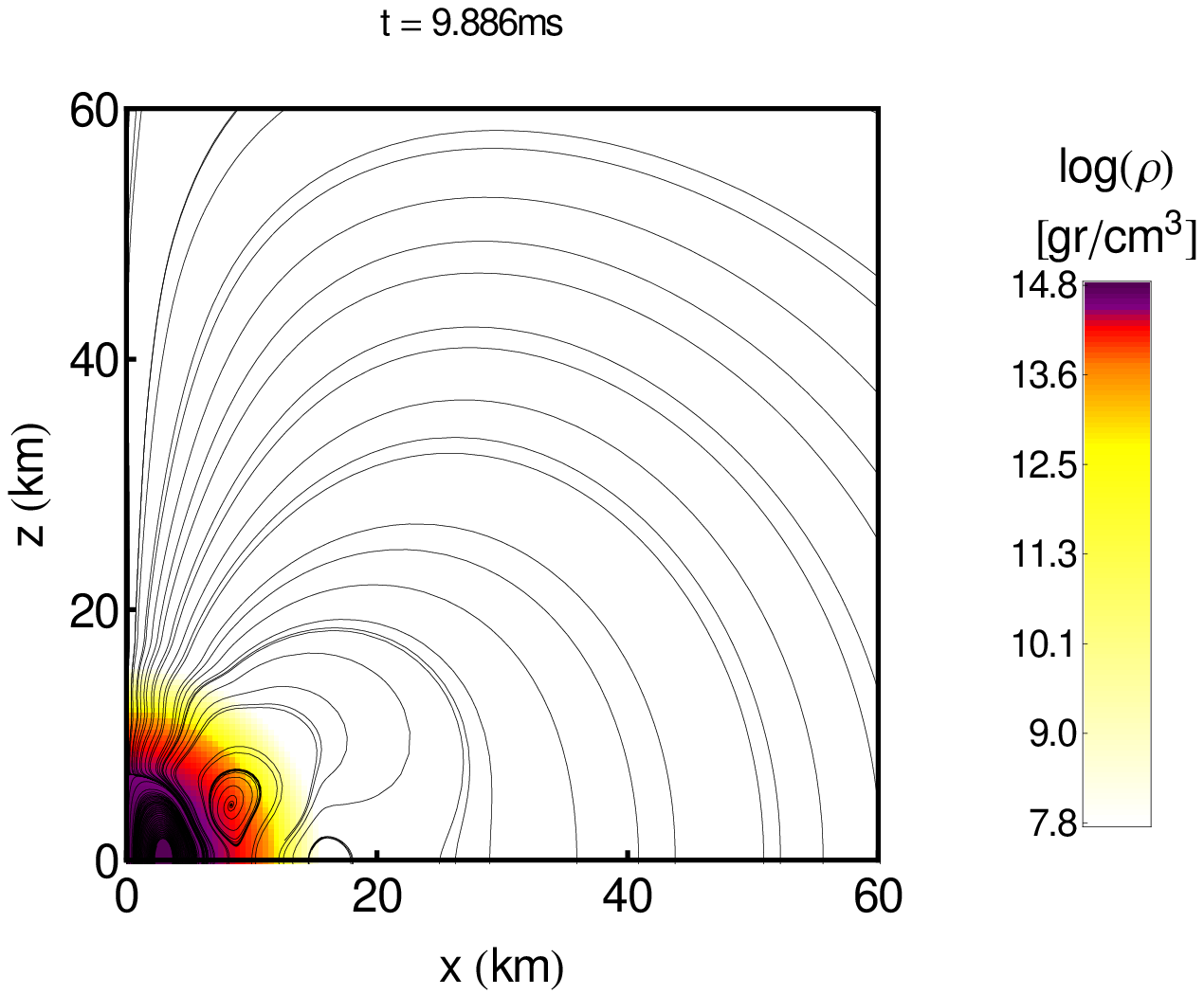}
 \includegraphics[angle=0,width=0.33\textwidth]{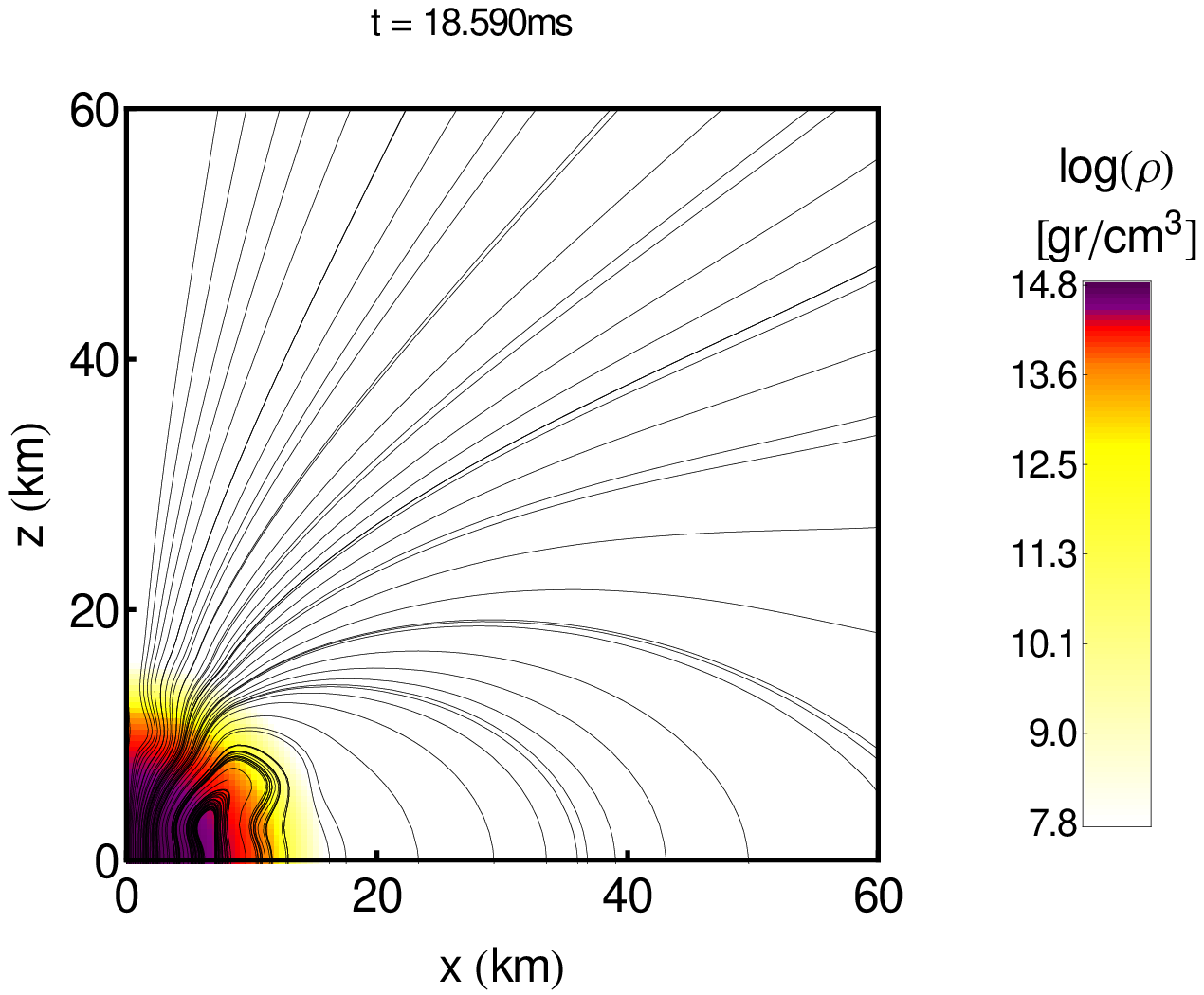}
 \end{center} 
\caption{Two-dimensional cuts on the $(x,z)$ plane of the solution of the
  rest-mass density (color code from white to red) and the magnetic
  field lines at times $t=0,\, 9.88,$ and $18.59\,\ms$. The
  evolution refers to a nonrotating star in a dynamical
  spacetime. Note that although the magnetic field is contained in the
  star initially, it diffuses out as a result of numerical and
  physical resistivity.}
\label{fig:tov_dissipation}
\end{figure*}

The middle panel of Fig.~\ref{fig:rhoc} displays instead the evolution of
the central value of the magnetic field, where lines of different color
refer to different resolutions (which have been checked to yield a
convergence order of $\sim 1.7$), while the thickness marks whether we
are considering a fixed or a dynamical spacetime (thin for the Cowling
approximation and thick for a full general-relativistic evolution).

On the other hand, Fig.~\ref{fig:tovmhdfourierrho} reports the power
spectral density computed from the evolution of the central rest-mass
density in the left panel of Fig.~\ref{fig:rhoc}. Different line types
refer to different resolutions and the dotted vertical lines mark the
eigenfrequencies obtained from linear perturbation theory. The match
between the numerical and perturbative results is clearly excellent and
the differences in the fundamental mode at the highest resolution are at
most $\lesssim 0.5\%$.

We also note that, as for the central rest-mass density, the evolution of
the central magnetic field is accompanied by a secular drift towards
lower values, and this is mostly the result of the intrinsic
numerical resistivity (we recall that these tests have been
performed with the resistive code but for very large conductivities and
hence in a virtual ideal-MHD regime). Clearly, the numerical resistivity
decreases with resolution and this is exactly what the behaviour in the
middle panel shows. It is interesting to note that while with sufficient
resolution the resistive losses saturate to about $20\%$ of the original
magnetic field over $\sim 12$ ms, these can be very large for low
resolution and dissipate up to $\sim 85\%$ of the initial magnetic field
over the same time-span. These numerical resistive losses should be
compared with the ones introduced instead by the physical
  resistivity and which can of course be much larger. This is shown in
the right panel of Fig.~\ref{fig:rhoc}, which is the same as the middle
one, but where we have used the highest resolution (i.e., $\Delta x =
0.222$ km) and varied the strength of the physical resistivity from
$\sigma_0 =10^6$ to $\sigma_0 =10^2$. Because the fluid velocities are
essentially zero at this resolution, the magnetic-field evolution follows
a simple diffusion equation with a Ohmic decay timescale which scales
linearly with $1/\sigma$. This is indeed what is shown in the right panel of
Fig.~\ref{fig:rhoc}, where, after the initial transient, the solution
settles to an exponential decay and where the magnetic field can be
reduced of almost two orders of magnitude over $12$ ms in the case of
$\sigma_0=10^2$.

Finally, we show in Fig.~\ref{fig:tov_dissipation} two-dimensional cuts
on the $(x,z)$ plane of the rest-mass density (shown in a color code from
white to red) and of the magnetic field lines for an oscillating star;
the three panels refer to times $t=0,\, 9.88,$ and $18.59\,\ms$,
respectively. It is important to remark that although we start with a
magnetic field that is initially confined inside the star, the inevitable
presence of a small but finite numerical resistivity and our choice of a
nonzero physical conductivity near the surface of the star [we recall
  that our conductivity follows the profile given in
  Eq.~\eqref{eq:sigma}], induce a slow but continuous ``leakage'' of the
magnetic field, which leaves the star and fills the computational
domain. Because the external magnetic field is essentially with a zero
divergence and with a vanishingly small Laplacian (we recall that in the
stellar exterior the resistivity is zero and the Maxwell equations tend
to the those in vacuum), it is to a very good approximation a potential
field, as shown by the clean dipolarlike structure. Clearly, the
numerical Ohmic diffusion timescale increases with resolution and
therefore the relaxation of the magnetic field to a stationary
dipolarlike structure takes place on longer timescales for the
high-resolution simulation. It is useful to point out that our star will
in general be subjected to the Tayler instability. However, in this test
the Alfv\'en timescale is much longer than the resistive and dynamical
timescales and therefore we can only capture resistive effects. Phenomena
that take place on the usual MHD timescales will be investigated in
future work.

We note that evolving these stars over long timescales while setting
the atmosphere velocities to zero leads to an artificial increase in
the central magnetic field. This is because the outer envelopes of the
star expand, while fluid elements in the inner part move towards the
center. Since the magnetic field lines in the ideal-MHD approximation
are tied to the fluid motion, the central magnetic field will
artificially increase even though the total electromagnetic energy of
the star remains conserved. Because the use of a zero velocity in the
atmosphere is inevitable to avoid spurious accretion of matter onto the
star, the results presented in Fig. 8 should be taken to be valid only
as long as the central magnetic field does not increase and this
happens at 7 ms in the worst case.

%%%%%%%%%%%%%%%%%%%%%%%%%%%%%%%%%%%%%%%%%%%%%%%%%%%%%%%%%%%%%%%%%%%%%%%%%%%%%%%5

\subsubsection{Stable star with extended magnetic fields}

\begin{figure*}
  \includegraphics[width=0.35\textwidth]{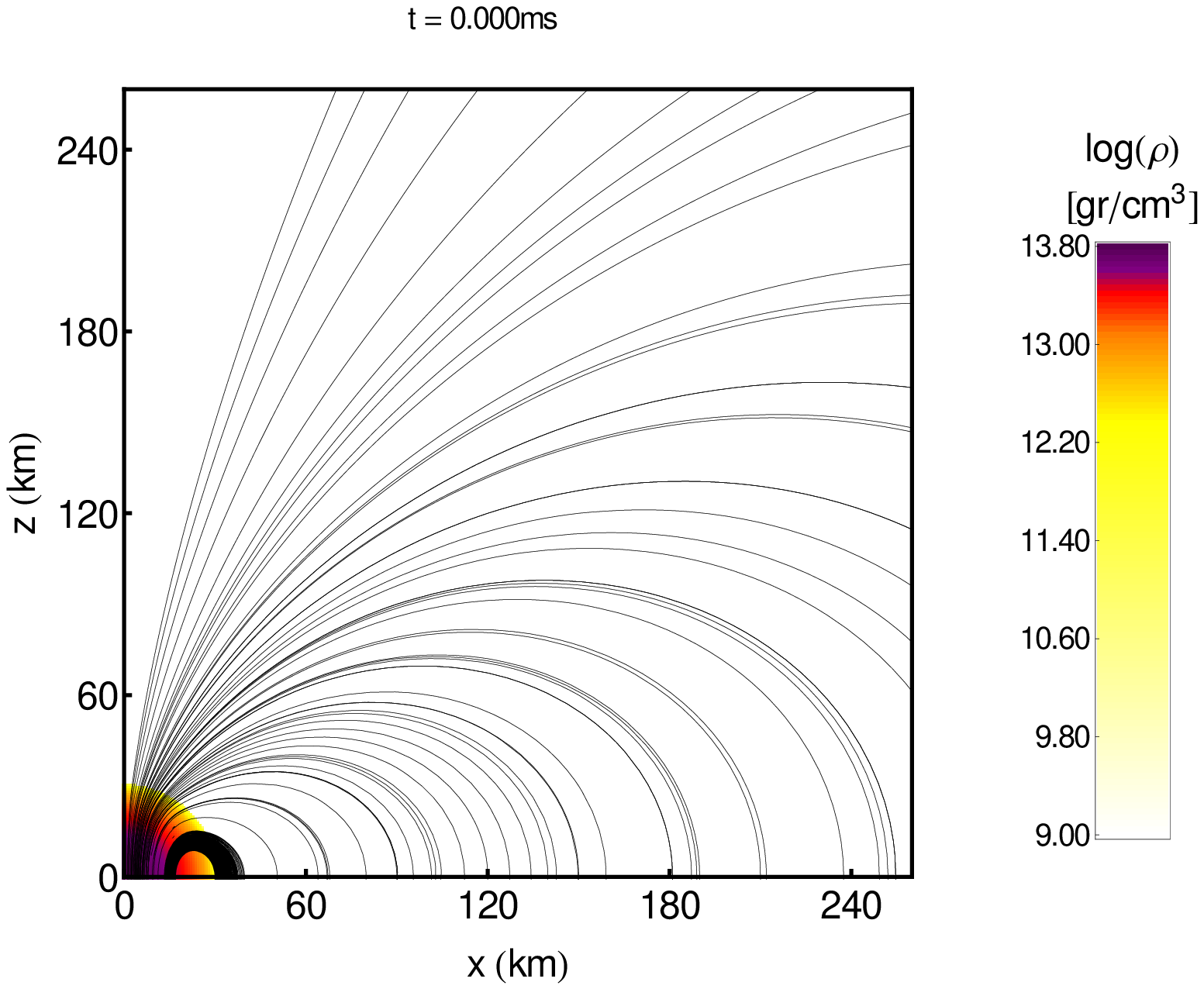}
  \includegraphics[width=0.35\textwidth]{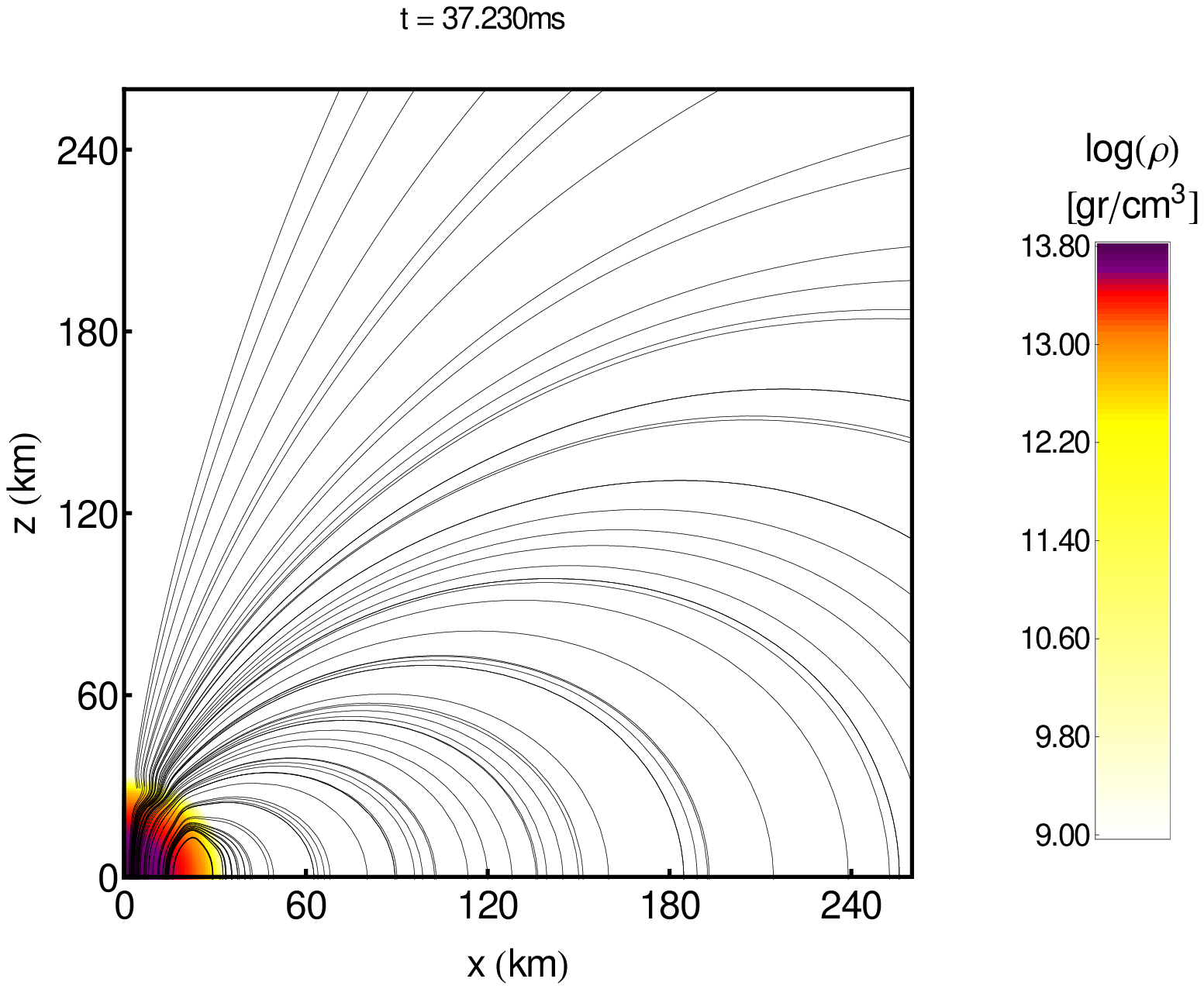}
  \includegraphics[width=0.285\textwidth]{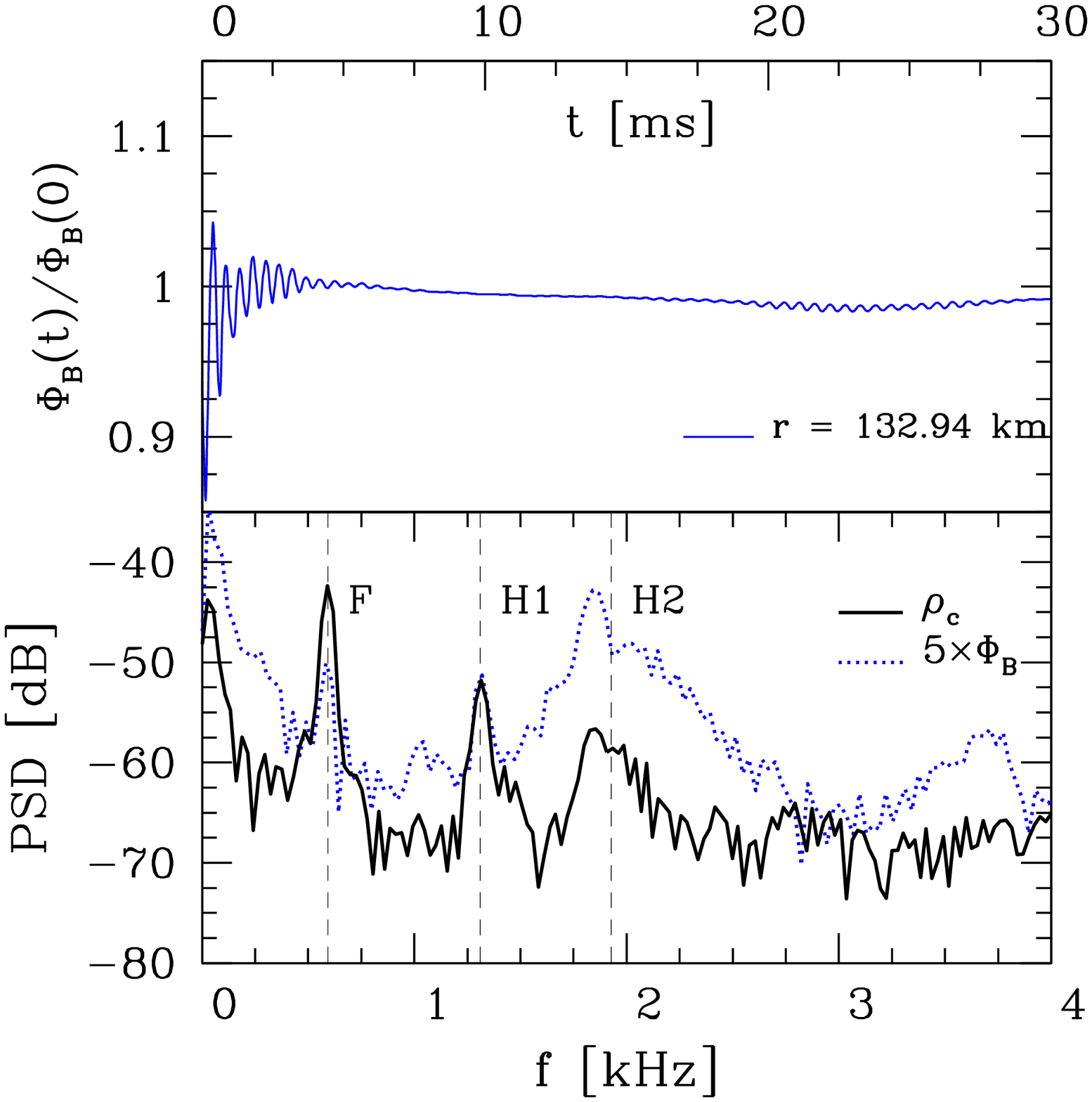}
  \caption{\textit{Left and middle panels:} Evolution of the magnetic
    field lines displayed at times $t=0\, \ms$ and $t=37.23\, \ms$. The
    rest mass density is also shown with purple-red-yellow
    colors. \textit{Right panel:} The top part shows the evolution of
    the magnetic flux computed across a hemispheric surface at a
    radius $r=132.94$ km, while the bottom part shows the power spectral
    density of the rest-mass density (black solid line) and of the
    magnetic flux (blue dotted line).}
\label{fig:magstarmhdrhoB}
  \end{figure*}
%%%%%%%%%%%%%%%%%%%%%%%%%%%%%%%%%%%%%%%%%%%%%%%%%%%%%%%%%%%%%%%%%%%%%%%%

We next consider initial data for a spherical magnetized star with a
poloidal magnetic field extending outside the star, as generated by
the \texttt{Magstar} code from \texttt{LORENE}
library~\cite{lorene41}. The external magnetic field is dipolar and is
computed by solving the Maxwell equations in vacuum, with boundary
conditions given by the interior poloidal magnetic field. This
solution is fully consistent with the Einstein equations and it
provides accurate measurements of the stellar deformations in response
to either rapid rotation or large magnetic
fields~\cite{Bocquet1995}. More specifically, we have considered a
nonrotating star modeled initially as polytrope with $\Gamma=2$ and
$K=372$, having a gravitational mass $M=1.33\, M_{\odot}$, and endowed
with a poloidal magnetic field of strength $B_c=2.4 \times 10^{14}$
G. The magnetic field in the atmosphere is given by the electrovacuum
solution, which has a dipolar structure. The evolutions have been
carried out in a computational domain with outer boundary at $R_{\rm
  out} = \pm 354.51$ km and a resolution of $\Delta x= 0.738$ km, corresponding
to $60$ points covering the positive part of finest grid which extends
up to $r=44.28$ km.

Figure~\ref{fig:magstarmhdrhoB} displays in its left and middle panels
two-dimensional cuts on the $(x,z)$ plane of the rest-mass density (shown
in a color code from white to red) at the initial and final times, i.e.,
$t=0\, \ms$ and $t=37.23\, \ms$. A rapid comparison among the two panels
clearly shows the ability of the code to reproduce stably over this
timescale the evolution of this oscillating star also when the magnetic
field extends in its exterior\footnote{We note that our star will in
  general be unstable to the Tayler instability but also that this will
  develop on a timescale that is $\gtrsim 10^4$ times larger than the one
  considered here~\cite{Lasky2011,Ciolfi2011}.}. The right panel of
Figure~\ref{fig:magstarmhdrhoB}, on the other hand, shows in its top part
the evolution of the magnetic flux computed across a hemispheric surface
at a radius $r=132.94$ km, which shows signs of oscillations. We
have computed the power spectrum of these oscillations and compared it
with the corresponding one obtained for the central rest-mass
density. The results of this comparison are shown in the bottom part of
the right panel, with a black solid line referring to the rest-mass
density a blue dotted line to the magnetic flux. The very good agreement
between the two implies that the oscillations observed in the magnetic
flux are essentially triggered by the oscillations in the rest-mass
density.

\begin{figure*}
 \begin{center}
 \includegraphics[angle=0,width=0.33\textwidth]{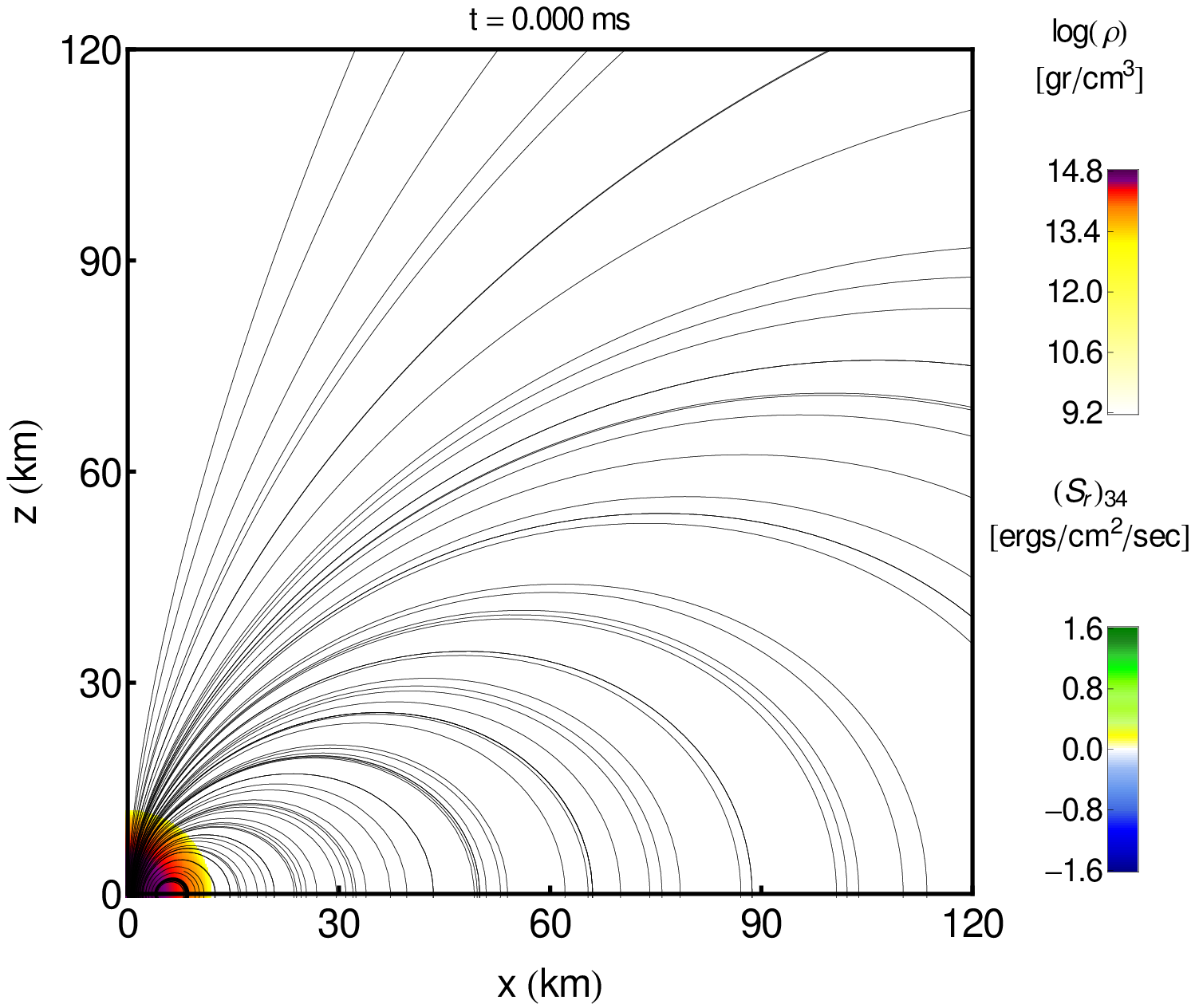}
 \includegraphics[angle=0,width=0.33\textwidth]{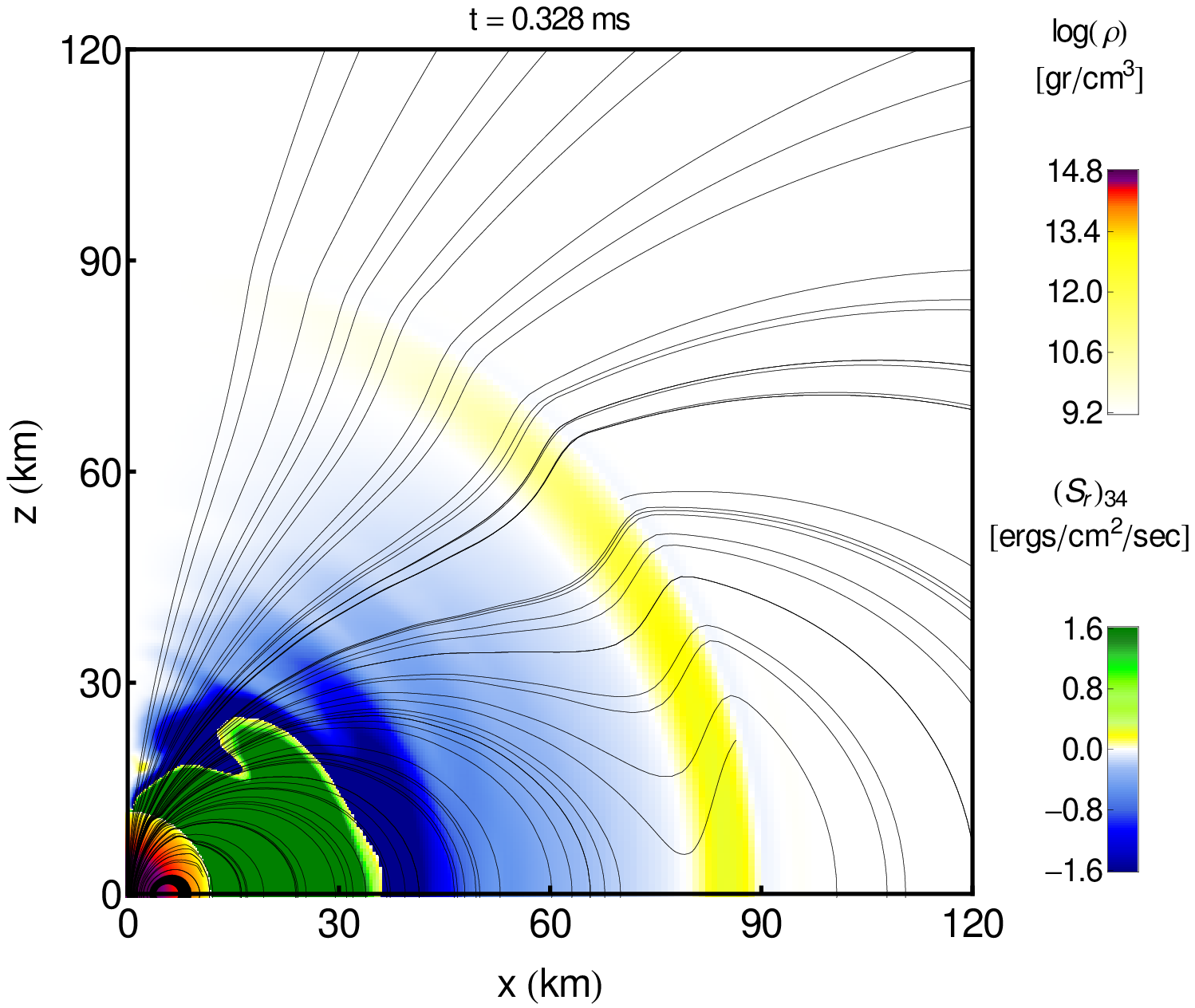}
 \includegraphics[angle=0,width=0.33\textwidth]{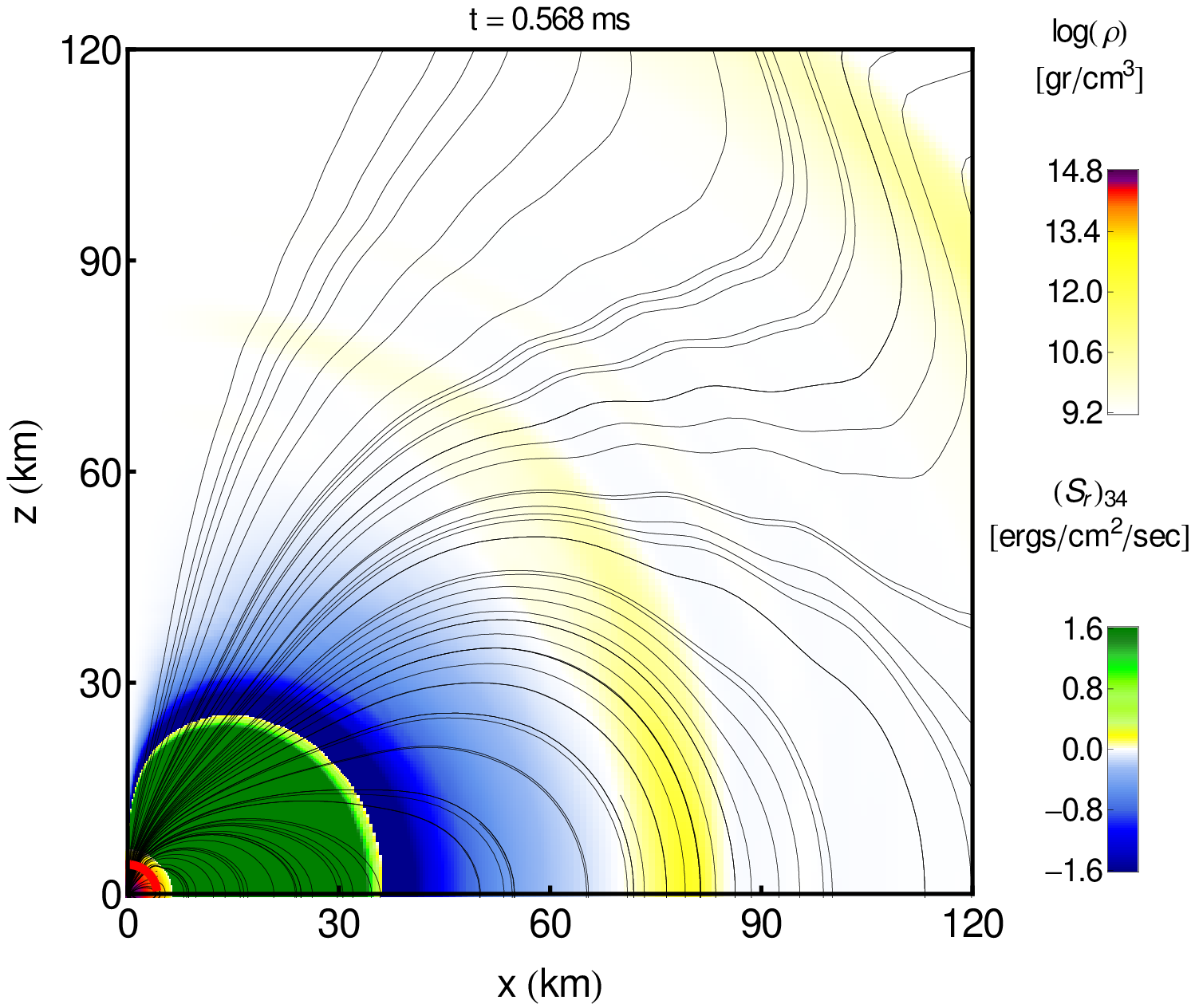}
 \includegraphics[angle=0,width=0.33\textwidth]{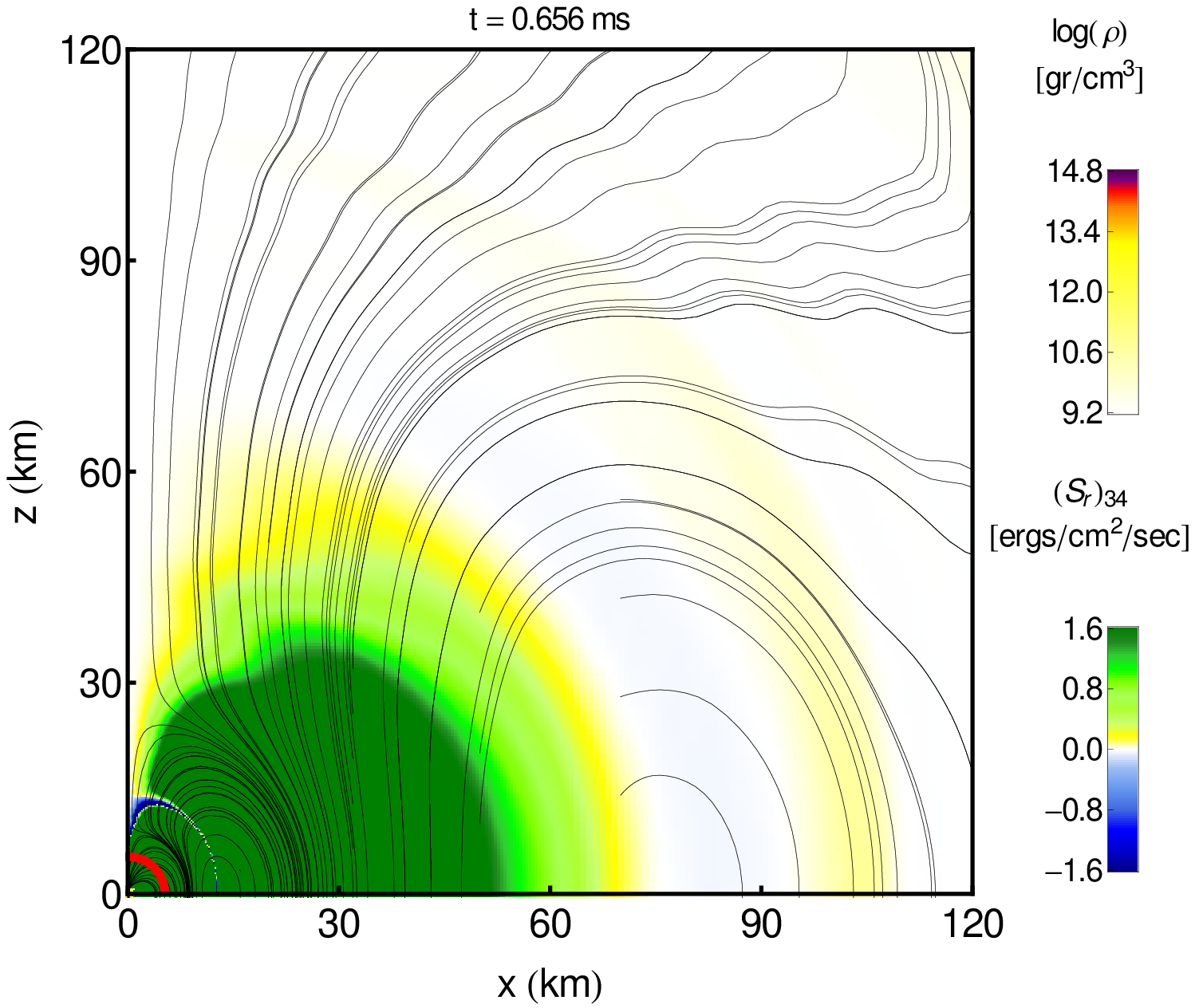}
 \includegraphics[angle=0,width=0.33\textwidth]{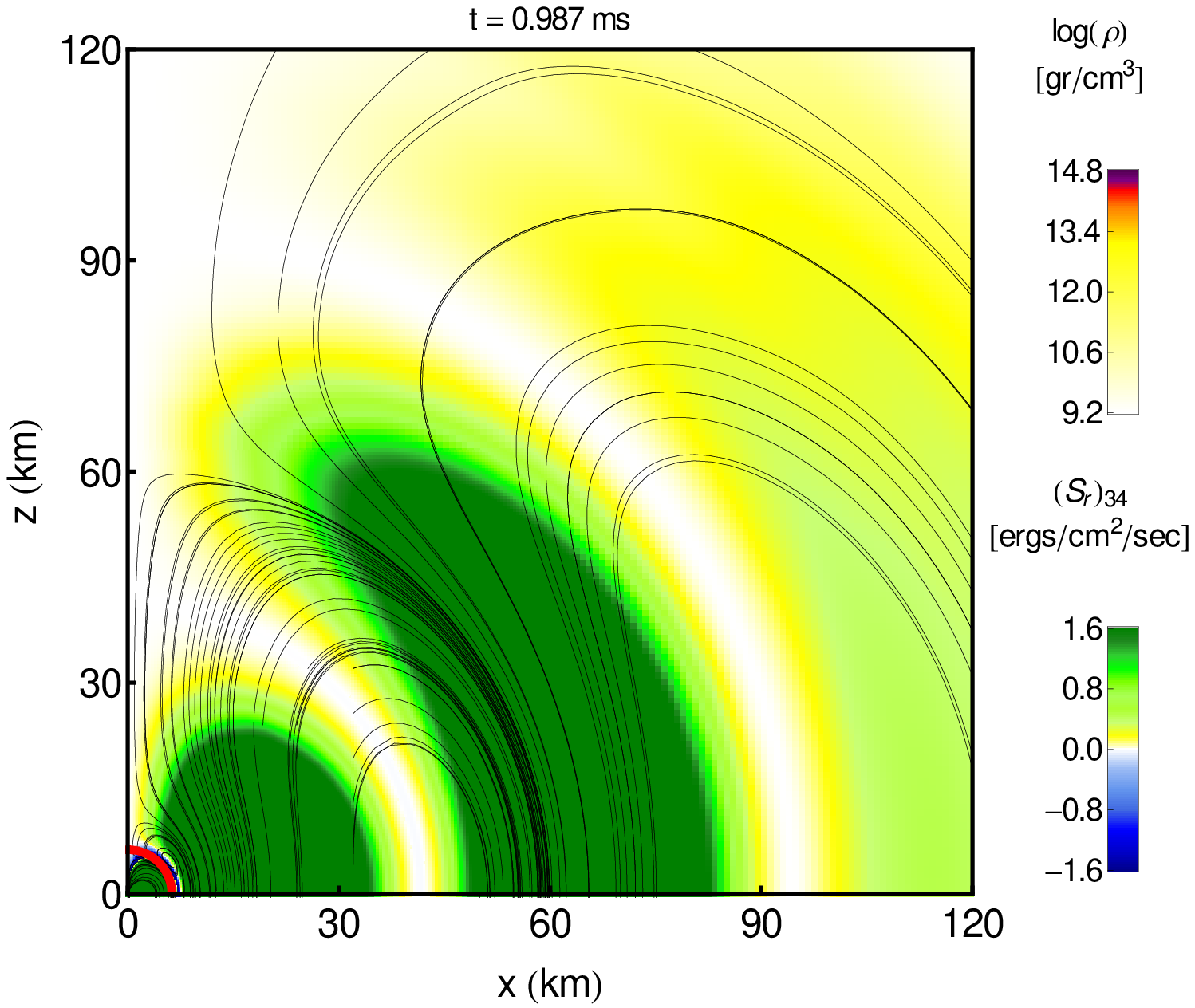}
 \includegraphics[angle=0,width=0.33\textwidth]{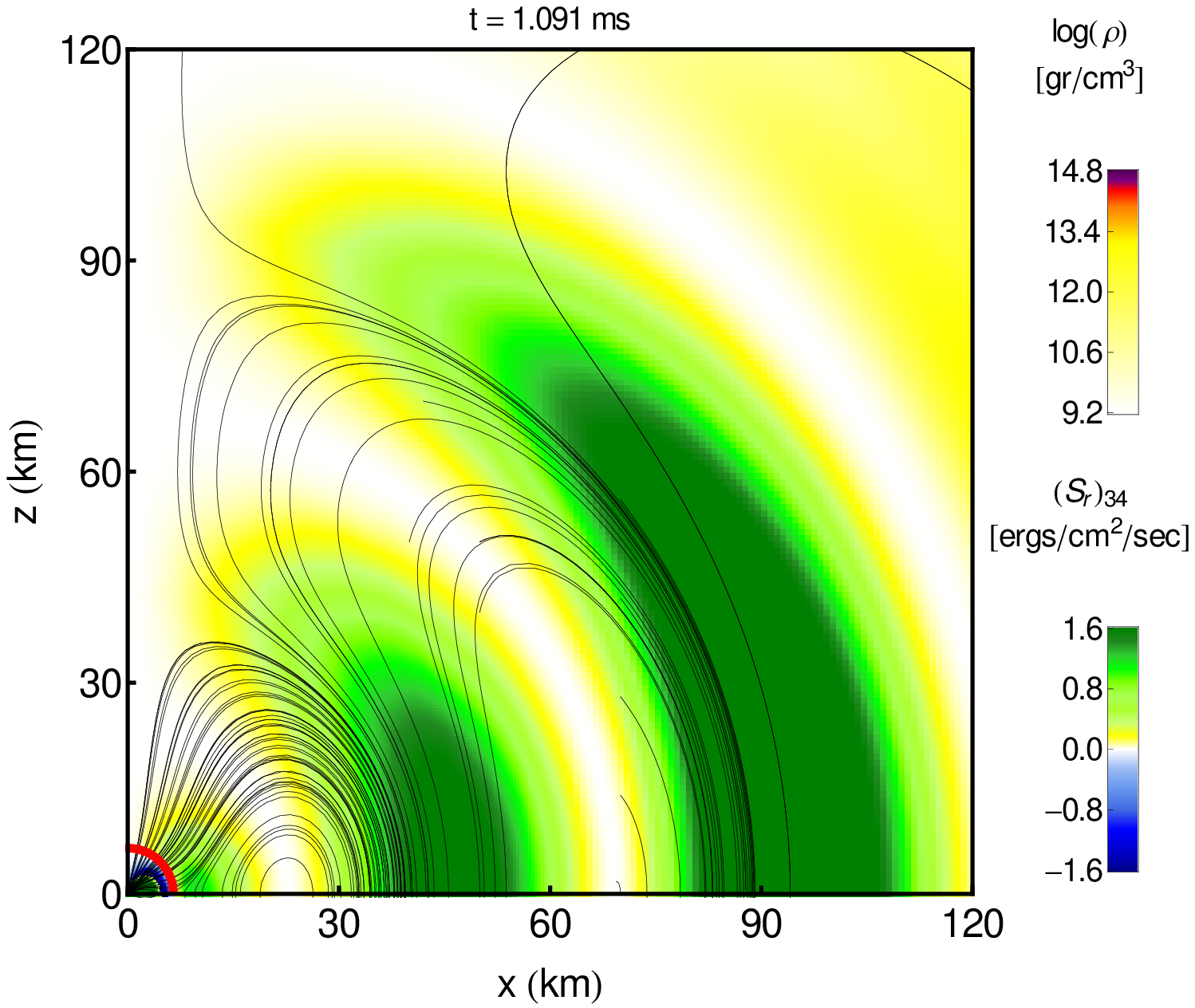}
 \end{center} 
\caption{Two-dimensional cuts on the $(x,z)$ plane of the collapse to
  a BH of a magnetized NS. Shown with colors are the rest-mass density
  (color code from white to red) and the radial Poynting vector
  (color code from blue to green) in units of $10^{34}$, while thin
  lines reproduce the magnetic-field lines. The different snapshots
  refer to times $t=0, 0.32, 0.57, 0.65, 1.0$ and $1.1$ ms, and an
  apparent horizon is marked with a thin red line starting from
  $t=0.57$ ms. Note that all the matter is accreted into the hole and
  that a quadrupolar QNM ringdown is clearly visible in the Poynting
  flux.}
\label{fig:collapse_rho_poynting}
\end{figure*}

\begin{figure*}
 \begin{center}
 \includegraphics[angle=0,width=0.33\textwidth]{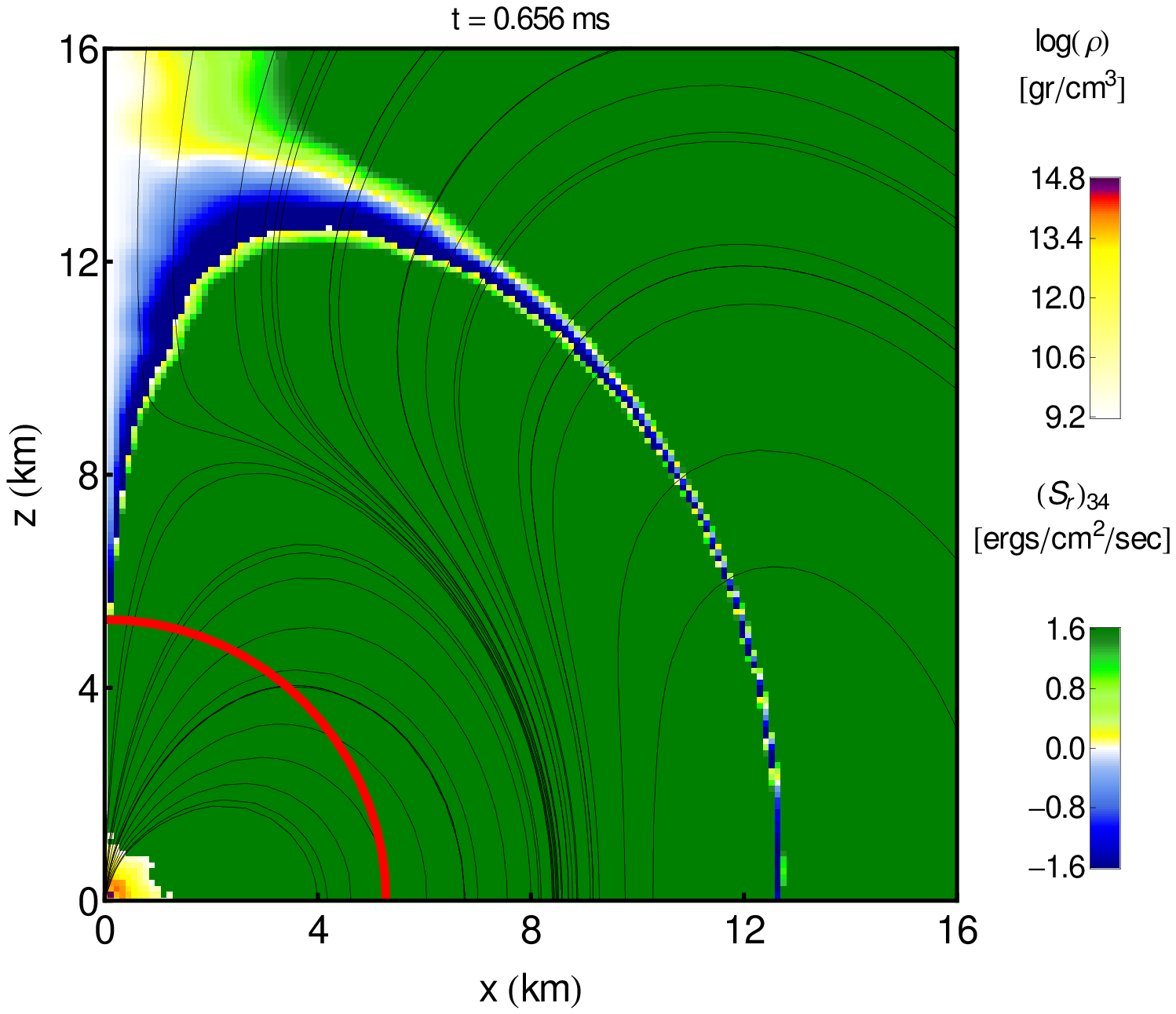}
 \includegraphics[angle=0,width=0.33\textwidth]{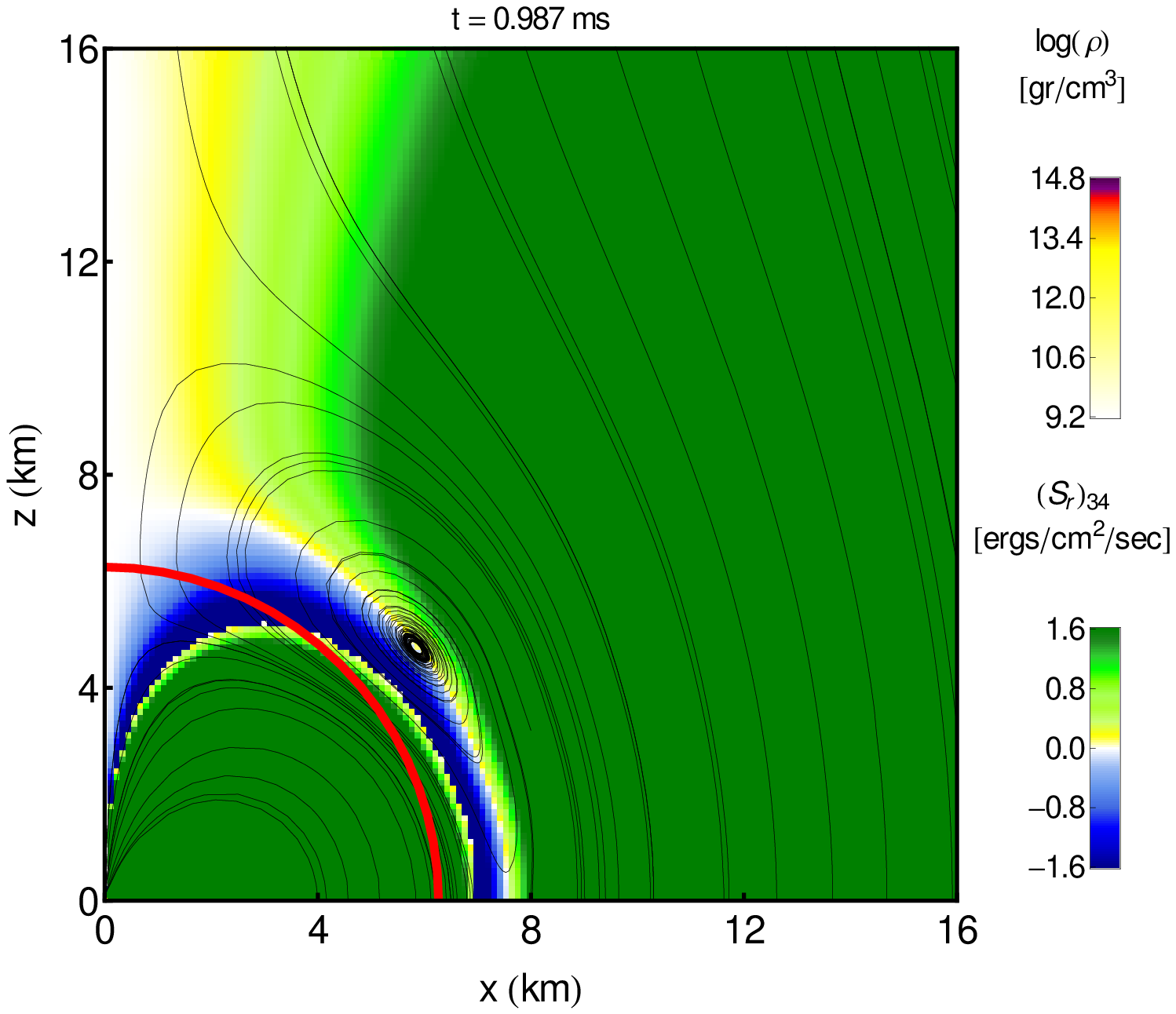}
 \includegraphics[angle=0,width=0.33\textwidth]{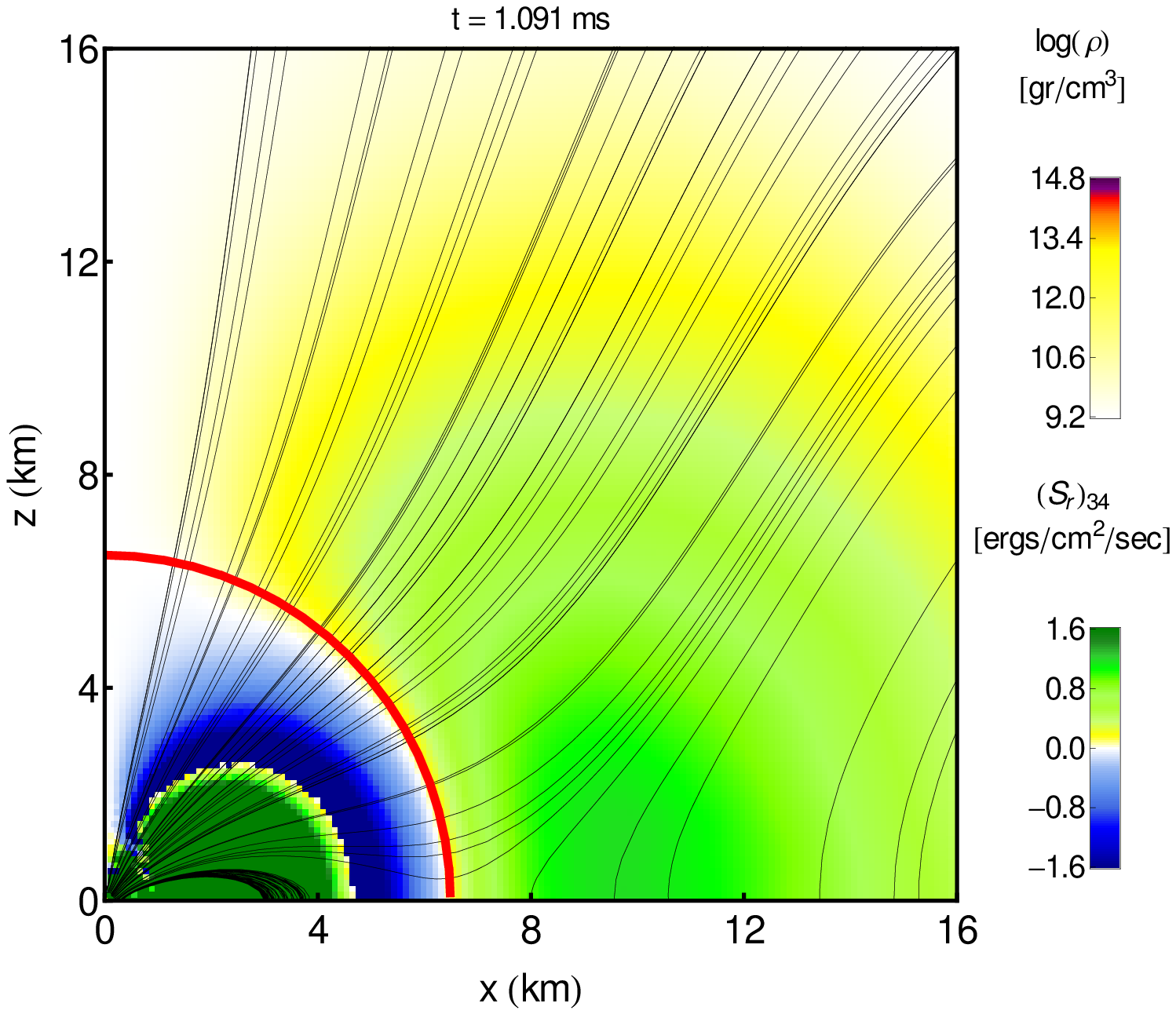}
 \end{center} 
\caption{The same as the three bottom panels of
  Fig.~\ref{fig:collapse_rho_poynting} but with a linear scale of $15.07$
  km to highlight the dynamics near the horizon. It is now very clear
  that a closed set of magnetic field lines is built just outside the
  horizon at $t=1.0$ ms, that is radiated away as QNM of the BH.}
\label{fig:collapse_rho_poynting_zoomed}
\end{figure*}

\begin{figure*}
 \begin{center}
 \includegraphics[angle=0,width=0.33\textwidth]{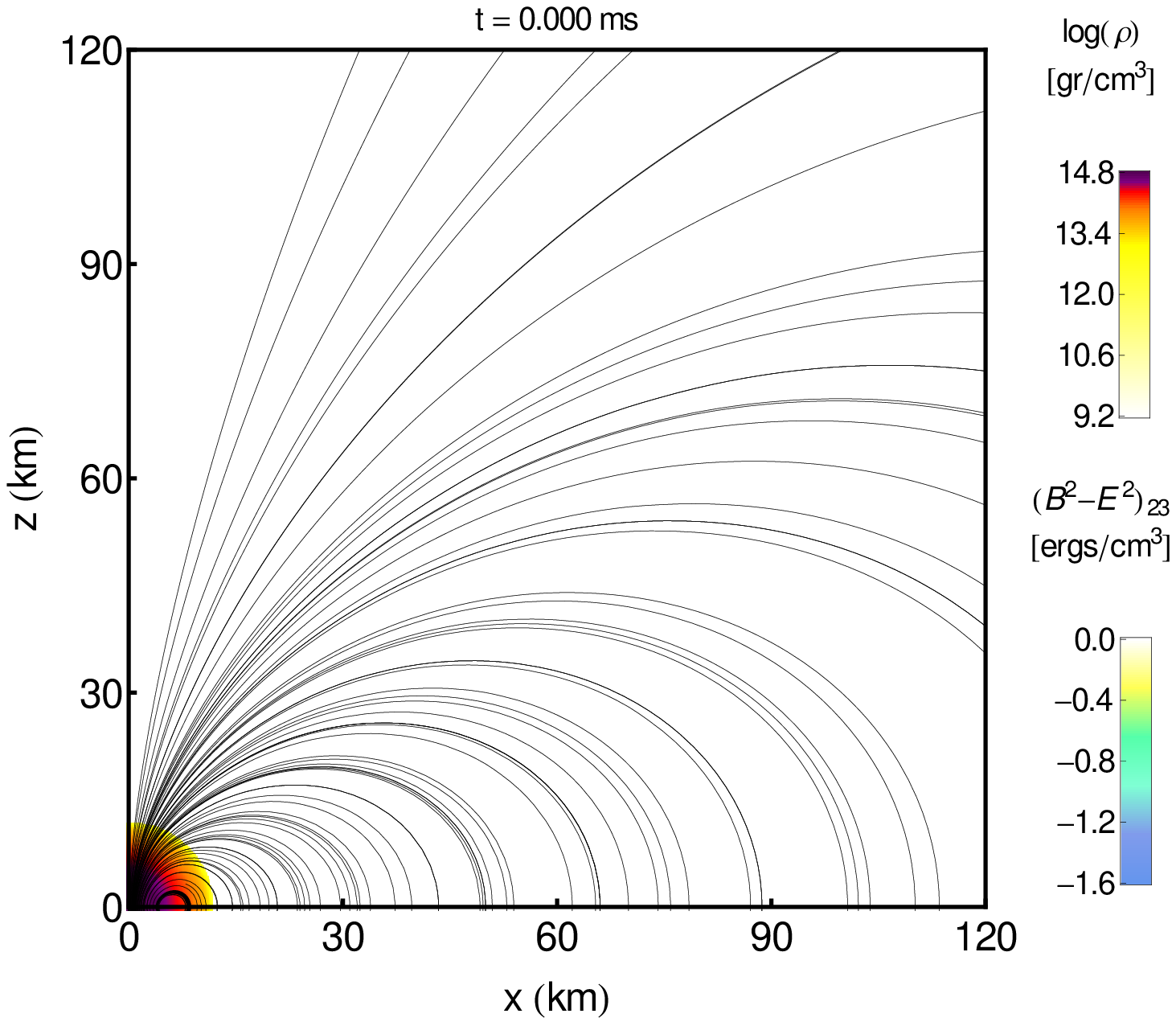}
 \includegraphics[angle=0,width=0.33\textwidth]{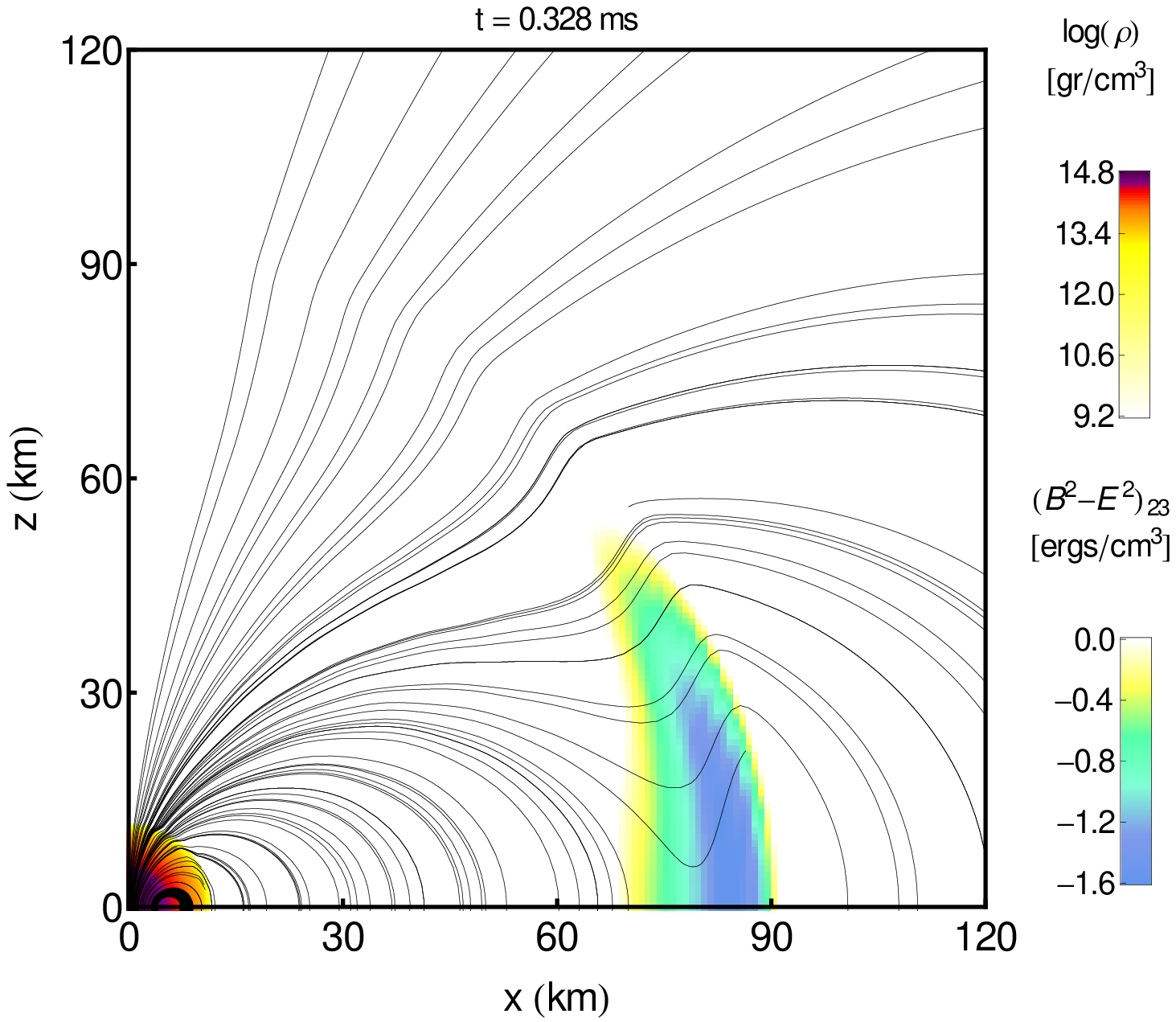}
 \includegraphics[angle=0,width=0.33\textwidth]{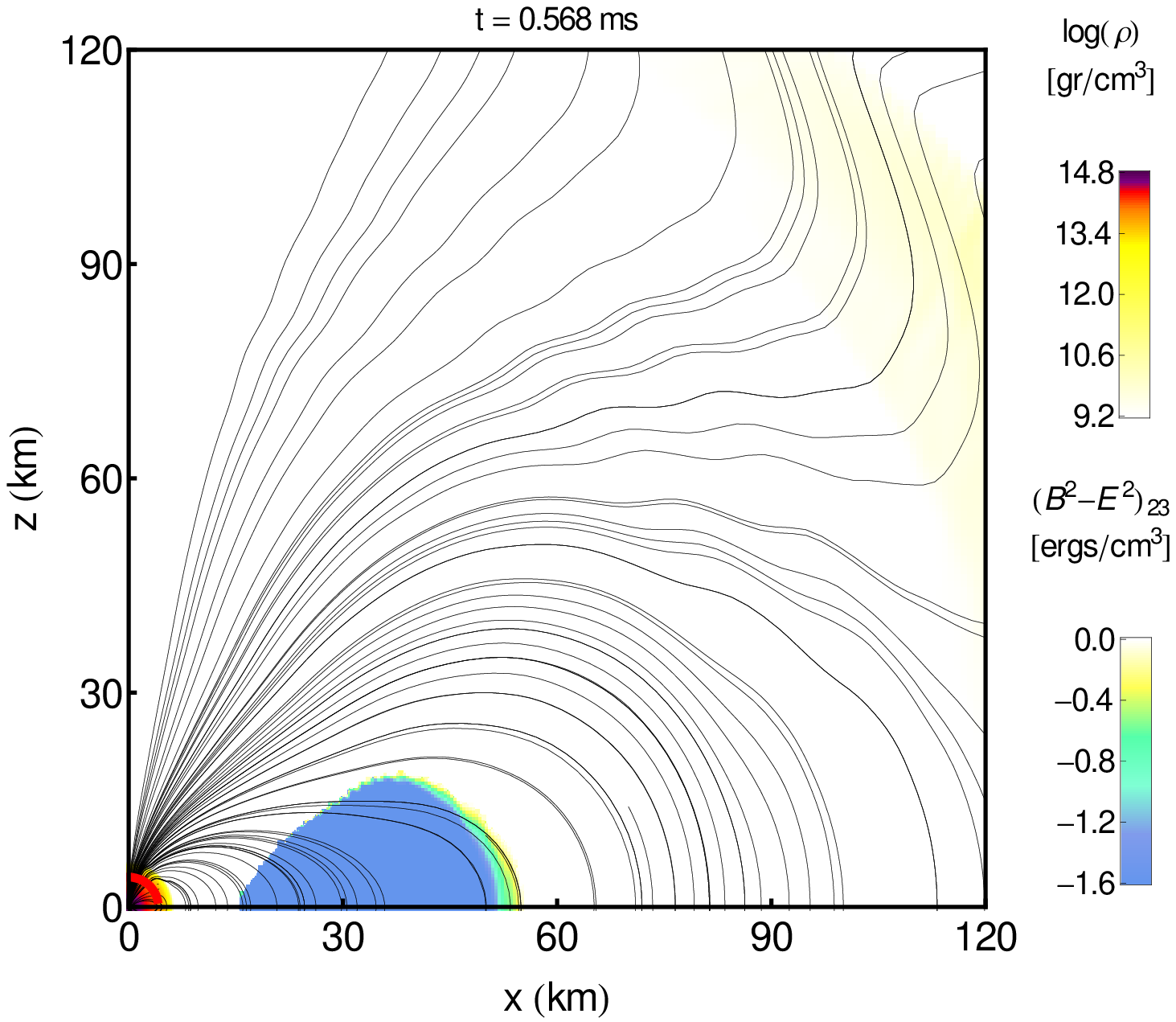}
 \includegraphics[angle=0,width=0.33\textwidth]{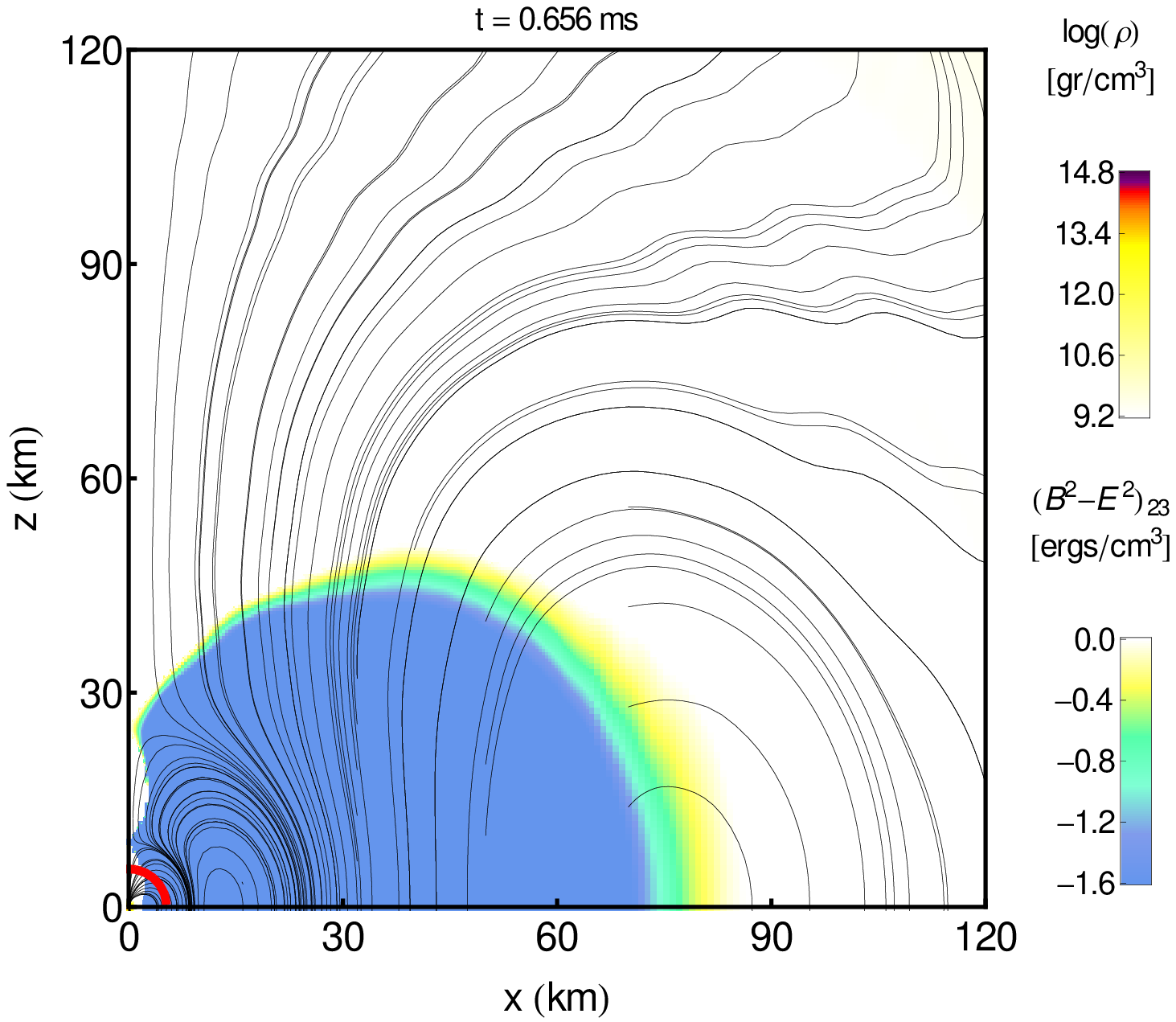}
 \includegraphics[angle=0,width=0.33\textwidth]{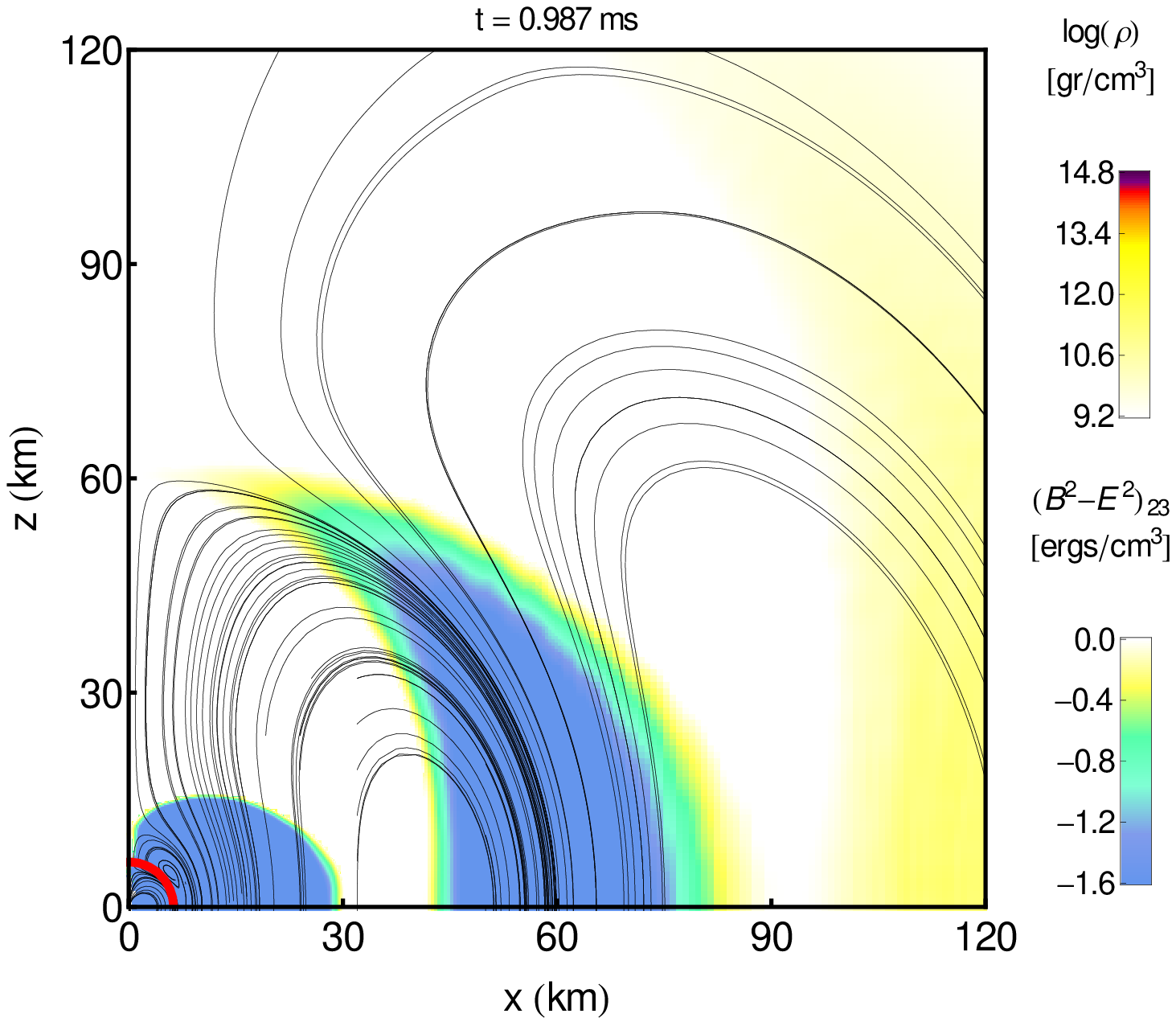}
 \includegraphics[angle=0,width=0.33\textwidth]{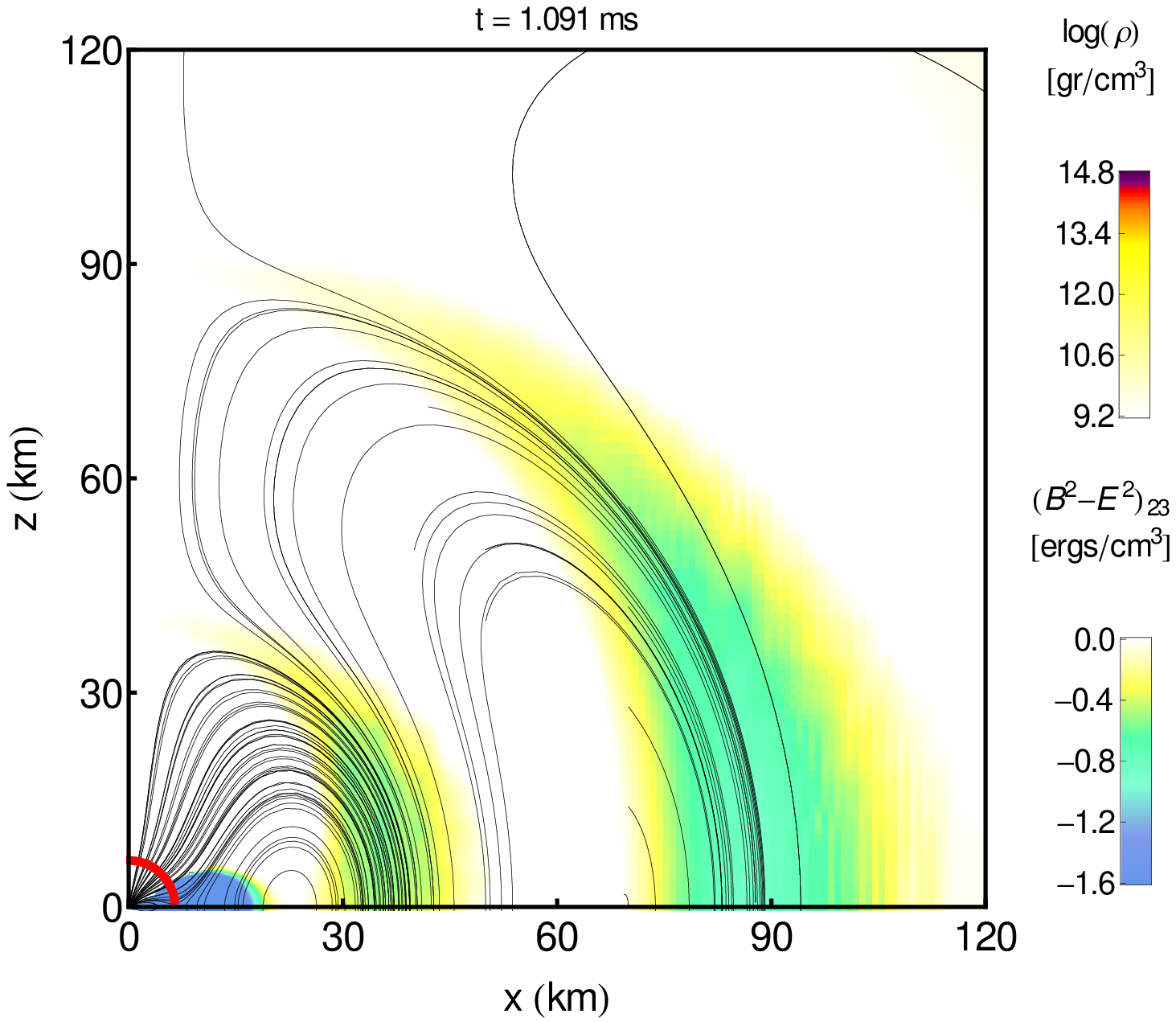}
 \end{center} 
\caption{The same as in Fig.~\ref{fig:collapse_rho_poynting}, but
  where in addition to the rest-mass density (color code from white to
  red) and the magnetic-field lines (thin solid lines) we show the
  electrically-dominated regions (i.e. $B^2 - E^2 < 0$, color code from
  light blue to white in units of $10^{23}$).}
\label{fig:collapse_rho_b2e2}
\end{figure*}

%%%%%%%%%%%%%%%%%%%%%%%%%%%%%%%%%%%%%%%%%%%%%%%%%%%%%%%%%%%%%%%%%%%%%%%%

\subsubsection{Magnetized collapse to a black hole}

Our final and most comprehensive test is represented by the collapse to a
BH of a magnetized nonrotating star. This is more than a purely numerical
test as it simulates a process that is expected to take place in
astrophysically realistic conditions, such as those accompanying the
merger of a binary system of magnetized neutron stars~\cite{Etienne08,
  Giacomazzo:2009mp}, or of an accreting magnetized neutron star. The
interest in this process lays in that the collapse will not only be a
strong source of gravitational waves, but also of electromagnetic
radiation, that could be potentially detectable (either directly or as
processed signal). The magnetized plasma and electromagnetic fields that
surround the star, in fact, will react dynamically to the rapidly
changing and strong gravitational fields of the collapsing star and
respond by emitting electromagnetic radiation. Of course, no
gravitational waves can be emitted in the case considered here of a
nonrotating star, but we can nevertheless explore the electromagnetic
emission and assess, in particular, the efficiency of the process and
thus estimate how much of the available binding energy is actually
radiated in electromagnetic waves. Our setup also allows us to
investigate the dynamics of the electromagnetic fields once a BH is
formed and hence to assess the validity of the no-hair theorem, which
predicts the exponential decay of any electromagnetic field in terms of
quasinormal mode (QNM) emission from the BH~\cite{Price72,Kokkotas99b}.

Ours is not the first detailed investigation of this process and relevant
previous studies are that in Ref.~\cite{Baumgarte02b2} and the more
recent one in Ref.~\cite{Lehner2011}. However, our approach differs from
previous ones in that it correctly describes the gravitational dynamics
of a collapsing fluid (the semianalytical work in
Ref.~\cite{Baumgarte02b2}, in fact, considered the more rapid collapse of
a dust sphere, for which the Oppenheimer-Snyder (OS) analytic solution
can be used~\cite{Oppenheimer39a}) and does not require any matching of
the solution near the stellar surface (the fully relativistic work in
Ref.~\cite{Lehner2011} had to resort to an ingenious matching between
the interior ideal-MHD solution and a force-free one in the
magnetosphere), leaving the complete evolution of the electromagnetic
fields to our prescription~\eqref{eq:sigma} of a nonuniform
conductivity. Indeed, our solution in the case of vanishing conductivity and charges
is expected to be locally that in electrovacuum, and thus to be very
similar to the force-free one with vanishing charges and
currents. However, these two limits will differ in regions where $B^2 -
E^2<0$ and where an anomalous resistivity appears, leading to different
global solutions at later times. Since we can handle such resistive regions, this test
illustrates the capabilities of our resistive implementation and serves
as an improved approach to this astrophysical scenario than the one
in~\cite{Lehner2011}, although it is still rather crude.

In practice, we have considered the evolution of a nonrotating neutron
star with a gravitational mass of $2.75 M_{\odot}$, which is chosen to
sit on the unstable branch of the equilibrium configurations and is
endowed with an initial poloidal magnetic field of strength $B_c=5
\times 10^{15}$ G extending also in the exterior space. As for the
previous stellar solutions, we use a polytropic EOS with $\Gamma=2$
and $K=364.7$ for the initial data and continue to use this isentropic
EOS also for the subsequent evolution. The evolutions have been
carried out in a computational domain with outer boundary at $R_{\rm
  out} = \pm 241.07$ km and a resolution of $\Delta x = 0.111$ km,
corresponding to $272$ points covering the finest grid which extends
up to $\pm 15.07$ km.

Because the magnetic energy is only a small fraction of the binding
energy, the hydrodynamical and spacetime evolution of the fluid star
as it collapses to a BH is very similar to the unmagnetized case and
this has been discussed in great detail in~\cite{Baiotti04}. The most
important difference, therefore, is in the dynamics of the magnetic
field, and this is shown in Fig.~\ref{fig:collapse_rho_poynting},
which reports two-dimensional cuts on the $(x,z)$ plane of the
collapse to a BH of a magnetized NS. Shown with colors are the
rest-mass density (color code from white to red) and the radial
Poynting vector (color code from blue to green), while thin solid lines
reproduce the magnetic-field lines.

At early times the star remains close to its initial state with the
exception of a small transient induced by truncation error, which
produces a small radiative outburst at $t\lesssim 0.3$ ms. As the
instability to gravitational collapse develops, there is a
rearrangement of the external electromagnetic fields, driven by a
toroidal electric field $E_{\phi} \approx -v_r B_{\theta}$ produced in
the interior of the perfectly conducting star, and which is continuous
across the stellar surface. As the collapse proceeds, the rest-mass
density in the center and the curvature of the spacetime increase
until an apparent horizon is found at $t=0.57$ ms and is marked with
a thin red line in Fig.~\ref{fig:collapse_rho_poynting} (we have used
the apparent-horizon finder described
in~\cite{Thornburg2003:AH-finding}).

\begin{figure}
\begin{center}
\includegraphics[width=0.45\textwidth]{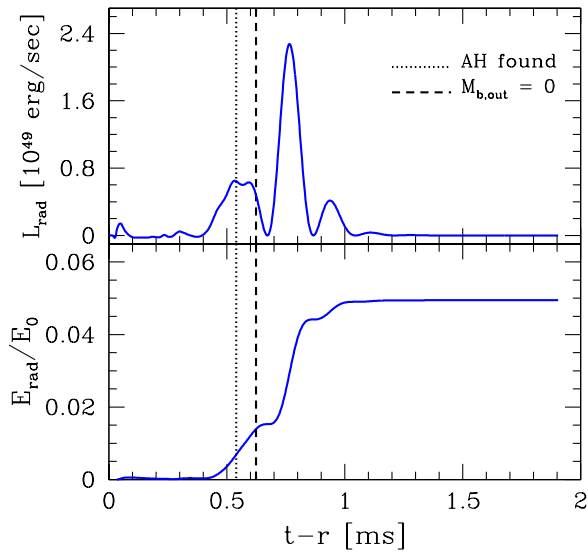}
\end{center} 
  \caption{\textit{Top panel:} Luminosity calculated at a distance
    $r=88.63$ km from the compact object. The black dotted line
    represents the time at which the apparent horizon is formed and
    the black dashed line corresponds to the time at which all the
    matter is well within the horizon. \textit{Bottom panel:}
    Evolution of the total radiated energy normalized to the initial
    magnetic energy.}
 \label{fig:collapse_luminosity}
\end{figure}

\begin{figure*}
\begin{center}
  \includegraphics[width=0.45\textwidth]{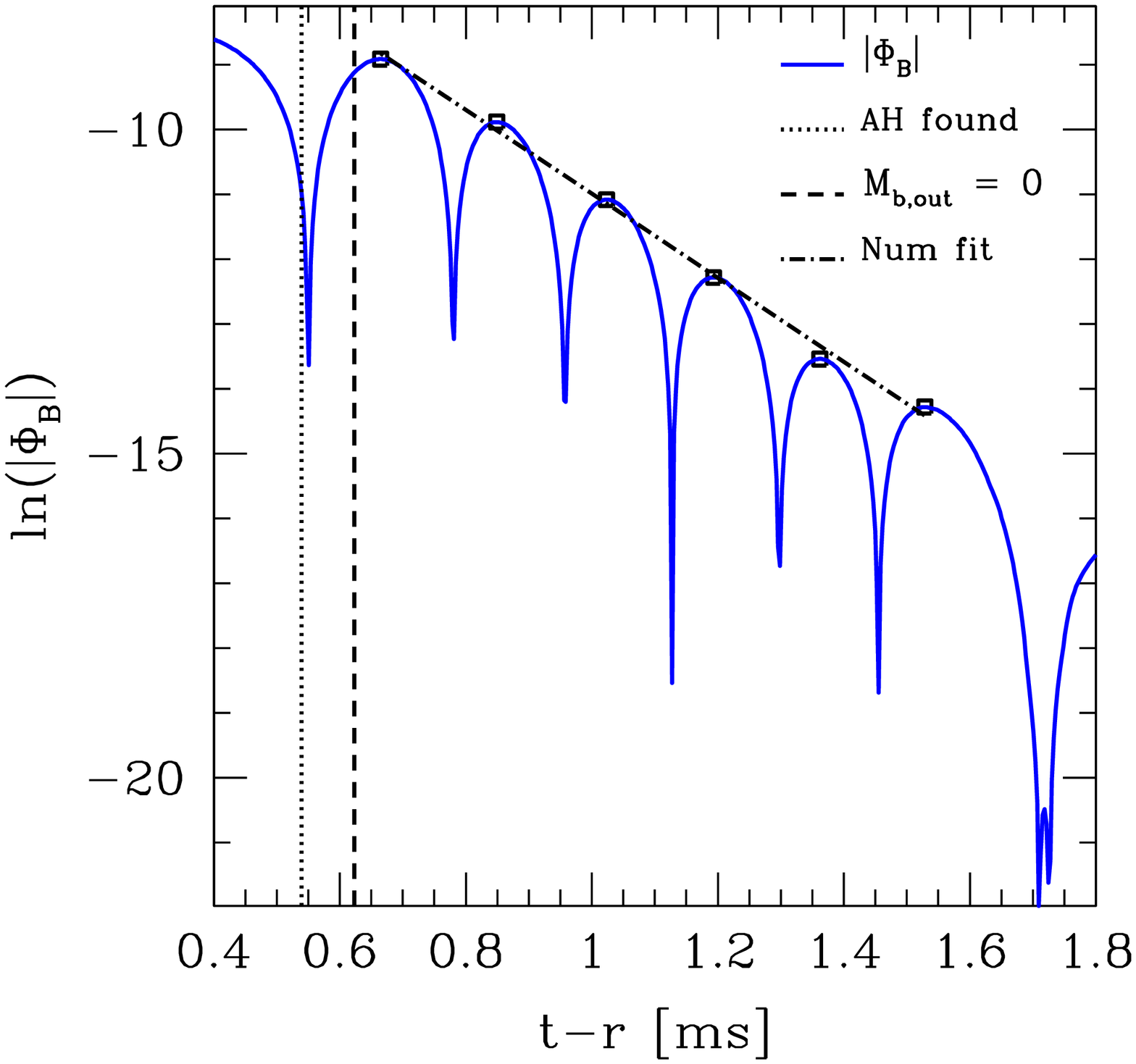}
  \includegraphics[width=0.45\textwidth]{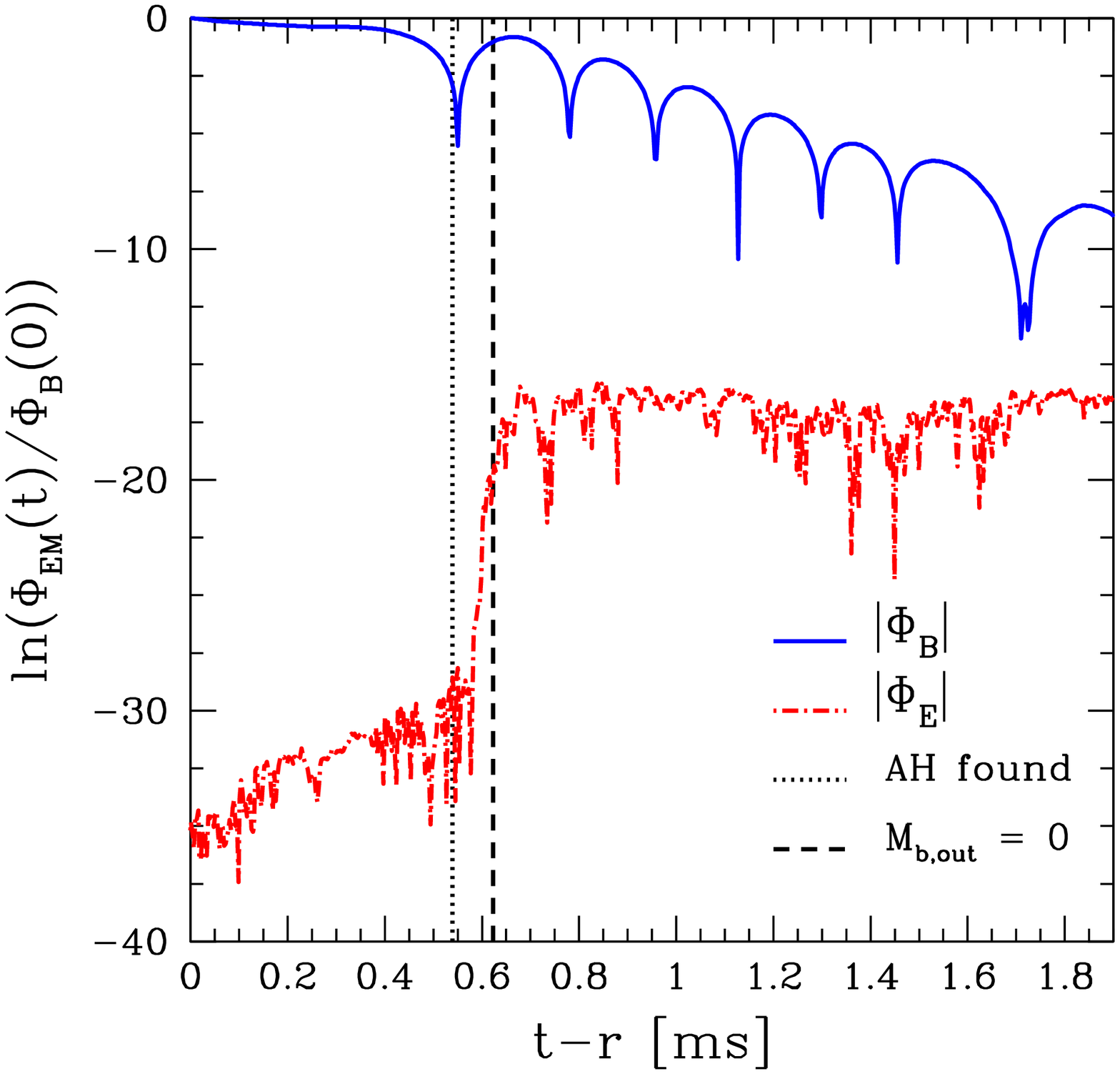}
\end{center} 
  \caption{\textit{Left panel:} QNM ringdown of the magnetic field as
    measured through the magnetic flux at $r=36.93$ km. Again, the black
    dotted line represents the time at which the apparent horizon is
    formed and the black dashed line corresponds to the time at which
    all the matter is well within the horizon; the dot-dashed line
    represents instead our fit to an exponential decay. \textit{Right
      panel:} Logarithm of the absolute values of the magnetic and
    electric fluxes as normalized to the initial magnetic flux.}
  \label{fig:collapse_QNM}
\end{figure*}

As the stellar matter is accreted onto the BH (the rest-mass outside the
horizon $M_{\rm b,\,out}=0$ is zero by $t \gtrsim 0.62$ ms), the external
magnetic field which was anchored on the stellar surface becomes
disconnected, forming closed magnetic-field loops which carry away the
electromagnetic energy mostly in the form of dipolar radiation. This
process, which has been described through a simplified nonrelativistic
analytical model in Ref.~\cite{Lehner2011}, predicts the presence of
regions where $|E|>|B|$ as the toroidal electric field propagates
outwards as a wave. This process can be observed very clearly in
Fig.~\ref{fig:collapse_rho_poynting_zoomed}, which displays the same
three bottom panels of Fig.~\ref{fig:collapse_rho_poynting} on a smaller
scale of only $15.07$ km to highlight the dynamics near the horizon. In
particular, it is now very clear that a closed set of magnetic field
lines is built just outside the horizon at $t=1.0$ ms, that is radiated
away. Note also that our choice of gauges (which are the same used
in~\cite{Baiotti06}) allows us to model without problems also the
solution inside the apparent horizon. While the left panel of
Fig.~\ref{fig:collapse_rho_poynting_zoomed} shows that most of the
rest-mass is dissipated away already by $t=0.65$ ms (see discussion
in~\cite{Thierfelder10} about why this happens), some of the matter
remains on the grid near the singularity, anchoring there the magnetic
field which slowly evolves as shown in the middle and right panels.  A
complementary view of the collapse process is also offered by
Fig.~\ref{fig:collapse_rho_b2e2}, which reports, in addition to the
rest-mass density (color code from white to red) and the magnetic-field
lines (thin solid lines), also the electrically-dominated regions
(i.e., $B^2 - E^2 < 0$, color code from light blue to white). The larger
scales used in this case makes it easier to follow the dynamics of the
closed field lines that once produced near the horizon, propagate as
dipolar radiation at infinity.

The total electromagnetic luminosity $L_{\rm rad}$ emitted during the
collapse and computed as a surface integral of the Poynting flux over a
spherical surface at $88.63$ km, computed through equation
\begin{equation}
  L_{rad}(r)= 4\int_{\theta=0}^{\pi/2}\int_{\phi=-\pi/2}^{\pi/2} -T^r_t \ r^2 \sin\theta \ d\theta d\phi,
\end{equation}
is shown in the top panel of
Fig.~\ref{fig:collapse_luminosity}. Note the presence of a rise during
the collapse and of several pulses after the stellar matter has been
accreted onto the black hole. The vertical dotted line represents the
time at which the apparent horizon is first found, while the vertical
dashed line corresponds to the time at which all the matter is within the
horizon (i.e., $M_{\rm b,\,out}=0$). The peaks in the electromagnetic
luminosity correspond to the closed magnetic-field loops that disconnect
from the star and transport electromagnetic energy. The bottom panel of
Fig.~\ref{fig:collapse_luminosity}, on the other hand, reports the
evolution of the total electromagnetic energy lost in radiation $E_{\rm
  rad}$ computed through
\begin{equation}
  E_{rad}(r)= \int_t L_{\rm{rad}} \ dt
\end{equation}
and when normalized to the value of the initial magnetic energy
outside the star, $E_0$. Our results indicate therefore, a total
electromagnetic efficiency which is $\simeq 5\%$; this result is smaller
than the estimate made in Ref.~\cite{Baumgarte02b2} (which was of $\simeq
20\%$), but, besides the different initial data used, this difference can
be easily accounted for by the fact that the gravitational collapse
simulated here is considerably slower (and hence inefficient) than the OS
one computed in ~\cite{Baumgarte02b2}, where matter is free falling. Our
efficiency is also smaller than the one computed in
Ref.~\cite{Lehner2011} and which is $\sim 16\%$ once the same definition
for $E_0$ is used. However, many other factors could be behind this
difference, e.g., differences in the initial data (use of a dipole
everywhere in contrast to a dipole only outside the star as in our case),
differences in the stellar models, differences in the numerical approach
(treatment of the surface of the star of the transition between ideal and
force free MHD), and that our stellar exterior is electrovacuum and not
force free. A closer comparison between the two approaches will be
carried out in a separate work.

After BH formation, the luminosity decreases exponentially in a
fashion which is typical of the QNM ringing of an electrovacuum
electromagnetic field in a Schwarzschild BH spacetime. These QNMs are
clearly visible also in the (absolute value of the) magnetic flux
\begin{equation}
  \Phi_{B|\rm{hem.}}(r)= 2\int_{\theta=0}^{\pi/2}\int_{\phi=-\pi/2}^{\pi/2} B^r \ r^2 \sin\theta \  d\theta d\phi
\end{equation}
shown in the left panel of Fig.~\ref{fig:collapse_QNM}, from which a
comparison with the perturbative expectations can be made. More
specifically, by fitting the harmonic oscillations of the ringdown and
the exponential decay we have computed the frequencies of the
``ringing-down'' magnetic-field flux for the $\ell=1$ mode to be $\omega=
0.344054 - i\, 6.46731\, \rm{kHz}$, corresponding to a nonrotating
black hole of $2.74\, M_{\odot}$. The agreement with the analytical
value is excellent, with a relative error of only $\sim 0.7\%$ for the
real part of the frequency and $\sim 5.6\%$ for the imaginary
one~\cite{Kokkotas99a}.

Finally, as a measure of the accuracy of our simulation we can compare
the magnetic flux with the corresponding electric flux, which should
vanish in the continuum limit since no net electric charge should be
present. This is indeed the case, as can be deduced from the right
panel of Fig.~\ref{fig:collapse_QNM}, which reports the two fluxes
normalized to the initial magnetic flux. Note that the electric flux
is about 30 orders of magnitude smaller than the magnetic flux before
BH formation, increasing after an apparent horizon is found, but
remaining 15--10 orders of magnitude smaller.

%%%%%%%%%%%%%%%%%%%%%%%%%%%%%%%%%%%%%%%%%%%%%%%%%%%%%%%%%%%%%%%%%%%%%%%%

\section{Conclusions}
\label{conclusions}

We have introduced a general-relativistic resistive MHD formalism as
an extension of the special relativistic resistive MHD formalism
reported in Ref.~\cite{Palenzuela:2008sf} for a 3+1 decomposition of
the spacetime. Our numerical implementation has been made within the
\texttt{Cactus} computational infrastructure as a continuation of the
already existing general-relativistic hydrodynamics code
\texttt{Whisky}~\cite{Baiotti03a,Baiotti04} and of the ideal-MHD code
\texttt{WhiskyMHD}~\cite{Giacomazzo:2007ti}. 

Our numerical approach exploits implicit-explicit (IMEX) methods and
allows us to treat astrophysical problems in which different spatial
regions fall into different regimes of conductivities. The flexibility
introduced by using the Runge-Kutta will allow us to consider not only
more general Ohm laws and a variety of astrophysical
dynamos~\cite{Bonanno:2003uw,Bucciantini2012}, but also to use better
dispersion relations as in ~\cite{Takamoto2011b}, to calculate the
velocities in the HLLE method and to describe more accurately also the
nonrelativistic limit.

Our implementation has been tested for a number of stringent tests and
its robustness has been verified. The special-relativistic tests
involved the propagation of circularly polarized Alfv\'en waves, the
evolution of current sheets and shock-tubes in one dimension,
cylindrical and spherical explosion tests in two and three dimensions
respectively, the evolution of stable and the collapse of unstable
magnetized stars in dynamical spacetime. We have compared our
numerical results either with the analytical solution (in the cases
where one exists), or with the numerical ideal-MHD solution (in the
limit of high conductivity), proving that our implementation is
suitable to describe regions with a wide range of conductivities, with
or without large discontinuities and shocks.

We have also considered genuinely general-relativistic tests in terms
of the evolution of nonrotating magnetized stars either with fixed or
fully dynamical spacetimes. Our stars have been endowed with magnetic
fields of varying strength, either confined in their interior or
permeating also the exterior space, and have been modeled with a
nonuniform conductivity that allows us to recover the ideal-MHD limit
in the interior of the star and such the electrovacuum limit outside
the star. All of our results indicate that the resistive
implementation is able to follow the evolution of the oscillations
triggered by the small truncation errors and that the associated
eigenfrequencies match well those either reported with other
hydrodynamics and ideal-MHD codes \cite{Baiotti04:shortal,
  Giacomazzo:2007ti} or from perturbation theory.

Finally, we have considered the challenging and comprehensive test
represented by the gravitational collapse of a magnetized nonrotating
star to a BH. This scenario has an astrophysical interest of its own
as it could lead to the emission of electromagnetic radiation,
potentially detectable. Indeed we have found that as the collapse
proceeds, electrically dominated regions develop and lead to the
development of magnetic-field loops that propagate at the speed of
light, carrying away electromagnetic energy. Up to $5\%$ of the
initial magnetic energy can be lost in this way and the following
evolution of the magnetic field follows a clean exponential decay, as
expected by an electromagnetic perturbation in a Schwarzschild
spacetime. The match of the measured QNMs and the perturbative
predictions is well of a few percent or less.

Our new code is now ready to be applied to study a variety of
astrophysical scenarios. These include the modeling of the
magnetosphere that could be produced after the merger of binary
neutron stars, or when the hypermassive neutron star collapses to a BH
and is surrounded by a hot torus. The work in
Ref.~\cite{Rezzolla:2011} has already reported that under these
conditions strong magnetic fields can be produced and that a jetlike
magnetic structure can develop. It is exciting to consider whether the
resistive losses that are expected in the process will provide
sufficient energy to launch of a powerful jet, not yet observed in
Ref.~\cite{Rezzolla:2011}. Also of great interest is to study BH
magnetospheres and the origin of jets so as to answer the question of
whether an ergosphere is critical for the development of the
Blandford-Znajek mechanism. Finally, our approach is also well suited
to study the properties of accretion disk onto BHs and to elucidate
the role that resistive losses play on the whole energetic budget. We
will report on these applications in forthcoming works.

%%%%%%%%%%%%%%%%%%%%%%%%%%%%%%%%%%%%%%%%%%%%%%%%%%%%%%%%%%%%%%%%%%%%

\begin{acknowledgments}

It is a pleasure to thank Wolfgang Kastaun and Kentaro Takami for their
help in the eigenfrequencies of the oscillating stars and Nikolaos
Stergioulas for helpful discussions. This work was supported in part by
the DFG grant SFB/Transregio~7 and by ``CompStar'', a Research Networking
Programme of the European Science Foundation. The computations were made
at the AEI and also on the cluster RANGER at the Texas Advanced Computing
Center (TACC) at The University of Texas at Austin through XSEDE grant
No. TG-PHY110027. BG acknowledges support from NASA Grant No. NNX09AI75G
and NSF Grant No. AST 1009396.

\end{acknowledgments}

%%%%%%%%%%%%%%%%%%%%%%%%%%%%%%%%%%%%%%%%%%%%%%%%%%%%%%%%%%%%%%%%%%%%
\begin{appendix}

\begin{figure}
  \includegraphics[width=0.45\textwidth]{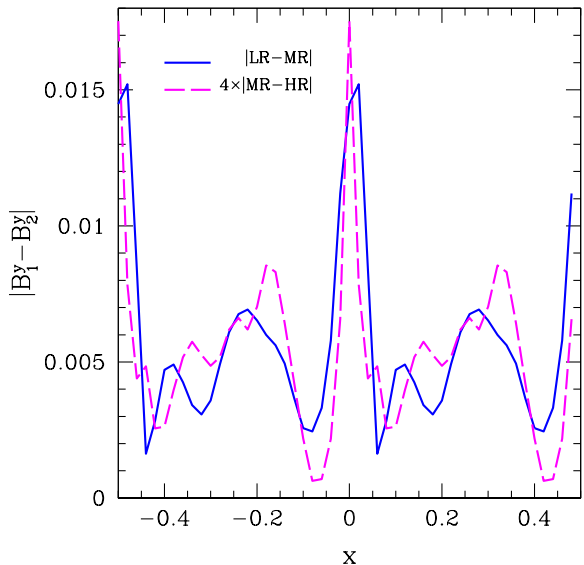}
  \caption{{\em Convergence tests.} Absolute errors of the solution of
    the $y$ component of the magnetic field $B^y$ along $x$ for different
    resolutions rescaled by a factor of four (as dictated by our scheme)
    regarding the one-dimensional Alfv\"{e}n-wave test we have performed in
    flat spacetime (Sec.~\ref{sec:Alfvenwave}) for a uniform conductivity
    $\sigma=10^6$.}
  \label{fig:convergencealfvenwave1}
\end{figure}

\begin{figure*}
  \includegraphics[width=0.45\textwidth]{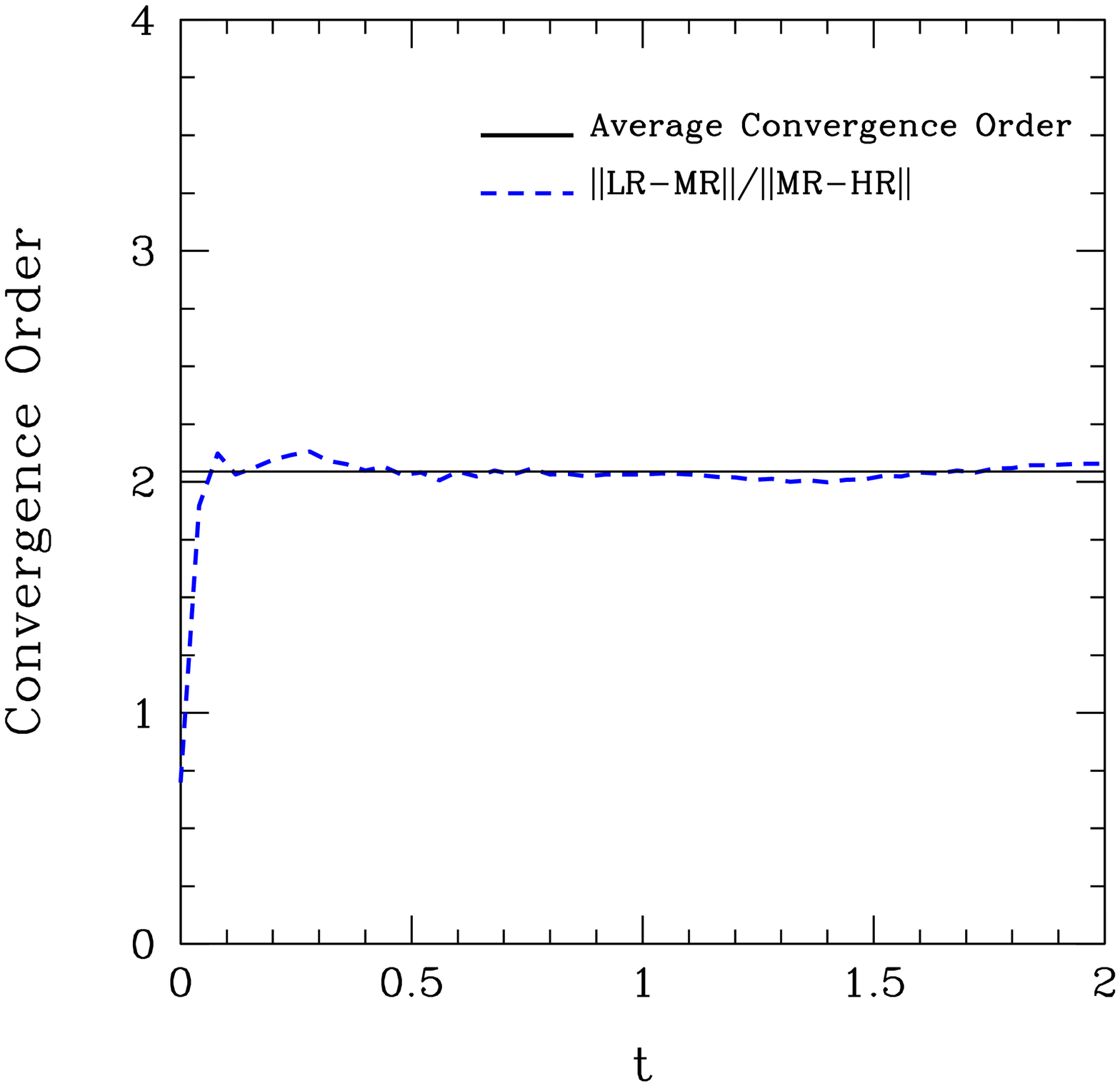}
  \hskip 0.5cm
  \includegraphics[width=0.45\textwidth]{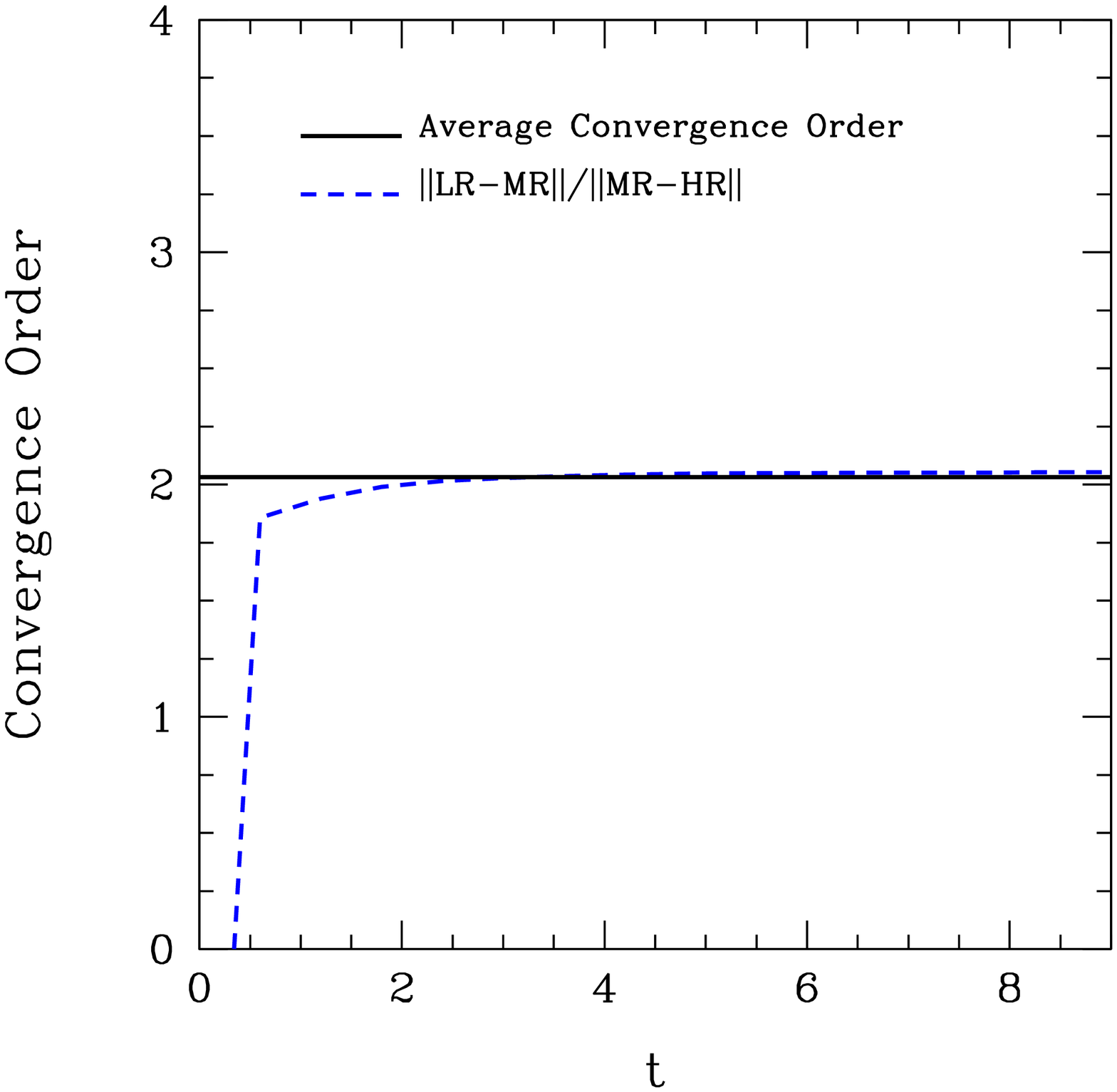}
  \caption{{\em Left panel:} Convergence order as a function of time for
    the one-dimensional Alfv\'en-wave test we have performed in flat
    spacetime (Sec.~\ref{sec:Alfvenwave}) for a uniform conductivity
    $\sigma=10^6$. The average convergence order is $2.05$. {\em Right
      panel:} The same as in the left one, but for the one-dimensional
    current-sheet test in Sec.~\ref{sec:CurrentSheet} for a uniform
    conductivity $\sigma=100$. The average convergence order is $2.03$.}
  \label{fig:convergencealfvenwave2}
\end{figure*}

\section{Convergence Tests}
In this section we study the convergence properties of our numerical
implementation. In general, for smooth data, one can study the
self-convergence order of a flux-conservative scheme by just comparing
the L1-norm of the relative errors of the solution for at least three
different resolutions and by checking that the rescaled absolute errors
lie almost on top of one another. In this way we can infer the
convergence order of our method by solving the following equation,

\begin{equation*}
  \frac{||B^y_{\rm{LR}}-B^y_{\rm{MR}}||_1}{||B^y_{\rm{MR}}-B^y_{\rm{HR}}||_1}=\frac{\Delta x_{\rm{LR}}^{\Gamma}-\Delta x_{\rm{MR}}^{\Gamma}}{\Delta x_{\rm{MR}}^{\Gamma}-\Delta x_{\rm{HR}}^{\Gamma}}
\end{equation*}
where $\Gamma$ is the convergence order, and plot it as a function of
time. Since in the tests considered here we always double the resolution
we expect the scaling factor to be equal to $2^{\Gamma}$. Since there
exist no well-posed convergence tests in resistive relativistic MHD we
choose to perform self-convergence tests for three different resolutions
for a variety of one-dimensional setups.

We report the convergence order of our scheme as a function of time for
some of the tests presented in (Sec.~\ref{sec:OneDtests}). More
specifically, we examine the evolution of a CP-Alfv\'en wave
(Sec.~\ref{sec:Alfvenwave}) to infer the convergence order of our method
in the high conductivity and nearly ideal-MHD regime and the evolution of
the self-similar current sheet presented in Sec.~\ref{sec:CurrentSheet})
to check the low-to-medium conductivity $\sigma=100$ regime for three
resolutions. Finally we repeat the same calculations for the shocktube
tests in the high/low conductivity regime presented in
Sec.~\ref{sec:ShocktubeTests} in order to check how the code behaves in
the presence of discontinuities.

In Fig.~\ref{fig:convergencealfvenwave1} we depict the absolute pointwise
error of the solution of the y-component of the magnetic field $B^y$
($|B^y_{\rm{1}}-B^y_{\rm{2}}|$) regarding the CP-Alfv\'en wave for
resolutions with $50$ and $100$ number of points along $x$, and $100$ and
$200$ with blue and magenta colors respectively, rescaled with the
appropriate factors. The convergence order as a function of time for this
test is shown in the left panel of
Fig.~\ref{fig:convergencealfvenwave2}. Taking the average over time we
conclude that our implementation in the high conductivity regime
converges at an order of $2.05$ and agrees with the anticipated value for
the linear reconstruction scheme employed (linear reconstruction with a
monotonized central-differences slope limiter~\cite{Leveque2002}).

In the right panel of Fig.~\ref{fig:convergencealfvenwave2} we show again
the convergence order of our scheme as a function of time in the
medium-low conductivity regime regarding the evolution of a current
sheet. The resolutions considered for this test are 50, 100 and 200
points along the $x$-axis. The average convergence order for the current
sheet test is $2.03$. Our method is once again second order accurate as
expected for a linear reconstruction scheme with a van Leer-type slope
limiter~\cite{Toro99} and smooth initial data.

\begin{figure*}
  \includegraphics[width=0.32\textwidth]{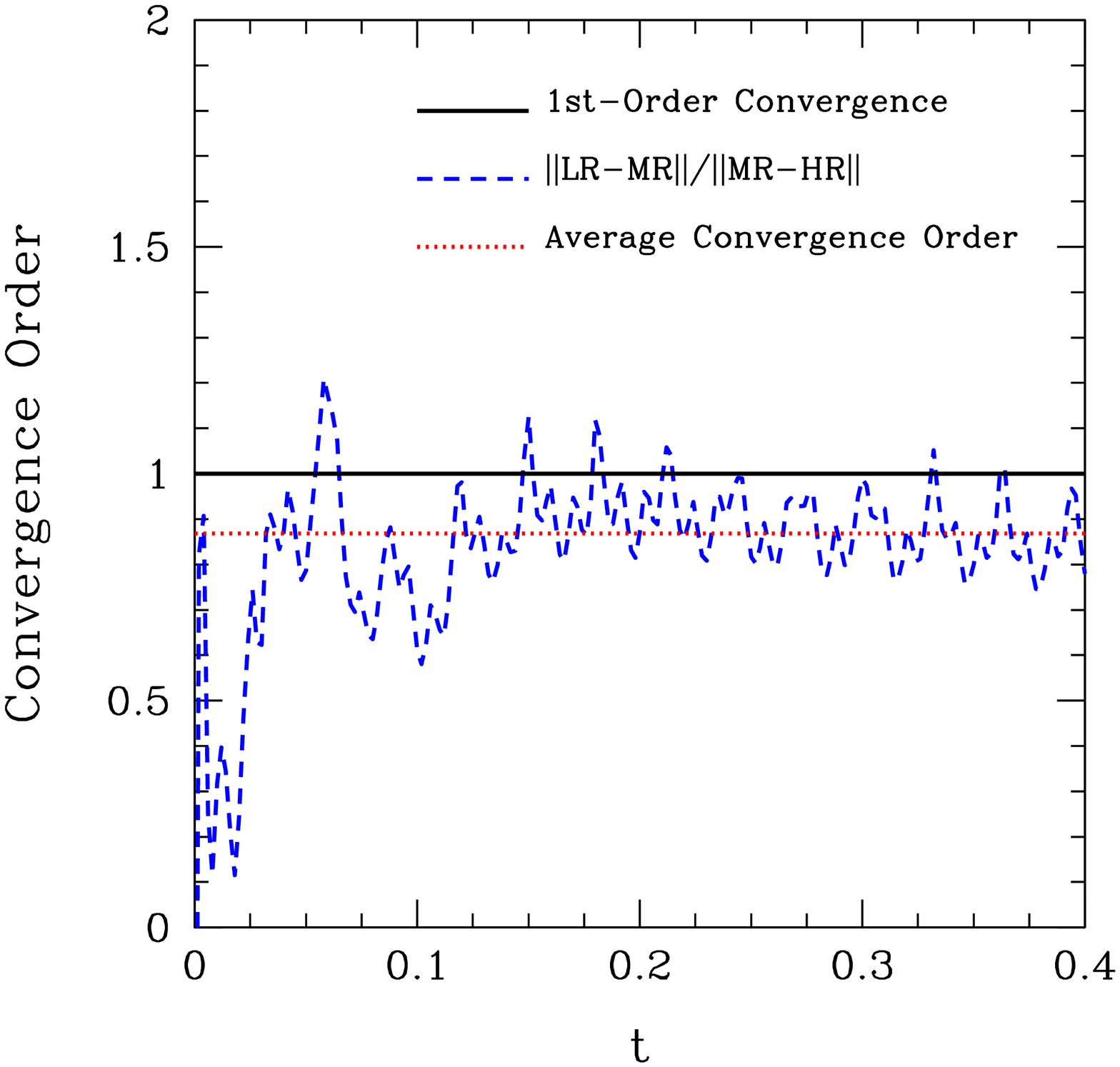}
  \hskip 0.25cm
  \includegraphics[width=0.32\textwidth]{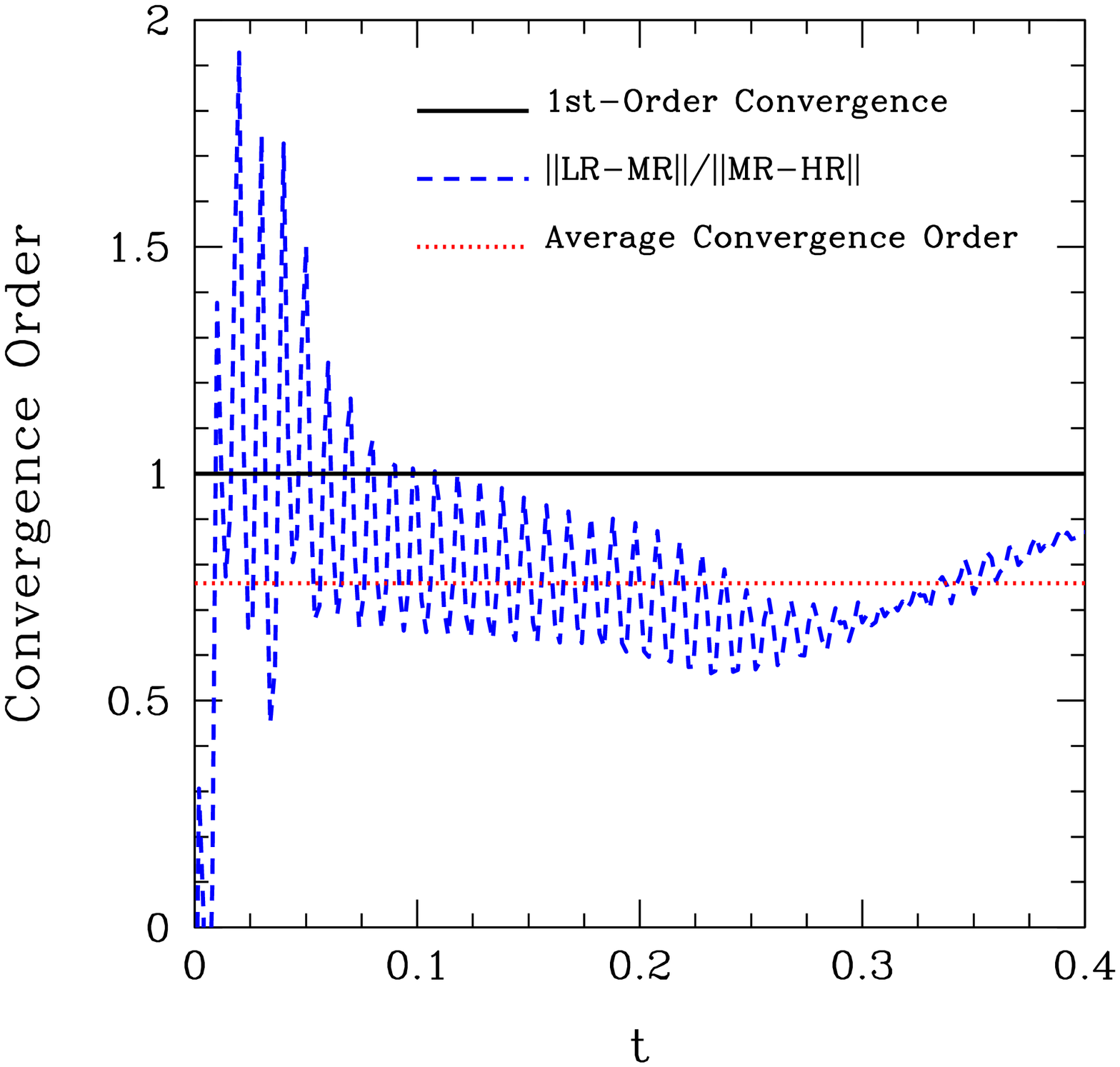}
  \hskip 0.25cm
  \includegraphics[width=0.32\textwidth]{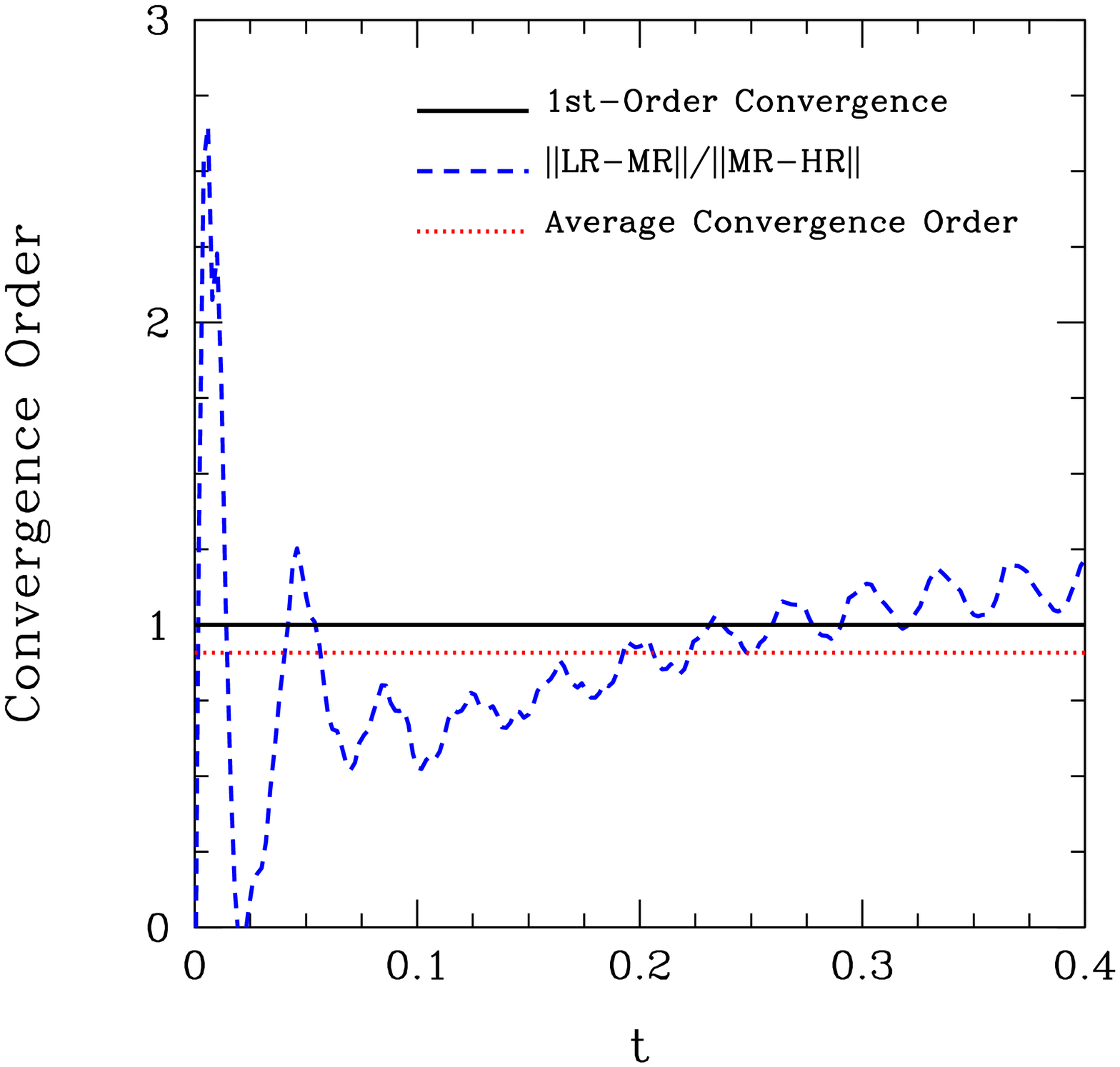}
  \caption{{\em Convergence Tests.} We report the convergence order as a
    function of time for the one-dimensional shocktube tests we have
    performed in flat spacetime (Sec.~\ref{sec:ShocktubeTests}) for
    $\sigma=10^6$, $\sigma=10$ and for a nonuniform power-law
    conductivity with $\sigma_0=10^6$ and $\gamma=9$, respectively. The
    convergence order as given by taking the average of the convergence
    order over time for these three different tests (from left to right)
    is $0.87$, $0.76$, $0.91$, respectively.}
  \label{fig:convergenceshocktubes}
\end{figure*}

We have also studied the convergence of our method for initial data
containing a contact discontinuity, namely the shocktubes presented in
(Sec.~\ref{sec:ShocktubeTests}). The fact that the location of the shock
strongly depends on the resolution adopted in the initial data makes it
impossible to recover exactly first order convergence at shocks, since
the test is not well-defined. Therefore, the convergence order of our
scheme drops to almost first order, as expected for the numerical
techniques adopted in this code as is shown in
Fig.~\ref{fig:convergenceshocktubes}. The tests involve the evolution of
a shocktube in the high conductivity regime $\sigma=10^6$ (left panel),
in the low conductivity regime (central panel) $\sigma=10$ and the
evolution of a nonuniform power-law conductivity with $\sigma_0=10^6$
and $\gamma=9$ (right panel). 

Since the initial data for the last two tests are not consistent with the
choice of conductivity, the errors introduced already at initial data
spoils the convergence order computed at the first timesteps. The
solution subsequently relaxes to a stable configuration with a more or
less constant convergence order. Thus, it makes sense to compute the average convergence order only after $t=0.05$ when the solution has already relaxed to a consistent solution of the equations. For these last three tests we have
considered resolutions of 100, 200 and 400 points along the x axis. The
average convergence order of the shocktube tests in the uniform high/low
conductivity regime is $0.87$/$0.76$, while adopting a conductivity
power-law prescription yields an average convergence order of $0.91$. The
reconstruction scheme adopted in this set of tests is linear with a
monotonized central-differences slope limiter function
(MC)~\cite{Leveque2002}).

\end{appendix}

\bibliography{aeireferences.bib,local.bib}

\end{document}